\documentclass[12pt,reqno]{amsart}

\usepackage{amsmath}
\usepackage{amssymb}
\usepackage{amsthm}
\usepackage{stmaryrd}
\usepackage{hyperref}
\usepackage{cleveref}
\usepackage{geometry}
\usepackage{arydshln}
\usepackage[linesnumbered,ruled]{algorithm2e}
\usepackage{color}
\usepackage{cases}
\allowdisplaybreaks

\usepackage{cite}      
\usepackage{natbib}    
\usepackage{subcaption}
\usepackage{graphicx}

\def \i {{\mathrm{i}}}
\newcommand{\ri}{{\rm i}}
\newcommand{\kr}{k_\rho}
\newcommand{\bs}[1]{\mathbf{#1}}
\newcommand{\bd}[1]{\mathbf{#1}}
\newcommand{\bh}[1]{\mathbb{#1}}
\newcommand{\J}[1]{\bh{J}_{#1}}

\newcommand{\lbr}{\llbracket}
\newcommand{\rbr}{\rrbracket}

\makeatletter
\@addtoreset{equation}{section}

\@addtoreset{figure}{section}

\@addtoreset{table}{section}

\makeatother

\theoremstyle{definition}

\theoremstyle{remark}
\newtheorem{rem}{Remark}

\title[On the derivation of LMDGs]{On the dyadic Green's functions for Maxwell equations in layered media}
\author{Heng Yuan\textsuperscript{1}}
\thanks{\textsuperscript{1} LCSM (MOE), School of Mathematics and Statistics,
	Hunan Normal University, Changsha, Hunan 410081, P. R. China; \texttt{yheng@hunnu.edu.cn}.
	This author acknowledges financial support provided by the Foundation Sciences of Hunan Province (Grant No.~2024JC0006).}

\author{Chunzhi Xiang\textsuperscript{2,1}}
\thanks{\textsuperscript{1} LCSM, Ministry of Education, School of Mathematics and Statistics,	Hunan Normal University, Changsha, Hunan 410081, P. R. China. \textsuperscript{2} School of Mathematical Sciences, Shanghai Jiao Tong University,
	Shanghai 200240, P. R. China; \texttt{chunzhixiang@sjtu.edu.cn}.}

\author{Wenzhong Zhang\textsuperscript{3}}
\thanks{\textsuperscript{3} Suzhou Institute for Advanced Research,
	University of Science and Technology of China,
	Suzhou, Jiangsu 215127, P. R. China; \texttt{wenzhong@ustc.edu.cn}.
	This author acknowledges financial support provided by NSF of China (Grant No.~12201603).}

\author{Bo Wang\textsuperscript{1,*}}
\thanks{\textsuperscript{*} Corresponding author.
	\textsuperscript{1} LCSM (MOE), School of Mathematics and Statistics,
	Hunan Normal University, Changsha, Hunan 410081, P. R. China; \texttt{bowang@hunnu.edu.cn}.
	This author acknowledges financial support provided by NSF of China
	(Grant Nos.~12371394, 12022104), the Major Program of Xiangjiang Laboratory
	(Grant No.~22XJ01013), and the Foundation Sciences of Hunan Province
	(Grant No.~2024JC0006).}

\begin{document}
	
	\maketitle
	
	\begin{abstract}
		In this paper, two formulations for the dyadic Green's functions (DGFs) of Maxwell's equations in layered media are revisited and compared. The first formulation, derived using the TE/TM decomposition \cite{TETM2002}, is well established and has been widely used in the engineering community. The second formulation, based on potential functions and a matrix basis, has recently been proposed in the development of a fast multipole method for electromagnetic problems in layered media. Given the complexity of these formulations, a detailed comparison is necessary to validate the new formulation and its methodology. We significantly simplify the derivation of the second formulation and demonstrate that it is equivalent to the TE/TM formulation. Nevertheless, the derivation of the second formulation is more straightforward due to the usage of the vector potential. The matrix basis used in the second formula unveils the algebraic nature of the TE/TM decomposition. This mathematical understanding paves the way for the investigation of other vector wave equations (e.g., elastic wave equation) in layered media.
		
	\end{abstract}

	\section{Introduction}
	
	Efficient and accurate computation of electromagnetic fields in layered media is of both fundamental scientific interest and considerable engineering value. It underpins a wide range of applications, including the analysis and design of microstrip circuits and antennas \cite{Michalski1997,Mosig1988,Chew1999}, geophysical exploration \cite{Ward1988,Constable2010,Key2009}, and the design of metamaterials \cite{Pendry2000,Fang2005,Poddubny2013}. Numerical methods based on the discretization of integral equations\cite{gibson2021method,colton2013integral,jin2015theory} primarily rely on the dyadic Green's functions (DGFs) in layered media. These functions satisfy the jump conditions on the interfaces and the out-going condition in the far field, enabling a substantial degree of freedom reduction in the resulted boundary integral equations.
	However, the computation of layered media dyadic Green's functions (LMDGFs) presents an intrinsic challenge: its $3\times3$ tensor structure necessitates the simultaneous solution of $9$ coupled components at all media interfaces. 
	
	In the derivation of the LMDGFs, the problem of a dipole over a half-space was first solved by
	Sommerfeld \cite{sommerfeld1909ausbreitung} in 1909 using the Hertz vector potential. After that, numerous authors have derived DGFs for layered media including both isotropic and anisotropic \cite{kay1963theory, kong1972electromagnetic, ali1979electromagnetic,sphicopoulos1985dyadic, barkeshli1992electromagnetic,TETM2002}. Note that Sommerfeld-type integrals appearing in the dyadic Green's functions of the potentials converge more rapidly and are easier to accelerate than those associated with the field forms that are, in effect, obtained by differentiation of the potentials. Various boundary integral equations have been derived using the potential kernels which leads to research interest in the formulations of the DGFs of the potentials (cf.   \cite{erteza1969nonuniqueness,michalski1990electromagnetic,michalski2003scalar}).
	

	
	A well-known strategy used in the derivation of the DGFs for a layered medium with arbitrary number of layers is the TE/TM decomposition (cf. \cite{TETM2002,chew2006matrix,xiong2009newly}). Namely, the electromagnetic field are decomposed into horizontally (xOy-plane) and vertically(z-axis) components, yielding coupled equations for Transverse Electric (TE) and Transverse Magnetic (TM) waves. By introducing a rotated coordinate system $(\hat{\bf u},\hat{\bf v},\hat{\bf z})$ in the frequency domain \cite{TETM2002}, the Maxwell equations are reduced into two independent scalar Helmholtz equations with decoupled interface conditions. This achieves a closed form for the LMDGs in the Fourier spectral domain. Since TE/TM decomposition is a property unique to electromagnetic waves, such methods are difficult to generalize to other vector wave equations such as elastic wave equation.
	
	Recently, we proposed a matrix basis \cite{bo2022maxwellDGF} and then established an algebraic framework to decouple the interface conditions. The matrix basis consists of nine $3\times3$ matrices  $\{\J{k}\}_{k=1}^9$ in the frequency domain. We prove that the vector potential can be represented $\J{1},\J{2},\cdots,\J{5}$ with radially symmetric coefficients in the Fourier $k_x-k_y$ plane. As in the TE/TM decomposition approach, we are able to reduce the problem into two independent Helmholtz problems in layered medium which can be solved by using the concept of general reflection and transmission coefficients, see Appendix A. The solutions lead to closed form of the LMDGFs in the Fourier spectral domain and then integral representations in physical domain can be obtained by applying inverse Fourier transform.

	The matrix basis approach does not rely on the TE/TM decomposition. The Helmholtz equations with decoupled interface conditions are directly derived for the coefficients of the electromagnetic fields under this new basis. Therefore, it is more straightforward and can be generalized to other vector wave equations (e.g., elastic wave equations) in layered media. Due to the complexity of the LMDGFs, it is difficult to clearly see the differences of the two approaches and the advantages of the new approach. In this paper, we significantly simplify the derivation in \cite{bo2022maxwellDGF} and present a detailed comparison of the two approaches. We will show that: i) The 9 basis matrices used in the matrix basis approach are essentially the interaction tensor basis of the rotated coordinate system used in the TE/TM decomposition; ii) The two decoupled scalar Helmholtz equations derived in both approaches are formally identical.
	Therefore, we not only unveil the equivalence of the two approaches but also demonstrate the algebraic nature of the TE/TM decomposition. 
	
	The rest of the paper is organized as follows. Section \ref{sect_notation} provides a systematic review of the TE/TM decomposition for solving Maxwell’s equations with point source in layered media. We elaborate on the TE/TM decomposition and the treatment of the interface conditions within layered media. In Section \ref{sect_fsgf}, a substantially simplified version of the derivation originally proposed in \cite{bo2022maxwellDGF} is presented. A comparative analysis between the resulting formulation and the well-known TE/TM formulation is also provided. The vector potential Green's function derived by our approach and the reliable C++ package Strata for a 3-layer problem are numerically compared in Section \ref{num_test}. Finally, Section \ref{sect_conclusion} concludes the paper with discussions on our on-going research on elastic wave equation.

	\section{Computation of the dyadic Green's function of Maxwell's equation in layered media using TE/TM decomposition}\label{sect_notation}
	In this section, we review the derivation of the DGFs of Maxwell’s equations in multilayered media using the TE/TM decomposition (cf. \cite{TETM2002}). The source and target are assumed to be located in any layers of the media.

	\subsection{The TE/TM decomposition of Maxwell’s equations}
	Consider the time-harmonic Maxwell’s equation 
	\begin{align}
		\nabla\times\mathbf{E}(\bs r)=&\ri\omega\mu\mathbf{H}(\bs r),\label{E}\\ 
		\nabla\times\mathbf{H}(\bs r)=&-\ri\omega\varepsilon\mathbf{E}(\bs r)+\mathbf{J}(\bs r),\label{H}\\
		\nabla\cdot\mathbf D(\bs r)=&{\rho}(\bs r),\\
		\nabla\cdot\mathbf B(\bs r)=& 0,
	\end{align}
	where $\mathbf{E}(\bs r)$, $\mathbf{H}(\bs r)$, $\mathbf{D}(\bs r)$, and $\mathbf{B}(\bs r)$ are the electric and magnetic field and flux densities, and $\rho(\bs r)$, $\mathbf{J}(\bs r)$ are the volume charge density and current density of any external source at given point $\mathbf r=(x, y, z)\in\mathbb R^3$. The time dependence $e^{-\ri\omega t}$ with angular frequency $\omega$ is assumed, and $\varepsilon,\mu$ denote the dielectric permittivity and magnetic permeability in a homogeneous medium. The electric and magnetic flux densities $\mathbf{D}$, $\mathbf{B}$ are related to the field intensities $\mathbf{E}$, $\mathbf{H}$ via constitutive relations:
	\begin{equation*}
		\mathbf D=\varepsilon\mathbf E,\quad\mathbf B=\mu\mathbf H.
	\end{equation*}

	Given any vector field $\mathbf A=[A_x, A_y, A_z]^{\rm T}$ in $\mathbb R^3$, we can decompose it into its horizontal ($x-y$ plane) and vertical components (along $z$ direction) as
	$$\mathbf A=\mathbf A_S+\mathbf A_z,\quad {\rm with}\quad \mathbf A_S=A_x\hat{\mathbf x}+A_y\hat{\mathbf{y}},\quad \mathbf A_z=A_z\hat{\mathbf{z}},$$
	where, $\hat{\mathbf x},\hat{\mathbf y},\hat{\mathbf z}$ are unit vectors in $(x, y, z)$-coordintes.
	Define horizontal gradient operator 
	\begin{equation*}
		\nabla_S:=\mathbf{\hat{x}}\displaystyle\frac{\partial}{\partial x}+\mathbf{\hat{y}}\displaystyle\frac{\partial}{\partial y}.
	\end{equation*}
	Then, the curl operator can be decomposed into
	\begin{equation}\label{xuanduEs}
		\nabla\times\mathbf{A}=(\nabla_S+\hat{\mathbf{z}}\frac{\partial}{\partial z})\times (\mathbf A_S+\mathbf A_z)=\nabla_S\times\mathbf{A}_S+\nabla_S\times\mathbf{A}_z+\hat{\mathbf{z}}\times\displaystyle\frac{\partial\mathbf{A}_S}{\partial z}.
	\end{equation}
	Apparently, the first term at the right end of the above formula is vertical (parallel to the $z$-direction), and the other two terms are horizontal (perpendicular to the $z$-axis).

	Applying the above deomposition to the electricmagnetic fields 
	\begin{equation*}
		\mathbf{E}=\mathbf{E}_S+\mathbf{E}_z,\quad\mathbf{H}=\mathbf{H}_S+\mathbf{H}_z,
	\end{equation*}
	the Farady's law \eqref{E} can be rewritten as
	\begin{equation*}
		\nabla_S\times\mathbf{E}_S+\nabla_S\times\mathbf{E}_z+\hat{\mathbf{z}}\times\displaystyle\frac{\partial\mathbf{E}_S}{\partial z}=\ri\omega\mu\mathbf H_S+\ri\omega\mu\mathbf H_{z}.
	\end{equation*}
	Matching the transverse and longitudinal components in the above equation gives
	\begin{subequations}\label{HEuniaxially}
		\begin{numcases}{}
			\nabla_S\times\mathbf{E}_S=\ri\omega\mu\mathbf{H}_{z},\label{HEuniaxially1}\\
			\displaystyle\nabla_S\times\mathbf{E}_z+\hat{\mathbf{z}}\times\frac{\partial\mathbf{E}_S}{\partial z}=\ri\omega\mu\mathbf{H}_S.\label{HEuniaxially2}
		\end{numcases}
	\end{subequations}
	Similarly, Ampere's law \eqref{H} has decomposition
	\begin{subequations}\label{EHuniaxially}
		\begin{numcases}{}
			\nabla_S\times\mathbf{H}_S=-\ri\omega\varepsilon\mathbf{E}_{z}+\mathbf{J}_z,\label{EHuniaxially1}\\
			\displaystyle\nabla_S\times\mathbf{H}_z+\hat{\mathbf{z}}\times\frac{\partial\mathbf{H}_S}{\partial z}=-\ri\omega\mathbf{E}_S+\mathbf{J}_S.\label{EHuniaxially2}
		\end{numcases}
	\end{subequations}
	By using the equations \eqref{HEuniaxially1}, \eqref{EHuniaxially1} in \eqref{EHuniaxially2} and \eqref{HEuniaxially2}, respectively, we can eliminate the vertical components to get
	\begin{subequations}\label{horizontalpart}
		\begin{numcases}{}
			\displaystyle\nabla_S\times\nabla_S\times\mathbf{H}_S-k^2\mathbf H_S-\ri\omega\varepsilon\hat{\mathbf{z}}\times\frac{\partial\mathbf{E}_S}{\partial z}=\nabla_S\times\mathbf{J}_z,\label{horizontalpart1}\\
			\displaystyle\nabla_S\times\nabla_S\times\mathbf{E}_S-k^2\mathbf E_S+\ri\omega\mu\hat{\mathbf{z}}\times\frac{\partial\mathbf{H}_S}{\partial z}=\ri\omega\mu\mathbf{J}_S,  \label{horizontalpart2}
		\end{numcases}
	\end{subequations}
	where $k=\omega\sqrt{\varepsilon\mu}$ is the wave number.
	Therefore, we have extracted the equations for the horizontal components $\mathbf H_S, \mathbf E_S$ out of the full Maxwell's equations. The vertical components $\mathbf H_z, \mathbf E_z$ then can be obtained by substituting the results back into \eqref{HEuniaxially1} and \eqref{EHuniaxially1}.

	In order to derive an analytic expression for the solutions of the Maxwell's equations, we shall use a partial Fourier transform to the decomposed Maxwell equations \eqref{HEuniaxially1}, \eqref{EHuniaxially1} and \eqref{horizontalpart1}-\eqref{horizontalpart2}. 
	The Fourier transform for any given vector field $\mathbf A$ in the $x-y$ plane and its inverse are defined as
	\begin{equation}\label{Fouriertransform}
		\begin{split}
			\mathcal{F}[\mathbf A]=&\int_{-\infty}^{+\infty}\int_{-\infty}^{+\infty}\mathbf A(x,y,z)e^{-\ri\mathbf{k}_{\rho}\cdot(\boldsymbol{\rho}-\boldsymbol\rho')}dxdy=:\hat{\mathbf A}(k_x, k_y,z),\\
			\mathcal{F}^{-1}[\hat{\mathbf A}]=&\displaystyle\frac{1}{4\pi^2}\int_{-\infty}^{+\infty}\int_{-\infty}^{+\infty}\hat{\mathbf A}(k_x, k_y,z)e^{\ri\mathbf{k}_{\rho}\cdot(\boldsymbol{\rho}-\boldsymbol\rho')}dk_xdk_y,
		\end{split}
	\end{equation}
	where
	\begin{equation*}
		\boldsymbol{\rho}=x\hat{\mathbf{x}}+y\hat{\mathbf{y}},\quad\mathbf{k}_{\rho}=k_x\mathbf{\hat{x}}+k_y\mathbf{\hat{y}},\quad k_{\rho}=|\mathbf{k}_{\rho}|=\sqrt{k_x^2+k_y^2}.
	\end{equation*}
	Given any scalar function $\psi(\mathbf r)$ and vector field $\mathbf A(\mathbf r)$, we have
	\begin{equation}\label{F2}
		\begin{aligned}
			&\mathcal{F}[\nabla_S\times\mathbf{A}]=\ri\mathbf{k}_{\rho}\times\hat{\mathbf{A}},\quad\mathcal{F}[\nabla_S \psi]=\ri\mathbf{k}_{\rho}\hat{\psi}.
		\end{aligned}
	\end{equation}
	
	Applying the Fourier transform to the equations in \eqref{horizontalpart} leads to an ODE system
	\begin{subequations}\label{horizontalpart_frequency}
		\begin{numcases}{}
			\displaystyle\ri \mathbf{k}_{\rho}\times\ri \mathbf{k}_{\rho}\times\widehat{\mathbf{H}}_S-k^2\widehat{\mathbf H}_S-\ri\omega\varepsilon\hat{\mathbf{z}}\times\frac{\partial\widehat{\mathbf{E}}_S}{\partial z}=\ri \mathbf{k}_{\rho}\times\hat{\mathbf{J}}_z,\\
			\displaystyle\ri \mathbf{k}_{\rho}\times\ri \mathbf{k}_{\rho}\times\widehat{\mathbf{E}}_S-k^2\widehat{\mathbf E}_S+\ri\omega\mu\hat{\mathbf{z}}\times\frac{\partial\widehat{\mathbf{H}}_S}{\partial z}=\ri\omega\mu\hat{\mathbf{J}}_S.   
		\end{numcases}
	\end{subequations}
	Left-multiplying $\mathbf{\hat{z}}\times$ on both sides of  \eqref{horizontalpart_frequency} and applying the identity 
	\begin{equation}
		\label{vectorfieldidentity1}
		(\mathbf{A}\times\mathbf{B})\cdot\mathbf{C}=\mathbf{A}\cdot (\mathbf{B}\times\mathbf{C}),\quad
		\mathbf{A}\times (\mathbf{B}\times \mathbf{C})=(\mathbf{A}\cdot \mathbf{C})\mathbf{B}-(\mathbf{A}\cdot \mathbf{B})\mathbf{C}.
	\end{equation}
	gives
	\begin{subequations}\label{horizontalpart_frequency_equivalent}
		\begin{numcases}{}
			\displaystyle\ri\omega\varepsilon\frac{\partial\widehat{\mathbf{E}}_S}{\partial z}=-\hat{\mathbf{z}}\times\ri \mathbf{k}_{\rho}\times\ri \mathbf{k}_{\rho}\times\widehat{\mathbf{H}}_S+k^2\hat{\mathbf{z}}\times\widehat{\mathbf H}_S+\hat{\mathbf{z}}\times\ri \mathbf{k}_{\rho}\times\hat{\mathbf{J}}_z,\\
			\displaystyle\ri\omega\mu\frac{\partial\widehat{\mathbf{H}}_S}{\partial z}=\hat{\mathbf{z}}\times\ri \mathbf{k}_{\rho}\times\ri \mathbf{k}_{\rho}\times\widehat{\mathbf{E}}_S-k^2\hat{\mathbf{z}}\times\widehat{\mathbf E}_S-\ri\omega\mu\hat{\mathbf{z}}\times\hat{\mathbf{J}}_S.   
		\end{numcases}
	\end{subequations}
	Note that 
	\begin{equation*}
		\hat{\mathbf{z}}\times\ri \mathbf{k}_{\rho}\times\ri \mathbf{k}_{\rho}\times{\mathbf{A}}=-\mathbf{k}_{\rho}\otimes\mathbf{k}_{\rho}^{\rm T}[{\mathbf{A}}\times\hat{\mathbf{z}}],\quad \mathbf{A}=\widehat{\mathbf{E}}_S,\widehat{\mathbf{H}}_S,
	\end{equation*}
	where $\otimes$ is Kronecker product. 
	Equation \eqref{horizontalpart_frequency_equivalent} can be written as
	\begin{subequations}
		\begin{numcases}{}
			\displaystyle\frac{\partial\widehat{\mathbf{E}}_S}{\partial z}=-\frac{1}{\ri\omega\varepsilon}\left[k^2{\bh I}-\mathbf k_{\rho}\otimes\mathbf{ k}_{\rho}^{\rm T}\right][\widehat{\mathbf{H}}_S\times\mathbf{\hat{z}}]+\frac{\hat{J}_z}{\omega\varepsilon}\mathbf{k}_{\rho}, \label{Esf}\\[1.2ex]
			\displaystyle\frac{\partial\widehat{\mathbf{H}}_S}{\partial z}=-\frac{1}{\ri\omega\mu}\left[k^2{\bh I}-\mathbf k_{\rho}\otimes\mathbf{ k}_{\rho}^{\rm T}\right][\mathbf{\hat{z}}\times\widehat{\mathbf{E}}_S]-\mathbf{\hat{z}}\times\hat{\mathbf{J}}_S,\label{Hsf1}
		\end{numcases}
	\end{subequations}
	where ${\bh I}$ is the $3\times 3$ identity matrix. 
	Similarly, the Fourier transform of the equations \eqref{HEuniaxially1} and \eqref{EHuniaxially1} gives the vertical components ${\mathbf H}_z, {\mathbf E}_z$ in the frequency domain, i.e.,
	\begin{equation}\label{EHzf}
		\widehat{\mathbf{H}}_{z}=\frac{1}{\omega\mu}\mathbf{k}_{\rho}\times\widehat{\mathbf{E}}_S,\quad
		\widehat{\mathbf{E}}_{z}=-\frac{1}{\omega\varepsilon}\mathbf{k}_{\rho}\times\widehat{\mathbf{H}}_S+\frac{\hat{\mathbf{J}}_z}{\ri\omega\varepsilon}. 
	\end{equation}
	
	In order to decouple the equations in \eqref{Esf}-\eqref{Hsf1}, we introduce orthogonal basis
	\begin{equation}\label{new_coordinate_uv}
		\mathbf{\hat{u}}=\frac{k_x}{k_{\rho}}\mathbf{\hat{x}}+\frac{k_y}{k_{\rho}}\mathbf{\hat{y}}=\frac{1}{k_{\rho}}\mathbf{k}_{\rho}, \quad\mathbf{\hat{v}}=-\frac{k_y}{k_{\rho}}\mathbf{\hat{x}}+\frac{k_x}{k_{\rho}}\mathbf{\hat{y}}=\frac{1}{k_{\rho}}\mathbf{\hat{z}}\times\mathbf{k}_{\rho}=\mathbf{\hat{z}}\times \mathbf{\hat{u}},
	\end{equation}
	in the $x-y$ plane. Let
	\begin{equation}\label{Newzb}
		\widehat{\mathbf{E}}_S=\mathbf{\hat{u}}V^e+\mathbf{\hat{v}}V^h,\quad \widehat{\mathbf{H}}_S\times\mathbf{\hat{z}}=\mathbf{\hat{u}}I^e+\mathbf{\hat{v}}I^h,\quad \hat{\mathbf{J}}_S=\mathbf{\hat{u}}\hat{J}_u+\mathbf{\hat{v}}\hat{J}_v,
	\end{equation}
	be the expressions of  the horizontal components $\widehat{\mathbf{E}}_S,\widehat{\mathbf{H}}_S,\hat{\mathbf{J}}_S$, where $\{V^e, V^h\}$, $\{I^e, I^h\}$ and $\{\hat J_ u,\hat J_{v}\}$ are their components in the $\hat{\mathbf u}$ and $\hat{\mathbf v}$ directions, respectively.
	Applying the cross product $\times\mathbf{\hat{z}}$ on both sides of  \eqref{Hsf1}, and then using the  fact 
	\begin{equation*}
		\big[(\mathbf{\hat{u}}\otimes\mathbf{\hat{u}}^{\rm T})(\mathbf{\hat{z}}\times\widehat{\mathbf{E}}_S)\big]\times\mathbf{\hat{z}}=(\mathbf{\hat{v}}\otimes\mathbf{\hat{v}}^{\rm T})\widehat{\mathbf{E}}_S,
	\end{equation*} 
	gives
	\begin{equation}\label{Hsf}
		\begin{aligned}
			\frac{\partial}{\partial z}[\widehat{\mathbf{H}}_S\times\mathbf{\hat{z}}]=&-\frac{1}{\ri\omega\mu}\left[k^2{\bh I}-k_{\rho}^2\hat{\mathbf u}\otimes\hat{\mathbf u}^{\rm T}\right][\mathbf{\hat{z}}\times\widehat{\mathbf{E}}_S]\times\mathbf{\hat{z}}-\hat{\mathbf{J}}_S\\
			=&-\frac{1}{\ri\omega\mu}\left[k^2{\bh I}-  k_{\rho}^2\mathbf{\hat{v}}\otimes\mathbf{\hat{v}}^{\rm T}\right]\widehat{\mathbf{E}}_S-\hat{\mathbf{J}}_S .
		\end{aligned}
	\end{equation}
	Substituting the expression into \eqref{Esf} and \eqref{Hsf}, we have
	\begin{subequations}
		\begin{numcases}{}
			\displaystyle\frac{\partial V^e}{\partial z}\hat{\mathbf u}+\frac{\partial V^h}{\partial z}\hat{\mathbf v}=-\frac{k^2}{\ri\omega\varepsilon}I^h\hat{\mathbf v}-\Big[\frac{k^2- k_{\rho}^2}{\ri\omega\varepsilon}I^e-\frac{k_{\rho}\hat J_z}{\omega\varepsilon}\Big]\hat{\mathbf u},\label{Euvcoord}\\
			\displaystyle	\frac{\partial I^e}{\partial z}\hat{\mathbf u}+\frac{\partial I^h}{\partial z}\hat{\mathbf v}=-\Big(\frac{k^2}{\ri\omega\mu}V^e+\hat J_u\Big)\hat{\mathbf u}-\Big[\frac{k^2- k_{\rho}^2}{\ri\omega\mu}V^h+\hat J_v\Big]\hat{\mathbf v},\label{Huvcoord}
		\end{numcases}
	\end{subequations}
	where the facts
	$$({\mathbf a}\otimes{\mathbf a}^{\rm T}){\mathbf a}=\mathbf a(\mathbf a\cdot\mathbf a)={\mathbf a},\quad ({\mathbf a}\otimes{\mathbf a}^{\rm T}){\mathbf b}=\mathbf a(\mathbf b\cdot\mathbf a)={\mathbf 0},$$
	for any orthogonal unit vectors ${\mathbf a},{\mathbf b}$ have been used. 
	Matching the components in \eqref{Euvcoord}-\eqref{Huvcoord}, we obtained two decoupled systems
	\begin{subequations}
		\begin{numcases}{}
			\frac{\partial V^e}{\partial z}=-\frac{ k_z^2}{\ri\omega\varepsilon}I^e+\frac{k_{\rho}}{\omega\varepsilon}\hat{J}_z, \label{DVe}\\[1.2ex]
			\frac{\partial I^e}{\partial z}=-\frac{k^2}{\ri\omega\mu}V^e-\hat{J}_u, \label{DIe}
		\end{numcases}
	\end{subequations}
	and
	\begin{subequations}
		\begin{numcases}{}
			\frac{\partial V^h}{\partial z}=-\frac{k^2}{\ri\omega\varepsilon}I^h, \label{DVh}\\[1.2ex]
			\frac{\partial I^h}{\partial z}=-\frac{ k_z^2}{\ri\omega\mu}V^h-\hat{J}_v. \label{DIh}
		\end{numcases}
	\end{subequations}
	where $k_z=\sqrt{k^2-k_{\rho}^2}$ with branch cut $\Im(k_z)\ge0$. Apparently, we can further reduce them into two Helmholtz equations as follows
	\begin{equation}\label{IeVh_helmholtzEq}
		\frac{\partial^2 I^e}{\partial z^2}+k_z^2I^e=\ri k_{\rho}\hat J_z-\frac{\partial \hat J_u}{\partial z},\qquad
		\frac{\partial^2 V^h}{\partial z^2}+k_z^2V^h=\frac{k^2}{\ri\omega\varepsilon}\hat J_v.
	\end{equation}
	The other two components $V^e$ and $I^h$ can be calculated via
	\begin{equation}\label{IhVe_relatedtoIeVh}
		V^e=-\frac{\ri\omega\mu}{k^2}\left[\frac{\partial I^e}{\partial z}+\hat{J}_u\right],\quad I^h=-\frac{\ri\omega\varepsilon}{k^2}\frac{\partial V^h}{\partial z}.
	\end{equation}
	Substituting \eqref{Newzb} into \eqref{EHzf} and using the identities in \eqref{vectorfieldidentity1} to simplify the results gives
	\begin{equation*}
		\widehat{E}_{z}=-\frac{k_{\rho}I^e}{\omega\varepsilon}+\frac{\hat{J}_z}{\ri\omega\varepsilon},\quad\widehat{H}_{z}=\frac{k_{\rho}V^h}{\omega\mu}.
	\end{equation*}
	As a result, the electromagnetic fields in the Fourier spectral domain are given by
	\begin{equation}\label{TETMformulation}
		\widehat{\mathbf{E}}=\widehat{\mathbf{E}}_S+\widehat{\mathbf{E}}_z=\mathbf{\hat{u}}V^e+\mathbf{\hat{v}}V^h-\mathbf{\hat{z}}\frac{\ri k_{\rho}I^e-\hat{J}_z}{\ri\omega\varepsilon},\quad
		\widehat{\mathbf{H}}=\widehat{\mathbf{H}}_S+\widehat{\mathbf{H}}_z=\mathbf{\hat{v}}I^e-\mathbf{\hat{u}}I^h+\mathbf{\hat{z}}\frac{\ri k_{\rho}V^h}{\ri\omega\mu}.
	\end{equation}
	
	\subsection{The DGFs of the Maxwell’s equations in free space}
	Consider the electromagnetic fields generated by a directed $\frac{1}{\ri\omega\mu}$-Hertz dipole of current moment located at $\bs r'$. The dyadic Green's function $\mathbf G_{\mathbf E}^{\hat{\bs t}}(\bs r, \bs r')$, $\mathbf G_{\mathbf H}^{\hat{\bs t}}(\bs r, \bs r')$ satisfy Maxwell equation \eqref{E}-\eqref{H} with
	\begin{equation}\label{Hertzdipolesource_free}
		\mathbf J=\dfrac{\delta(\bs r- \bs r')}{\ri\omega\mu}\hat{\mathbf t},\quad\hat{\mathbf t}=\hat{\mathbf x}, \hat{\mathbf y}, \hat{\mathbf z}
	\end{equation}
	or accordingly \eqref{horizontalpart_frequency} with
	\begin{equation}
		\hat{\mathbf J}=\dfrac{1}{\ri\omega\mu}\delta(z-z')\hat{\mathbf t},\quad\hat{\mathbf t}=\hat{\mathbf x}, \hat{\mathbf y}, \hat{\mathbf z}
	\end{equation}
	in the frequency domain. Here, $\delta(\bs r-\bs r')$ and $\delta(z-z')$ are the 3-dimensional and 1-dimensional Dirac functions, respectively. Following the analysis above, we have solutions given by
	\begin{equation}\label{Green_EH_frequency_vector}
		\widehat{\mathbf G}_{\mathbf E}^{\hat{\bs t}}=\mathbf{\hat{u}}\widehat G_{V^e}^{\hat{\bs t}}+\mathbf{\hat{v}}\widehat G_{V^h}^{\hat{\bs t}}-\mathbf{\hat{z}}\frac{\ri k_{\rho}\widehat G_{I^e}^{\hat{\bs t}}}{\ri\omega\varepsilon}-\mathbf{\hat{z}}\dfrac{\mathbf{\hat{z}}\cdot\mathbf{\hat{t}}}{k^2}\delta(z-z'),\quad \widehat{\mathbf{G}}_{\mathbf{H}}^{\hat{\bs t}}=\mathbf{\hat{v}}\widehat G_{I^e}^{\hat{\bs t}}-\mathbf{\hat{u}}\widehat G_{I^h}^{\hat{\bs t}}+\mathbf{\hat{z}}\frac{\ri k_{\rho}\widehat G_{V^h}^{\hat{\bs t}}}{\ri\omega\mu},
	\end{equation}
	while the coefficients $\widehat{G}_{I^e}^{\hat{\bs t}}, \widehat{G}_{V^h}^{\hat{\bs t}}$ satisfy
	\begin{equation}\label{Green_VhIe}
		\begin{split}
			\frac{\partial^2 \widehat G_{I^e}^{\hat{\bs t}}}{\partial z^2}+k_z^2\widehat G_{I^e}^{\hat{\bs t}}=&\dfrac{k_{\rho}}{\omega\mu}\hat{\mathbf z}\cdot\hat{\mathbf t}\delta(z-z')-\dfrac{\hat{\mathbf u}\cdot\hat{\mathbf t}}{\ri\omega\mu}\delta'(z-z'),\\
			\frac{\partial^2 \widehat G_{V^h}^{\hat{\bs t}}}{\partial z^2}+k_z^2\widehat G_{V^h}^{\hat{\bs t}}=&-\hat{\mathbf v}\cdot\hat{\mathbf t}\delta(z-z'),
		\end{split}
	\end{equation}
	and the other two can be calculated via
	\begin{equation}\label{Green_VeIh}
		\widehat G_{V^e}^{\hat{\bs t}}=-\dfrac{\ri\omega\mu}{k^2}\left[\frac{\partial \widehat G_{I^e}^{\hat{\bs t}}}{\partial z}+\frac{\hat{\mathbf u}\cdot\hat{\mathbf t}}{\ri\omega\mu}\delta(z-z')\right],\quad 
		\widehat G_{I^h}^{\hat{\bs t}}=-\dfrac{\ri\omega\varepsilon}{k^2}\frac{\partial \widehat G_{V^h}^{\hat{\bs t}}}{\partial z}.
	\end{equation}
	
	Obviously, equations in \eqref{Green_VhIe} are the Fourier transform of the 3-D Helmholtz equations. In fact, the Sommerfeld identity 
	\begin{equation*}
		G^f(\bs r, \bs r')= \frac{e^{\ri k |\bs r-\bs r'|}}{4\pi|\bs r-\bs r'|}=\frac{\ri}{8\pi^2}\int_{-\infty}^{+\infty}\int_{-\infty}^{+\infty}\frac{e^{\ri k_{z}|z-z'|}}{k_z}e^{\ri\mathbf{k}_{\rho}\cdot(\boldsymbol{\rho}-\boldsymbol{\rho}')}dk_xdk_y
	\end{equation*}
	shows that the Fourier transform of the 3-D Helmholtz Green's function $G^f(\bs r, \bs r')$ is given by
	\begin{equation}\label{HelmholtzGreenfunfreespectra}
		\widehat{G}^{f}(k_{\rho},z,z')=\frac{\ri e^{\ri k_{z}|z-z'|}}{2k_{z}},
	\end{equation}
	which satisfies
	\begin{equation}\label{helmholtz_Green_frequency}
		\partial_{zz}\widehat{G}^{f}(k_{\rho},z,z')+k_z^2\widehat{G}^{f}(k_{\rho},z,z')=-\delta(z-z').
	\end{equation}
	Taking the derivative with respect to $z$ on both sides of \eqref{helmholtz_Green_frequency}, we can simply verify that $\phi(k_{\rho}, z, z'):=\partial_z\widehat{G}^f(k_{\rho}, z, z')$ satisfies
	\begin{equation}
		\partial_{zz}\phi(k_{\rho},z,z')+k_z^2\phi(k_{\rho},z,z')=-\delta'(z-z').
	\end{equation}
	Consequently, the principle of superposition implies that equations \eqref{Green_VhIe} have solutions:
	\begin{equation}\label{Green_VeIh_frequency_vector}
		\widehat G_{I^e}^{\hat{\bs t}}=-\dfrac{1}{\ri\omega\mu}\left[{\ri k_{\rho}}\hat{\mathbf z}\cdot\hat{\mathbf t}-{\hat{\mathbf u}\cdot\hat{\mathbf t}}\partial_z\right]\widehat{G}^{f}(k_{\rho},z,z'),\quad
		\widehat G_{V^h}^{\hat{\bs t}}=\hat{\mathbf v}\cdot\hat{\mathbf t}\widehat{G}^{f}(k_{\rho},z,z').
	\end{equation}
	Substituting \eqref{Green_VeIh_frequency_vector} into \eqref{Green_VeIh}, and using equation \eqref{helmholtz_Green_frequency} to eliminate the second order derivative gives
	\begin{equation}\label{Green_VhIe_frequency_vector}
		\widehat G_{V^e}^{\hat{\bs t}}=\dfrac{1}{k^2}\left[{\ri k_{\rho}}\hat{\mathbf z}\cdot\hat{\mathbf t}{\partial_z}+k_z^2\hat{\mathbf u}\cdot\hat{\mathbf t}\right]\\ \widehat{G}^{f}(k_{\rho},z,z'),\quad 
		\widehat G_{I^h}^{\hat{\bs t}}=-\dfrac{\ri\omega\varepsilon}{k^2}\hat{\mathbf v}\cdot\hat{\mathbf t}{\partial_z}\widehat{G}^{f}(k_{\rho},z,z').
	\end{equation}
	Further, using the expressions \eqref{Green_VeIh_frequency_vector}-\eqref{Green_VhIe_frequency_vector} in \eqref{Green_EH_frequency_vector} and then applying the identity $({\bs a}\otimes{\bs b}^{\rm T}){\bs c}={\bs a}({\bs b}\cdot{\bs c})$, we obtain
	\begin{equation}\label{Green_EH_frequency_vector_equivalent}
		\begin{split}
			\widehat{\mathbf G}_{\mathbf E}^{\hat{\bs t}}=&\dfrac{1}{k^2}\left[k_z^2\hat{\mathbf u}\otimes\hat{\mathbf u}^{\rm T}+k^2\hat{\mathbf v}\otimes\hat{\mathbf v}^{\rm T}-\ri k_{\rho}\left({\ri k_{\rho}}\hat{\mathbf z}\otimes\hat{\mathbf z}^{\rm T}-({\hat{\mathbf z}\otimes\hat{\mathbf u}^{\rm T}}+\hat{\mathbf u}\otimes\hat{\mathbf z}^{\rm T})\partial_z\right)\right](\widehat{G}^{f}\hat{\bf t})\\
			&-\dfrac{\mathbf{\hat{z}}\otimes\mathbf{\hat{z}}^{\rm T}}{k^2}\mathbf{\hat{t}}\delta(z-z'),\\
			\widehat{\mathbf{G}}_{\mathbf{H}}^{\hat{\bs t}}=&-\dfrac{1}{\ri\omega\mu}\left[{\ri k_{\rho}}(\hat{\mathbf v}\otimes\hat{\mathbf z}^{\rm T}-\hat{\mathbf z}\otimes\hat{\mathbf v}^{\rm T})-{\hat{\mathbf v}\otimes\hat{\mathbf u}^{\rm T}}\partial_z+\hat{\mathbf u}\otimes\hat{\mathbf v}^{\rm T}{\partial_z}\right](\widehat{G}^{f}\hat{\bf t}).
		\end{split}
	\end{equation}
	Therefore, the dyadic Green functions 
	\begin{equation}
		\widehat{\bh G}_{\mathbf E}^f=\left[\widehat{\mathbf G}_{\mathbf E}^{\hat{\bs x}},\widehat{\mathbf G}_{\mathbf E}^{\hat{\bs y}},\widehat{\mathbf G}_{\mathbf E}^{\hat{\bs z}}\right],\quad\widehat{\bh G}_{\mathbf H}^f=\left[\widehat{\mathbf G}_{\mathbf H}^{\hat{\bs x}},\widehat{\mathbf G}_{\mathbf H}^{\hat{\bs y}},\widehat{\mathbf G}_{\mathbf H}^{\hat{\bs z}}\right]
	\end{equation}
	are given by
	\begin{align}
		\widehat{\bh G}_{\mathbf E}^f=&\dfrac{1}{k^2}\left[{\ri k_{\rho}}(\hat{\mathbf u}\otimes\hat{\mathbf z}^{\rm T}+\hat{\mathbf z}\otimes\hat{\mathbf u}^{\rm T}){\partial_z}+k_z^2\hat{\mathbf u}\otimes\hat{\mathbf u}^{\rm T}+k^2\hat{\mathbf v}\otimes\hat{\mathbf v}^{\rm T}+k_{\rho}^2\hat{\mathbf z}\otimes\hat{\mathbf z}^{\rm T}\right]\widehat{G}^{f}\nonumber\\
		&-\dfrac{\mathbf{\hat{z}}\otimes\mathbf{\hat{z}}^{\rm T}}{k^2}\delta(z-z'),\label{Green_E_free_frequency}\\
		\widehat{\bh{G}}_{\mathbf{H}}^f=&-\dfrac{1}{\ri\omega\mu}\left[{\ri k_{\rho}}(\hat{\mathbf v}\otimes\hat{\mathbf z}^{\rm T}-\hat{\mathbf z}\otimes\hat{\mathbf v}^{\rm T})-({\hat{\mathbf v}\otimes\hat{\mathbf u}^{\rm T}}-{\hat{\mathbf u}\otimes\hat{\mathbf v}^{\rm T}})\partial_z\right]\widehat{G}^{f},\label{Green_H_free_frequency}
	\end{align}
	Moreover, using Eq.\eqref{helmholtz_Green_frequency} to replace $\delta(z-z')$ in \eqref{Green_E_free_frequency} and simplifying the resulting equation by the identity 
	$${\bh I}=\mathbf{\hat{u}}\otimes\mathbf{\hat{u}}^{\rm T}+\mathbf{\hat{v}}\otimes\mathbf{\hat{v}}^{\rm T}+\mathbf{\hat{z}}\otimes\mathbf{\hat{z}}^{\rm T},$$
	we obtain
	\begin{equation}\label{Green_E_free_frequency_equivalent}
		\widehat{\bh G}_{\mathbf E}^f={\bh I}\widehat{G}^{f}+\dfrac{1}{k^2}\left[\hat{\mathbf z}\otimes\hat{\mathbf z}^{\rm T}{\partial_{zz}}+{\ri k_{\rho}}(\hat{\mathbf u}\otimes\hat{\mathbf z}^{\rm T}+\hat{\mathbf z}\otimes\hat{\mathbf u}^{\rm T}){\partial_z}-k_{\rho}^2\hat{\mathbf u}\otimes\hat{\mathbf u}^{\rm T}\right]\widehat{G}^{f}.
	\end{equation}
	
	The expressions \eqref{Green_H_free_frequency} and \eqref{Green_E_free_frequency_equivalent} are the spectral-domain DGFs of the Maxwell's equations. We now transform them back to the physical domain. By the following calculations
	\begin{equation}\label{fourieridentity}
		\begin{aligned}
			\mathcal{F}\left[\nabla{G}^{f}\right]=&\left(\ri k_{\rho}\hat{\bf u}+\hat{\bf z}\partial_z\right)\widehat{G}^{f},\quad\mathcal{F}\left[\nabla\times({G}^{f}\hat{\bf u})\right]=\hat{\bf v}\partial_z\widehat{G}^{f}\\
			\mathcal{F}\left[\nabla\times({G}^{f}\hat{\bf v})\right]=&\left(\ri k_{\rho}\hat{\bf z}-\hat{\bf u}\partial_z\right)\widehat{G}^{f},\quad\mathcal{F}\left[\nabla\times({G}^{f}\hat{\bf z})\right]=-\ri k_{\rho}\hat{\bf v}\widehat{G}^{f}\\
			\mathcal{F}[\nabla\nabla{G}^{f}]=&\left(\ri k_{\rho}\hat{\bf u}+\hat{\bf z}\partial_z\right)\otimes\left(\ri k_{\rho}\hat{\bf u}^{\rm T}+\hat{\bf z}^{\rm T}\partial_z\right)\widehat{G}^{f}\\
			=&\left[\hat{\mathbf z}\otimes\hat{\mathbf z}^{\rm T}{\partial_{zz}}+{\ri k_{\rho}}(\hat{\mathbf u}\otimes\hat{\mathbf z}^{\rm T}+\hat{\mathbf z}\otimes\hat{\mathbf u}^{\rm T}){\partial_z}-k_{\rho}^2\hat{\mathbf u}\otimes\hat{\mathbf u}^{\rm T}\right]\widehat{G}^{f}\\
			\mathcal{F}[\nabla\times({G}^{f}{\bh I})]=&\sum_{\hat{\bf t}=\hat{\bf u},\hat{\bf v},\hat{\bf z}}\mathcal{F}[\nabla\times({G}^{f}\hat{\bf t})]\otimes\hat{\bf t}^{\rm T}\\
			=&\left[-{\ri k_{\rho}}(\hat{\mathbf v}\otimes\hat{\mathbf z}^{\rm T}-\hat{\mathbf z}\otimes\hat{\mathbf v}^{\rm T})+({\hat{\mathbf v}\otimes\hat{\mathbf u}^{\rm T}}-{\hat{\mathbf u}\otimes\hat{\mathbf v}^{\rm T}})\partial_z\right]\widehat{G}^{f}.
		\end{aligned}
	\end{equation}
	the DGFs given by \eqref{Green_H_free_frequency} and \eqref{Green_E_free_frequency_equivalent} can be written as 
	\begin{equation}
		\widehat{\bh{G}}_{\bs E}^{f}=\mathcal{F}\left[\left({\bh I}+\dfrac{\nabla\nabla}{k^2}\right){G}^{f}\right],\quad \widehat{\bh{G}}_{\bs H}^{f}=\dfrac{1}{\ri\omega\mu}\mathcal{F}\left[\nabla\times({G}^{f}{\bh I})\right],
	\end{equation}
	that is the DGFs in physical domain are given by
	\begin{align}\label{GEandGH}
		\bh{G}_{\bs E}^f(\bs{r},\bs{r}') = \ri\omega\left({\bh I} + \frac{\nabla\nabla}{k^2} \right){\bh G}_{\bs A}^f(\bs r, \bs r') ,
		\quad \bh{G}_{\bs H}^f(\bs{r},\bs{r}') = \frac{1}{\mu} \nabla \times {\bh G}_{\bs A}^f(\bs r, \bs r').
	\end{align}
	and
	\begin{equation}\label{GAF}
		\bh{G}_{\bs A}^f(\bs r,\bs r') = \frac{1}{\ri \omega} G^f(\bs r,\bs r') {\bh I},
	\end{equation}
	They are the solution of Maxwell’s equation with a directed $\frac{1}{\ri\omega\mu}$-Hertz dipole of current moment \eqref{Hertzdipolesource_free} and the Lorenz gauge (cf \cite{caiwei2013elecphe}).
	
	As discussed in this section, the essence of the derivation lies in reducing the vector electromagnetic problem (Eqs. \eqref{E}–\eqref{H}) to a set of scalar equations (Eqs. \eqref{DVe}–\eqref{DIh}). The procedure is summarized as follows:
	\begin{itemize}
		\item {\bf TE/TM decomposition:} The electromagnetic fields are decomposed into horizontal and vertical components and the Maxwell's equations are rewritten into \eqref{HEuniaxially}–\eqref{EHuniaxially}.
		\item {\bf Derivative reduction:} Apply the partial Fourier transform  to obtain ODE systems \eqref{Esf}–\eqref{Hsf1} in the frequency domain.
		\item {\bf ODE system decoupling:} Introduce the rotated coordinate system \eqref{new_coordinate_uv}  to decouple the ODE system \eqref{Esf} and \eqref{Hsf1} to Helmholtz equations \eqref{IeVh_helmholtzEq}-\eqref{IhVe_relatedtoIeVh}.
	\end{itemize}
	Due to the symmetry of the free space problem, the TE/TM decomposition is not necessary here. However, we shall not have full symmetry if we consider a layered media problem which is the main topic of this paper. The TE/TM decomposition will be used to decompose the solution into a symmetric component in the horizontal plane and a vertical component along the axis of the layered structure.

	\subsection{The DGFs of the Maxwell’s equations in layered media}\label{tem_layer}
	Consider a multi-layered medium with $L+1$ layers along the z-direction, where the interface is located at $z = d_{\ell}$ for $\ell = 0, 1, \ldots, L-1$. We will also use the notation $d_{-1}=+\infty$, $d_{L}=-\infty$ throughout this paper. Each layer has a dielectric constant and magnetic permeability denoted by $\{\varepsilon_{\ell},\mu_{\ell}\}_{\ell=0}^{L}$. 
	Denote by $k_\ell = \omega\sqrt{ \varepsilon_\ell \mu_\ell}$
	the wave numbers in the $\ell$-th layer for all $\ell=0,1,\cdots,L$.
	Assume there is a directed Hertzian current source \eqref{Hertzdipolesource_free} in the $\ell'$-th layer, i.e., $d_{\ell'}<z'<d_{\ell'-1}$.
	The Green's functions $\bs G_{\bs E}^{\hat{\bs t}}(\bs r, \bs r')$ and $\bs G_{\bs H}^{\hat{\bs t}}(\bs r, \bs r')$ corresponding to the point source are piecewise smooth vector fields that satisfy
	\begin{equation}\label{maxwelleqlayer}
		\begin{split}
			\nabla\times \bs G_{\bs E}^{\hat{\bs t}}(\bs r, \bs r')=&\ri \omega\mu_{\ell}\bs G_{\bs H}^{\hat{\bs t}}(\bs r, \bs r'), \quad d_{\ell}<z<d_{\ell-1},\quad \ell=0,\cdots,L,\\
			\nabla\times \bs G_{\bs H}^{\hat{\bs t}}(\bs r, \bs r')=&-\ri\omega\varepsilon_{\ell}\bs G_{\bs E}^{\hat{\bs t}}(\bs r, \bs r')+\frac{\delta(\bs r, \bs r')}{\ri\omega\mu_{\ell'}}\widehat{\bs t}.  \quad d_{\ell}<z<d_{\ell-1},\quad \ell=0,\cdots,L,
		\end{split}
	\end{equation}
	in each layer and the transmission conditions
	\begin{equation}\label{interfacecondphysical_vector}
		\llbracket\bs n\times \bs G_{\bs E}^{\hat{\bs t}}\rrbracket=\bs 0, \quad \llbracket\bs{n}\times \bs G_{\bs H}^{\hat{\bs t}}\rrbracket=\bs 0,\quad \llbracket\bs{n}\cdot \varepsilon\bs G_{\bs E}^{\hat{\bs t}}\rrbracket=0,\quad \llbracket\bs{n}\cdot \mu\bs G_{\bs H}^{\hat{\bs t}}\rrbracket=0,
	\end{equation}
	across the interfaces $\{z=d_{\ell}\}_{\ell=0}^{L-1}$, where $\bs n=\hat{\bf z}$, and $\llbracket\cdot\rrbracket$ represents the jump of the piece-wise smooth function across the interface, i.e. 
	\begin{equation*}
		\llbracket f\rrbracket = \lim_{z\to d_{\ell}+0}f - \lim_{z\to d_{\ell}-0}f.
	\end{equation*}
	We would like to mention that the transmission conditions on the normal components of the fields are redundant. They are implied by the  transmission conditions on the tangential components and the Maxwell equations.

	Apparently, we can apply the Fourier transform and TE/TM decomposition technique to the Maxwell's equations \eqref{maxwelleqlayer} in each layer. According to the formulas in \eqref{IeVh_helmholtzEq}–\eqref{IhVe_relatedtoIeVh} and \eqref{Green_EH_frequency_vector}, and following an analysis similar to that for the free-space problem, the Fourier transform of the Green's functions in each layer can be represented as
	\begin{align}
		&\widehat{\bs G}_{\bs E}^{\hat{\bs t}}(k_x, k_y,z, z')=\mathbf{\hat{u}}{V}^{e,\hat{\bs t}}_{\ell\ell'}+\mathbf{\hat{v}}{V}^{h,\hat{\bs t}}_{\ell\ell'}-\mathbf{\hat{z}}\frac{\ri k_{\rho}{I}^{e,\hat{\bs t}}_{\ell\ell'}}{\ri\omega_{\ell}}-\dfrac{\mathbf{\hat{z}}\otimes\mathbf{\hat{z}}^{\rm T}}{k_{\ell}^2}\hat{\bf t}\delta(z-z'),\quad d_{\ell}<z<d_{\ell-1},\label{Hat_E_layered}\\
		&  \widehat{\bs G}_{\bs H}^{\hat{\bs t}}(k_x, k_y,z, z')=\mathbf{\hat{v}}{I}^{e,\hat{\bs t}}_{\ell\ell'}-\mathbf{\hat{u}}{I}^{h,\hat{\bs t}}_{\ell\ell'}+\mathbf{\hat{z}}\frac{k_{\rho}{V}^{h,\hat{\bs t}}_{\ell\ell'}}{\omega\mu_{\ell}},\quad d_{\ell}<z<d_{\ell-1},\label{Hat_H_layered}  
	\end{align}
	where $\{{V}^{e,\hat{\bs t}}_{\ell\ell'}, {I}^{e,\hat{\bs t}}_{\ell\ell'}, {V}^{h,\hat{\bs t}}_{\ell\ell'}, {I}^{h,\hat{\bs t}}_{\ell\ell'}\}_{\ell=0}^L$
	are functions defined in each layer and satisfy 
	\begin{equation}\label{IeVh_vector_layered}
		\begin{split}
			&\dfrac{\partial^2 {I}^{e,\hat{\bs t}}_{\ell\ell'}}{\partial z^2}+k_{\ell z}^2{I}^{e,\hat{\bs t}}_{\ell\ell'}=\dfrac{k_{\rho}}{\omega\mu_{\ell'}}\hat{\mathbf z}\cdot\hat{\mathbf t}\delta(z-z')-\dfrac{\hat{\mathbf u}\cdot\hat{\mathbf t}}{\ri\omega\mu_{\ell'}}\delta'(z-z'),\\
			& \dfrac{\partial^2 {V}^{h,\hat{\bs t}}_{\ell\ell'}}{\partial z^2}+k_{\ell z}^2{V}^{h,\hat{\bs t}}_{\ell\ell'}=-\hat{\mathbf v}\cdot\hat{\mathbf t}\delta(z-z'),
		\end{split}
	\end{equation}
	and
	\begin{equation}\label{VeIh_vector_layered}
		{V}^{e,\hat{\bs t}}_{\ell\ell'}=-\dfrac{\ri\omega\mu_{\ell}}{k_{\ell}^2}\left[\dfrac{\partial {I}^{e,\hat{\bs t}}_{\ell\ell'}}{\partial z}+\frac{\hat{\mathbf u}\cdot\hat{\mathbf t}}{\ri\omega\mu_{\ell'}}\delta(z-z')\right],\quad {I}^{h,\hat{\bs t}}_{\ell\ell'}=-\dfrac{\ri\omega_{\ell}}{k_{\ell}^2}\dfrac{\partial {V}^{h,\hat{\bs t}}_{\ell\ell'}}{\partial z},
	\end{equation}
	for $d_{\ell}<z<d_{\ell-1}$. Throughout this paper, we use notation
	\begin{equation}
		k_{\ell z}=\sqrt{k_{\ell}^2-k_{\rho}^2},
	\end{equation}
	with branch cut $\Im(k_{\ell z})\ge0$, subscripts $\ell$ and $\ell'$ denote the indices of the target and source layers, respectively.

	Now, we use the interface conditions \eqref{interfacecondphysical_vector} to derive coupled equations for $\{{I}^{e,\hat{\bs t}}_{\ell\ell'}\}_{\ell=0}^L$ and $\{{V}^{h,\hat{\bs t}}_{\ell\ell'}\}_{\ell=0}^L$. Only the interface conditions on the tangential components will be used as the other two are redundant. The frequency domain counterparts of the first two equations in \eqref{interfacecondphysical_vector} are given by
	\begin{equation}\label{interfacecondfreq}
		\lbr\bs n\times \widehat{\bs G}_{\bs E}^{\hat{\bs t}}\rbr=\bs 0, \quad \lbr\bs{n}\times \widehat{\bs G}_{\bs H}^{\hat{\bs t}}\rbr=\bs 0,\quad{\rm at}\quad z=d_{\ell},
	\end{equation}
	for all $\ell=0, 1, \cdots, L-1$.
	Using the expression \eqref{Hat_E_layered} in the interface conditions for $\widehat{\bs G}_{\bs E}^{\hat{\bs t}}$, we obtain
	\begin{equation*}
		\hat{\bf z} \times \Big({V}^{e,\hat{\bs t}}_{\ell-1,\ell'}\hat{\bf u}+{V}^{h,\hat{\bs t}}_{\ell-1,\ell'}\hat{\bf v}-\frac{k_{\rho}}{\omega_{\ell-1}}{I}^{e,\hat{\bs t}}_{\ell-1,\ell'}\hat{\bf z}-{V}^{e,\hat{\bs t}}_{\ell\ell'}\hat{\bf u}-{V}^{h,\hat{\bs t}}_{\ell\ell'}\hat{\bf v}+\frac{k_{\rho}}{\omega_{\ell}}{I}^{e,\hat{\bs t}}_{\ell\ell'}\hat{\bf z}\Big)=0,
	\end{equation*}
	i.e.
	\begin{equation*}
		{V}^{e,\hat{\bs t}}_{\ell-1,\ell'}\hat{\bf v}-{V}^{h,\hat{\bs t}}_{\ell-1,\ell'}\hat{\bf u}-{V}^{e,\hat{\bs t}}_{\ell\ell'}\hat{\bf v}+{V}^{h,\hat{\bs t}}_{\ell\ell'}\hat{\bf u}=0,\quad \ell=1, 2, \cdots, L.
	\end{equation*}
	Apparently, the above interface conditions are decoupled. Define piece-wise smooth functions \begin{equation*}
		\begin{split}
			{V}^{e,\hat{\bs t}}(k_x, k_y, z, z') ={V}^{e,\hat{\bs t}}_{\ell\ell'}(k_x, k_y, z, z') ,\quad {I}^{e,\hat{\bs t}}(k_x, k_y, z, z') ={I}^{e,\hat{\bs t}}_{\ell\ell'}(k_x, k_y, z, z') ,\quad d_{\ell}<z<d_{\ell-1},\\
			{V}^{h,\hat{\bs t}}(k_x, k_y, z, z') ={V}^{h,\hat{\bs t}}_{\ell\ell'}(k_x, k_y, z, z') ,\quad {I}^{h,\hat{\bs t}}(k_x, k_y, z, z') ={I}^{h,\hat{\bs t}}_{\ell\ell'}(k_x, k_y, z, z') ,\quad d_{\ell}<z<d_{\ell-1},
		\end{split}
	\end{equation*}
	We have interface conditions for $V^{e,\hat{\bs t}}$ and $V^{h,\hat{\bs t}}$ as follows
	\begin{equation}\label{VEVHinterfacecondv}
		\llbracket {V}^{e,\hat{\bs t}} \rrbracket =0, \quad \llbracket {V}^{h,\hat{\bs t}} \rrbracket =0,\quad z=d_{\ell},\quad\ell=0, 1, \cdots, L-1.
	\end{equation}
	Similarly, the transimission conditions for $\widehat{\bs G}_{\bs H}^{\hat{\bs t}}$ gives us
	\begin{equation*}
		\hat{\bf z} \times \Big(I^{e,\hat{\bs t}}_{\ell-1,\ell'}\hat{\bf v}-{I}^{h,\hat{\bs t}}_{\ell-1,\ell'}\hat{\bf u}+\frac{k_{\rho}}{\omega\mu_{\ell-1}}{V}^{h,\hat{\bs t}}_{\ell-1,\ell'}\hat{\bf z}-{I}^{e,\hat{\bs t}}_{\ell\ell'}\hat{\bf v}+{I}^{h,\hat{\bs t}}_{\ell\ell'}\hat{\bf u}-\frac{k_{\rho}}{\omega\mu_{\ell}}{V}^{h,\hat{\bs t}}_{\ell\ell'}\hat{\bf z}\Big)=0,
	\end{equation*}
	i.e.
	\begin{equation}\label{IEIHinterfacecondv}
		\llbracket {I}^{h,\hat{\bs t}} \rrbracket =0,\quad \llbracket {I}^{e,\hat{\bs t}} \rrbracket =0,\quad{\rm at}\quad z=d_{\ell},\quad\ell=0, 1, \cdots, L-1.
	\end{equation}
	Further, the definition \eqref{VeIh_vector_layered} combined with the interface condition on $I^{e,\hat{\bs t}}$, $V^{h,\hat{\bs t}}$ in \eqref{VEVHinterfacecondv},\eqref{IEIHinterfacecondv} gives
	\begin{equation}\label{IEVHinterfacecondv}
		\left\llbracket \dfrac{1}{\mu}\dfrac{\partial {V}^{h,\hat{\bs t}}}{\partial z}\right\rrbracket=0,\ \left\llbracket \dfrac{1}{\varepsilon}\dfrac{\partial {I}^{e,\hat{\bs t}}}{\partial z}\right\rrbracket=0,\quad{\rm at}\quad z=d_{\ell},\quad\ell=0, 1, \cdots, L-1.
	\end{equation}
	In summary, we obtain two interface problems 
	\begin{equation}\label{Ie_layered}
		\begin{cases}
			\dfrac{\partial^2 {I}^{e,\hat{\bs t}}}{\partial z^2}+k_{\ell z}^2{I}^{e,\hat{\bs t}}=\dfrac{k_{\rho}}{\omega\mu_{\ell'}}\hat{\mathbf z}\cdot\hat{\mathbf t}\delta_{\ell\ell'}\delta(z-z')-\dfrac{\hat{\mathbf u}\cdot\hat{\mathbf t}}{\ri\omega\mu_{\ell'}}\delta_{\ell\ell'}\delta'(z-z'),\quad d_{\ell}<z<d_{\ell-1},\\[1.2ex]
			\llbracket {I}^{e,\hat{\bs t}}\rrbracket=0,\quad \left\llbracket \dfrac{1}{}\dfrac{\partial {I}^{e,\hat{\bs t}}}{\partial z}\right\rrbracket=0,\quad {\rm at}\;\; z=d_{\ell},\;\;\ell=0, 1, \cdots, L-1,
		\end{cases}
	\end{equation}
	and
	\begin{equation}\label{Vh_layered}
		\begin{cases}
			\dfrac{\partial^2 {V}^{h,\hat{\bs t}}}{\partial z^2}+k_{\ell z}^2{V}^{h,\hat{\bs t}}=-\hat{\mathbf v}\cdot\hat{\mathbf t}\delta_{\ell\ell'}\delta(z-z'),\quad d_{\ell}<z<d_{\ell-1}\\[1.2ex]
			\llbracket {V}^{h,\hat{\bs t}}\rrbracket=0,\quad \left\llbracket \dfrac{1}{\mu}\dfrac{\partial {V}^{h,\hat{\bs t}}}{\partial z}\right\rrbracket=0,\quad {\rm at}\;\; z=d_{\ell},\;\;\ell=0, 1, \cdots, L-1,
		\end{cases}
	\end{equation}
	for $I^{e,\hat{\bs t}}$ and $V^{h,\hat{\bs t}}$ with outgoing condition on the upper and lower most layers.

	To solve the problems \eqref{Ie_layered}-\eqref{Vh_layered}, we introduce the following interface problems:
	\begin{equation}\label{G_f1_frequency}
		\begin{cases}
			\dfrac{\partial^2 \widehat{G}_{1}(k_x, k_y, z, z')}{\partial z^2}+k_{\ell z}^2\widehat{G}_{1}(k_x, k_y, z, z')=-\delta_{\ell\ell'}\delta(z-z'),\quad d_{\ell}<z<d_{\ell-1},\\
			\llbracket \widehat{G}_{1}(k_x, k_y, z, z')\rrbracket=0,\quad \left\llbracket \dfrac{1}{\mu}\dfrac{\partial \widehat{G}_{1}(k_x, k_y, z, z')}{\partial z}\right\rrbracket=0,\quad{\rm at}\quad z=\{d_{\ell}\}_{\ell=0}^{L-1},
		\end{cases}
	\end{equation}
	\begin{equation}\label{G_f2_frequency}
		\begin{cases}
			\dfrac{\partial^2 \widehat{G}_{2}(k_x, k_y, z, z')}{\partial z^2}+k_{\ell z}^2\widehat{G}_{2}(k_x, k_y, z, z')=-\delta_{\ell\ell'}\delta(z-z'),\quad d_{\ell}<z<d_{\ell-1},\\
			\llbracket \widehat{G}_{2}(k_x, k_y, z, z')\rrbracket=0,\quad \left\llbracket \dfrac{1}{\varepsilon}\dfrac{\partial \widehat{G}_{2}(k_x, k_y, z, z')}{\partial z}\right\rrbracket=0,\quad{\rm at}\quad z=\{d_{\ell}\}_{\ell=0}^{L-1},
		\end{cases}
	\end{equation}
	\begin{equation}\label{G_f3_frequency}
		\begin{cases}
			\dfrac{\partial^2 \widehat{G}_{3}(k_x, k_y, z, z')}{\partial z^2}+k_{\ell z}^2\widehat{G}_{3}(k_x, k_y, z, z')=-\delta_{\ell\ell'}\delta'(z-z'),\quad d_{\ell}<z<d_{\ell-1},\\
			\llbracket \widehat{G}_{3}(k_x, k_y, z, z')\rrbracket=0,\quad \left\llbracket \dfrac{1}{\varepsilon}\dfrac{\partial \widehat{G}_{3}(k_x, k_y, z, z')}{\partial z}\right\rrbracket=0,\quad{\rm at}\quad z=\{d_{\ell}\}_{\ell=0}^{L-1}.
		\end{cases}
	\end{equation}
	It is clear that problems \eqref{G_f1_frequency} and \eqref{G_f2_frequency} are the Fourier transform of the Helmholtz equation with point source in layered media. Analytic solution can be obtained, see appendix \ref{helmholtzGreenFun} for detailed derivation. Moreover, the solution for the problem \eqref{G_f3_frequency} can be derived from the solution of \eqref{G_f2_frequency}. In fact, taking derivative with respect to $z'$ on both sides of equation and the jump conditions in \eqref{G_f2_frequency} gives
	\begin{equation}
		(\partial_{zz}  + k_{\ell z}^2) (\partial_{z'}\widehat{G}_2(k_x, k_y, z, z'))=\delta'(z-z')
	\end{equation}
	and
	\begin{equation}
		\llbracket\partial_{z'}\widehat{G}_2(k_x, k_y, z, z')\rrbracket=0,\quad \Big\llbracket\frac{1}{\varepsilon}\partial_z\partial_{z'}\widehat{G}_2(k_x, k_y, z, z')\Big\rrbracket=0,
	\end{equation}
	which implies that
	$$\widehat{G}_3(k_x, k_y, z, z')=-\partial_{z'}\widehat{G}_2(k_x, k_y, z, z').$$ 
	In general, we have
	\begin{equation}\label{threedensity}
		\begin{split}
			\widehat{G}_{1}(k_x, k_y, z, z')&=\delta_{\ell\ell'}\widehat{G}^f(k_{\rho}, z, z')+\frac{\ri}{2 k_{\ell' z}}\sum_{\ast,\star=\uparrow,\downarrow}\phi_{\ell\ell'}^{\ast\star}(\kr){\mathcal Z}_{\ell\ell'}^{\ast\star}(k_{\rho}, z,z'),\\
			\widehat{G}_{2}(k_x, k_y, z, z')&=\delta_{\ell\ell'}\widehat{G}^f(k_{\rho}, z, z')+\frac{\ri}{2k_{\ell' z}}\sum_{\ast,\star=\uparrow,\downarrow}\psi_{\ell\ell'}^{\ast\star}(\kr){\mathcal Z}_{\ell\ell'}^{\ast\star}(k_{\rho}, z,z'),\\
			\widehat{G}_{3}(k_x, k_y, z, z')&=\delta_{\ell\ell'}\partial_{z}\widehat{G}^f(k_{\rho}, z, z')+\frac{1}{2}\sum_{\ast,\star=\uparrow,\downarrow}s_{2}^{\star}\psi_{\ell\ell'}^{\ast\star}(\kr){\mathcal Z}_{\ell\ell'}^{\ast\star}(k_{\rho}, z,z'),
		\end{split}
	\end{equation}
	for $d_{\ell}<z<d_{\ell-1}$ where the formulations for the exponential functions ${\mathcal Z}_{\ell\ell'}^{\ast\star}(k_{\rho}, z,z')$, the stable calculation of the densities $\phi^{\ast\star}_{\ell\ell'}(k_{\rho})$, $\psi_{\ell\ell'}^{\ast\star}(k_{\rho})$ are summarized in appendix \ref{helmholtzGreenFun}, and the sign 
	\begin{equation*}
		s_{2}^{\uparrow}=-1,\quad s_{2}^{\downarrow}=1,
	\end{equation*}
	comes from the derivative
	\begin{equation*}
		\partial_{z'}{\mathcal Z}_{\ell\ell'}^{\ast\star}(k_{\rho}, z,z')=\ri s_{2}^{\star}k_{\ell'z}{\mathcal Z}_{\ell\ell'}^{\ast\star}(k_{\rho}, z,z').
	\end{equation*}
	The Kronecker symbol $\delta_{\ell\ell'}$ is due to the fact that the free space component $\widehat G^f$ only exists in the $\ell'$-th layer.
	
	By the principle of superposition, we have
	\begin{equation*}
		{I}^{e,\hat{\bs t}}_{\ell\ell'}=-\dfrac{1}{\ri\omega\mu_{\ell'}}\left[\ri k_{\rho}\mathbf{\hat{z}}\cdot\hat{\mathbf t}\widehat{G}_{2}-\mathbf{\hat{u}}\cdot\hat{\mathbf t}\widehat{G}_{3}\right],\quad {V}^{h,\hat{\bs t}}=\mathbf{\hat{v}}\cdot\hat{\mathbf t}\widehat{G}_{1}.
	\end{equation*}
	Substituting into equation \eqref{VeIh_vector_layered} gives
	\begin{equation*}
		{I}^{h,\hat{\bs t}}_{\ell\ell'}=-\mathbf{\hat{v}}\cdot\hat{\mathbf t}\dfrac{\ri\omega_{\ell}}{k_{\ell}^2}\dfrac{\partial \widehat{G}_{1}}{\partial z},\quad {V}^{e,\hat{\bs t}}_{\ell\ell'}=\dfrac{\mu_{\ell}}{\mu_{\ell'}k_{\ell}^2}\left[\ri k_{\rho}\mathbf{\hat{z}}\cdot\hat{\mathbf t}\dfrac{\partial \widehat{G}_{2}}{\partial z}-\mathbf{\hat{u}}\cdot\hat{\mathbf t}\dfrac{\partial \widehat{G}_{3}}{\partial z}\right]-\hat{\bf u}\cdot\hat{\mathbf t}\dfrac{\delta(z-z')}{k_{\ell'}^2}.
	\end{equation*}
	Then, using these coefficients in \eqref{Hat_E_layered}-\eqref{Hat_H_layered} gives
	\begin{equation*}\label{Hat_E_layered_vector}
		\begin{aligned}
			\widehat{\bs G}_{\bs E}^{\hat{\bs t}}
			=&\dfrac{\mu_{\ell}}{\mu_{\ell'}k_{\ell}^2}\left[-\mathbf{\hat{u}}\otimes\mathbf{\hat{u}}^{\rm T}\dfrac{\partial \widehat{G}_{3}}{\partial z}+\ri k_{\rho}\mathbf{\hat{u}}\otimes\mathbf{\hat{z}}^{\rm T}\dfrac{\partial \widehat{G}_{2}}{\partial z}+\ri k_{\rho}\mathbf{\hat{z}}\otimes\mathbf{\hat{u}}^{\rm T}\widehat{G}_{3}+k_{\rho}^2\mathbf{\hat{z}}\otimes\mathbf{\hat{z}}^{\rm T}\widehat{G}_{2}\right]\hat{\mathbf t}\\
			&+(\mathbf{\hat{v}}\otimes\mathbf{\hat{v}}^{\rm T})\hat{\mathbf t}\widehat{G}_{1}-\left[\mathbf{\hat{u}}\otimes\mathbf{\hat{u}}^{\rm T}+\mathbf{\hat{z}}\otimes\mathbf{\hat{z}}^{\rm T}\right]\hat{\mathbf t}\frac{\delta(z-z')}{k_{\ell}^2},\\
			\widehat{\bs G}_{\bs H}^{\hat{\bs t}}
			=&-\dfrac{1}{\ri\omega\mu_{\ell'}}\left[-\mathbf{\hat{v}}\otimes\mathbf{\hat{u}}^{\rm T}\widehat{G}_{3}+\ri k_{\rho}\mathbf{\hat{v}}\otimes\mathbf{\hat{z}}^{\rm T}\widehat{G}_{2}\right]\hat{\mathbf t}+\dfrac{1}{\ri\omega\mu_{\ell}}\left[-\mathbf{\hat{u}}\otimes\mathbf{\hat{v}}^{\rm T}\dfrac{\partial \widehat{G}_{1}}{\partial z}+\ri k_{\rho}\mathbf{\hat{z}}\otimes\mathbf{\hat{v}}^{\rm T}\widehat{G}_{1}\right]\hat{\mathbf t}.
		\end{aligned}
	\end{equation*}
	Further, the dyadic Green functions
	\begin{equation*}
		\widehat{\bh G}_{\bs E}=\begin{bmatrix}
			\widehat{\mathbf G}_{\mathbf E}^{\hat{\bs x}} & \widehat{\mathbf G}_{\mathbf E}^{\hat{\bs y}} & \widehat{\mathbf G}_{\mathbf E}^{\hat{\bs z}}
		\end{bmatrix},\quad\widehat{\bh G}_{\bs H}=\left[\widehat{\mathbf G}_{\mathbf H}^{\hat{\bs x}},\widehat{\mathbf G}_{\mathbf H}^{\hat{\bs y}},\widehat{\mathbf G}_{\mathbf H}^{\hat{\bs z}}\right]
	\end{equation*}
	have expressions 
	\begin{equation}\label{Hat_E_layeredv}
		\begin{aligned}
			\widehat{\bh G}_{\bs E}
			=&\dfrac{\mu_{\ell}}{\mu_{\ell'}k_{\ell}^2}\left[-\mathbf{\hat{u}}\otimes\mathbf{\hat{u}}^{\rm T}\dfrac{\partial \widehat{G}_{3}}{\partial z}+\ri k_{\rho}\mathbf{\hat{u}}\otimes\mathbf{\hat{z}}^{\rm T}\dfrac{\partial \widehat{G}_{2}}{\partial z}+\ri k_{\rho}\mathbf{\hat{z}}\otimes\mathbf{\hat{u}}^{\rm T}\widehat{G}_{3}+k_{\rho}^2\mathbf{\hat{z}}\otimes\mathbf{\hat{z}}^{\rm T}\widehat{G}_{2}\right]\\
			&+\mathbf{\hat{v}}\otimes\mathbf{\hat{v}}^{\rm T}\widehat{G}_{1}-\left[\mathbf{\hat{u}}\otimes\mathbf{\hat{u}}^{\rm T}+\mathbf{\hat{z}}\otimes\mathbf{\hat{z}}^{\rm T}\right]\frac{\delta(z-z')}{k_{\ell'}^2},\\
			\widehat{\bh G}_{\bs H}=&-\dfrac{1}{\ri\omega\mu_{\ell'}}\left[\ri k_{\rho}\mathbf{\hat{v}}\otimes\mathbf{\hat{z}}^{\rm T}\widehat{G}_{2}-\mathbf{\hat{v}}\otimes\mathbf{\hat{u}}^{\rm T}\widehat{G}_{3}\right]+\dfrac{1}{\ri\omega\mu_{\ell}}\left[\ri k_{\rho}\mathbf{\hat{z}}\otimes\mathbf{\hat{v}}^{\rm T}\widehat{G}_{1}-\mathbf{\hat{u}}\otimes\mathbf{\hat{v}}^{\rm T}\dfrac{\partial \widehat{G}_{1}}{\partial z}\right],
		\end{aligned}
	\end{equation}
	for $d_{\ell}<z<d_{\ell-1}$. Note that 
	\begin{equation}\label{helmholtz_Green_frequency_zzprime}
		\partial_{zz}\widehat{G}^f+k_{\ell z}^2\widehat{G}^f=-\delta(z-z'),
	\end{equation}
	Together with the expressions \eqref{threedensity}, we have
	\begin{equation*}
		\begin{split}
			\widehat{\bh G}_{\bs E}
			=&\delta_{\ell\ell'}\Big[\bh{I}+\frac{1}{k_{\ell}^2}(\hat{\mathbf z}\otimes\hat{\mathbf z}^{\rm T})\partial_{zz}+\frac{\mu_{\ell}}{\mu_{\ell'}k_{\ell}^2}[\ri k_{\rho}[(\hat{\mathbf u}\otimes\hat{\mathbf z}^{\rm T})+(\hat{\mathbf z}\otimes\hat{\mathbf u}^{\rm T})]\partial_z-k_{\rho}^2(\hat{\mathbf u}\otimes\hat{\mathbf u}^{\rm T})]\Big]\widehat{G}^f   \\
			&+\frac{\ri}{2k_{\ell' z}}\sum_{\ast,\star=\uparrow,\downarrow}\Big\{\frac{\mu_{\ell}}{\mu_{\ell'}k_{\ell}^2}\Big[k_{\ell'z}(s_{2}^{\star}k_{\rho}(\hat{\mathbf z}\otimes\hat{\mathbf u}^{\rm T})-s_{1}^{\ast}s_2^{\star}k_{\ell z}(\hat{\mathbf u}\otimes\hat{\mathbf u}^{\rm T}))\\
			&+k_{\rho}[k_{\rho}(\hat{\mathbf z}\otimes\hat{\mathbf z}^{\rm T})-s_{1}^{\ast}k_{\ell z}(\hat{\mathbf u}\otimes\hat{\mathbf z}^{\rm T})]\Big]\psi_{\ell\ell'}^{\ast\star}+(\hat{\mathbf v}\otimes\hat{\mathbf v}^{\rm T}){\phi}_{\ell\ell'}^{\ast\star}\Big\}{\mathcal Z}_{\ell\ell'}^{\ast\star}\\
			=&\delta_{\ell\ell'}[\widehat G^f\bh{I}+k_{\ell}^{-2}\mathcal F[\nabla\nabla G^f]]+\dfrac{\ri}{2k_{\ell'z}}\sum_{\ast,\star=\uparrow,\downarrow}\bs\Theta_{\bs E,\ell\ell'}^{\ast\star}(k_x, k_y){\mathcal Z}_{\ell\ell'}^{\ast\star}(k_{\rho}, z,z'),
		\end{split}
	\end{equation*}
	and
	\begin{equation*}
		\begin{split}
			\widehat{\bh G}_{\bs H}=&-\dfrac{\delta_{\ell\ell'}}{\ri\omega\mu_{\ell'}}\left[\ri k_{\rho}\left(\mathbf{\hat{v}}\otimes\mathbf{\hat{z}}^{\rm T}-\mathbf{\hat{z}}\otimes\mathbf{\hat{v}}^{\rm T}\right)-\left(\mathbf{\hat{v}}\otimes\mathbf{\hat{u}}^{\rm T}-\mathbf{\hat{u}}\otimes\mathbf{\hat{v}}^{\rm T}\right)\partial_z\right]\widehat{G}^f\\
			&-\dfrac{1}{2\omega\mu_{\ell} k_{\ell'z}}\sum_{\ast,\star=\uparrow,\downarrow}\Big[\left(\ri s_{1}^{\ast}k_{\ell z}\mathbf{\hat{u}}\otimes\mathbf{\hat{v}}^{\rm T}-\ri k_{\rho}\mathbf{\hat{z}}\otimes\mathbf{\hat{v}}^{\rm T}\right)\phi_{\ell\ell'}^{\ast\star}(\kr)\\
			&+\dfrac{\mu_{\ell}}{\mu_{\ell'}}\left(\ri s_{2}^{\star}k_{\ell' z}\mathbf{\hat{v}}\otimes\mathbf{\hat{u}}^{\rm T}+\ri k_{\rho}\mathbf{\hat{v}}\otimes\mathbf{\hat{z}}^{\rm T}\right)\psi_{\ell\ell'}^{\ast\star}\Big]{\mathcal Z}_{\ell\ell'}^{\ast\star}(k_{\rho}, z,z')\\
			=&\delta_{\ell\ell'}\widehat{\bh G}_{\bs H}^f+\dfrac{1}{2\omega\mu_{\ell} k_{\ell'z}}\sum_{\ast,\star=\uparrow,\downarrow}\Theta_{\bs H,\ell\ell'}^{\ast\star}(k_x,k_y){\mathcal Z}_{\ell\ell'}^{\ast\star}(k_{\rho}, z,z'),
		\end{split}
	\end{equation*}
	where the signs $s_1^{\uparrow}=1$, $s_1^{\downarrow}=-1$
	are introduced due to the derivative
	$$\partial_{z}{\mathcal Z}_{\ell\ell'}^{\ast\star}(k_{\rho}, z,z')=\ri s_{1}^{\ast}k_{\ell z}{\mathcal Z}_{\ell\ell'}^{\ast\star}(k_{\rho}, z,z'),$$
	and the densities for the electromagnetic fields are defined as follows
	\begin{equation}\label{density_uvz}
		\begin{aligned}
			\Theta_{\bs E,\ell\ell'}^{\ast\star}=&\mathbf{\hat{v}}\otimes\mathbf{\hat{v}}^{\rm T}\phi_{\ell\ell'}^{\ast\star}+\dfrac{\mu_{\ell}}{\mu_{\ell'}k_{\ell}^2}\big[k_{\ell'z}\left(-s_{1}^{\ast}s_2^{\star}k_{\ell z}\mathbf{\hat{u}}\otimes\mathbf{\hat{u}}^{\rm T}+s_2^{\star}k_{\rho}\mathbf{\hat{z}}\otimes\mathbf{\hat{u}}^{\rm T}\right)\\
			&-s_{1}^{\ast}k_{\rho}k_{\ell z}\mathbf{\hat{u}}\otimes\mathbf{\hat{z}}^{\rm T}+k_{\rho}^2\mathbf{\hat{z}}\otimes\mathbf{\hat{z}}^{\rm T}\big]\psi_{\ell\ell'}^{\ast\star},\\
			\Theta_{\bs H,\ell\ell'}^{\ast\star}=&-\left(\ri k_{\ell z}s_{1}^{\ast}\mathbf{\hat{u}}\otimes\mathbf{\hat{v}}^{\rm T}-\ri k_{\rho}\mathbf{\hat{z}}\otimes\mathbf{\hat{v}}^{\rm T}\right)\phi_{\ell\ell'}^{\ast\star}-\dfrac{\mu_{\ell}}{\mu_{\ell'}}\left(\ri k_{\ell' z}s_{2}^{\star}\mathbf{\hat{v}}\otimes\mathbf{\hat{u}}^{\rm T}+\ri k_{\rho}\mathbf{\hat{v}}\otimes\mathbf{\hat{z}}^{\rm T}\right)\psi_{\ell\ell'}^{\ast\star},
		\end{aligned}
	\end{equation}
	for $\ast,\star=\uparrow,\downarrow$. Taking inverse Fourier transform gives
	\begin{equation}\label{electromagnetic_Green_fun_layer}
		{\bh G}_{\bs E}(\bs r,\bs r')=\delta_{\ell\ell'}{\bh G}_{\bs E}^f(\bs r,\bs r')+{\bh G}_{\bs E}^r(\bs r,\bs r'),\quad
		{\bh G}_{\bs H}(\bs r,\bs r')=\delta_{\ell\ell'}{\bh G}_{\bs H}^f(\bs r,\bs r')+{\bh G}_{\bs H}^r(\bs r,\bs r'),
	\end{equation}
	where
	\begin{equation}\label{dyasicsinphys}
		\begin{split}
			{\bh G}_{\bs E}^r(\bs r,\bs r')= &\dfrac{\ri}{8\pi^2}\sum_{\ast,\star=\uparrow,\downarrow}\int_{-\infty}^{+\infty}\int_{-\infty}^{+\infty}\Theta_{\bs E,\ell\ell'}^{\ast\star}(k_x,k_y){\mathcal Z}_{\ell\ell'}^{\ast\star}(k_{\rho}, z,z')\dfrac{e^{\ri\mathbf{k}_{\rho}\cdot(\boldsymbol{\rho-\rho'})}}{k_{\ell'z}}dk_xdk_y,\\
			{\bh G}_{\bs H}^r(\bs r,\bs r')=&\dfrac{1}{8\pi^2\omega\mu_{\ell}}\sum_{\ast,\star=\uparrow,\downarrow}\int_{-\infty}^{+\infty}\int_{-\infty}^{+\infty}\Theta_{\bs H,\ell\ell'}^{\ast\star}(k_x,k_y){\mathcal Z}_{\ell\ell'}^{\ast\star}(k_{\rho}, z,z')\dfrac{e^{\ri\mathbf{k}_{\rho}\cdot(\boldsymbol{\rho-\rho'})}}{k_{\ell'z}}dk_xdk_y,
		\end{split}
	\end{equation}
	
	\begin{rem}
		Denoted by $\widehat{\nabla}=\ri\mathbf{k}_{\rho}+\hat{\mathbf z}\partial_z$, and $\widehat{\nabla}'=-\ri\mathbf{k}_{\rho}+\hat{\mathbf z}\partial_{z'}$, direct calculation gives
		\begin{equation*}
			\begin{split}
				&\frac{1}{k_{\rho}^2}(\widehat{\nabla}\times\widehat{\nabla}\times\hat{\mathbf z})\otimes(\widehat{\nabla}'\times\widehat{\nabla}'\times\hat{\mathbf z}){\mathcal Z}_{\ell\ell'}^{\ast\star}(k_{\rho}, z,z')
				=\big[k_{\ell'z}\left(-s_{1}^{\ast}s_2^{\star}k_{\ell z}\mathbf{\hat{u}}\otimes\mathbf{\hat{u}}^{\rm T}+s_2^{\star}k_{\rho}\mathbf{\hat{z}}\otimes\mathbf{\hat{u}}^{\rm T}\right)\\
				&\qquad\qquad\qquad\qquad\qquad\qquad\qquad\qquad\qquad-s_{1}^{\ast}k_{\rho}k_{\ell z}\mathbf{\hat{u}}\otimes\mathbf{\hat{z}}^{\rm T}
				+k_{\rho}^2\mathbf{\hat{z}}\otimes\mathbf{\hat{z}}^{\rm T}\big]{\mathcal Z}_{\ell\ell'}^{\ast\star}(k_{\rho}, z,z'),\\
				&\frac{1}{k_{\rho}^2}(\widehat{\nabla}\times\hat{\mathbf z})\otimes(\widehat{\nabla}'\times\hat{\mathbf z}){\mathcal Z}_{\ell\ell'}^{\ast\star}(k_{\rho}, z,z')=\mathbf{\hat{v}}\otimes\mathbf{\hat{v}}^{\rm T}{\mathcal Z}_{\ell\ell'}^{\ast\star}(k_{\rho}, z,z'),\\
				&\frac{1}{k_{\rho}^2}(\widehat{\nabla}\times\widehat{\nabla}\times\hat{\mathbf z})\otimes(\widehat{\nabla}'\times\hat{\mathbf z}){\mathcal Z}_{\ell\ell'}^{\ast\star}(k_{\rho}, z,z')
				=-\left(\ri s_{1}^{\ast}k_{\ell z}\mathbf{\hat{u}}\otimes\mathbf{\hat{v}}^{\rm T}-\ri k_{\rho}\mathbf{\hat{z}}\otimes\mathbf{\hat{v}}^{\rm T}\right){\mathcal Z}_{\ell\ell'}^{\ast\star}(k_{\rho}, z,z'),\\
				&\frac{1}{k_{\rho}^2}(\widehat{\nabla}\times\hat{\mathbf z})\otimes(\widehat{\nabla}'\times\widehat{\nabla}'\times\hat{\mathbf z}){\mathcal Z}_{\ell\ell'}^{\ast\star}(k_{\rho}, z,z')
				=-\left(\ri s_{2}^{\star}k_{\ell' z}\mathbf{\hat{v}}\otimes\mathbf{\hat{u}}^{\rm T}+\ri k_{\rho}\mathbf{\hat{v}}\otimes\mathbf{\hat{z}}^{\rm T}\right){\mathcal Z}_{\ell\ell'}^{\ast\star}(k_{\rho}, z,z').
			\end{split}
		\end{equation*}
		Therefore, the formulations \eqref{dyasicsinphys} in the physical domain can also be written as
		\begin{equation*}
			\begin{split}
				{\bh G}_{\bs E}^r(\bs r,\bs r')= &({\nabla}\times\hat{\mathbf z})\otimes({\nabla}'\times\hat{\mathbf z})g^{\rm TE}(\bs r,\bs r')+\frac{1}{\omega^2\varepsilon_{\ell}\mu_{\ell'}}({\nabla}\times{\nabla}\times\hat{\mathbf z})\otimes({\nabla}'\times{\nabla}'\times\hat{\mathbf z})g^{\rm TM}(\bs r,\bs r'),\\
				{\bh G}_{\bs H}^r(\bs r,\bs r')=&\frac{1}{\ri\omega}\Big[\frac{1}{\mu_{\ell}}({\nabla}\times{\nabla}\times\hat{\mathbf z})\otimes({\nabla}'\times\hat{\mathbf z}) g^{\rm TE}(\bs r,\bs r')+\frac{1}{\mu_{\ell'}}({\nabla}\times\hat{\mathbf z})\otimes({\nabla}'\times{\nabla}'\times\hat{\mathbf z}) g^{\rm TM}(\bs r,\bs r')\Big],
			\end{split}
		\end{equation*}
		where 
		\begin{equation*}
			\begin{split}
				& g^{\rm TE}(\bs r,\bs r')=\dfrac{\ri}{8\pi^2}\sum_{\ast,\star=\uparrow,\downarrow}\int_{-\infty}^{+\infty}\int_{-\infty}^{+\infty}\dfrac{e^{\ri\mathbf{k}_{\rho}\cdot(\boldsymbol{\rho}-\boldsymbol{\rho}')}}{k_{\ell'z}}{\mathcal Z}_{\ell\ell'}^{\ast\star}(k_{\rho}, z,z')\frac{\phi_{\ell\ell'}^{\ast\star}(\kr)}{k_{\rho}^2}dk_xdk_y,\\
				& g^{\rm TM}(\bs r,\bs r')=\dfrac{\ri}{8\pi^2}\sum_{\ast,\star=\uparrow,\downarrow}\int_{-\infty}^{+\infty}\int_{-\infty}^{+\infty}\dfrac{e^{\ri\mathbf{k}_{\rho}\cdot(\boldsymbol{\rho}-\boldsymbol{\rho}')}}{k_{\ell'z}}{\mathcal Z}_{\ell\ell'}^{\ast\star}(k_{\rho}, z,z')\frac{\psi_{\ell\ell'}^{\ast\star}(\kr)}{k_{\rho}^2}dk_xdk_y.
			\end{split}
		\end{equation*}
		These are the formulations used in the method of moments simulation (cf. \cite{xiong2009newly,chew2006matrix}). 
	\end{rem}
	
	\section{Computation of the dyadic Green's function of Maxwell's equations in layered media using a matrix basis}\label{sect_fsgf}

	In this section, we present a simplified version of the derivation in \cite{bo2022maxwellDGF} for the dyadic Green's function of Maxwell's equations in layered media, and demonstrate that it leads to the same formulations obtained via the conventional TE/TM decomposition reviewed in the last section. The approach is more straightforward, as it introduces the dyadic vector potential analogously to the free-space case. Instead of relying on TE/TM decomposition, the derivation employs a matrix basis to decompose the problem into three layered media problems of Helmholtz equation.
	
	\subsection{Dyadic vector potential} Consider the dyadic form of the interface problem \eqref{maxwelleqlayer}-\eqref{interfacecondphysical_vector}. The dyadic vector potential
	$$
	\bh G_{\bd A}=\begin{bmatrix}
		\bd G_{\bd A}^{\hat{\bd x}} & \bd G_{\bd A}^{\hat{\bd y}} & \bd G_{\bd A}^{\hat{\bd z}}
	\end{bmatrix}
	$$
	satisfies
	\begin{equation}\label{eq-GA-eqn-layer}
		\nabla^2 \bh{G}_{\bs A}(\bs r,\bs r') + k_{\ell}^2 \bh{G}_{\bs A}(\bs r,\bs r') = -\frac{1}{\ri \omega} \delta(\bs r-\bs r')\bh{I},\quad d_{\ell}<z<d_{\ell-1}, \;\ell=0, 1, \cdots, L.
	\end{equation}
	Further, by the Lorenz gauge, we have
	\begin{equation}\label{eq-GE-GH-repre1}
		\bh{G}_{\bs E} = \ri \omega \left(\bh{I} + \frac{\nabla\nabla}{k_{\ell}^2}\right)\bh{G}_{\bs A},\quad \bh{G}_{\bs H} = \frac{1}{\mu_{\ell}} \nabla \times \bh{G}_{\bs A},\quad d_{\ell}<z<d_{\ell-1}, \;\ell=0, 1, \cdots, L.
	\end{equation}
	Recall that the right-hand side of the equation \eqref{eq-GA-eqn-layer} is nontrivial if and only if $\bd{r}$ is in the same layer as $\bd{r}'$, i.e. $\ell=\ell'$. Define
	\begin{equation}\label{eq-GAf}
		{\bh{G}}_{\bs A}^r(\bd{r},\bd{r}') = \begin{cases}
			{\bh{G}}_{\bs A}(\bd{r},\bd{r}') - {\bh{G}}_{\bs A}^f(\bd{r},\bd{r}') & {\rm if}\;\; \ell=\ell',\\[7pt]
			{\bh{G}}_{\bs A}(\bd{r},\bd{r}') & {\rm otherwise},
		\end{cases}
	\end{equation}
	where ${\bh{G}}_{\bs A}^f(\bd{r},\bd{r}')$ is the free space dyadic Green's function of the vector potential defined in \eqref{GAF}. Then, ${\bh{G}}_{\bs A}^r$ satisfies the \emph{homogeneous} Helmholtz equation
	\begin{equation}
		\nabla^2 {\bh{G}}_{\bs A}^r(\bd{r},\bd{r}') + k^2_{\ell} {\bh{G}}_{\bs A}^r(\bd{r},\bd{r}') = \bd{0},\quad d_{\ell}<z<d_{\ell-1},
	\end{equation}
	in each layer. In the Fourier spectral domain, the equation is transformed to 
	\begin{equation}\label{eq-Helmholtz-freq}
		\partial_{zz} \widehat{\bh{G}}_{\bs A}^r(k_x, k_y, z, z') + k_{\ell z}^2\widehat{\bh{G}}_{\bs A}^r(k_x, k_y, z, z') = \bd{0},\quad d_{\ell}<z<d_{\ell-1}.
	\end{equation}
	The general solution to \eqref{eq-Helmholtz-freq}, when treated as an second order ODE of $z$, is given by
	\begin{equation}\label{eq-GAr}
		\widehat{\bh{G}}_{\bd A}^r(k_x, k_y, z, z') =  \widehat{\bh{G}}_{\ell\ell'}^{\uparrow}(k_x, k_y, z')e^{\i k_{\ell z} (z-d_{\ell})} +  \widehat{\bh{G}}_{\ell\ell'}^{\downarrow}(k_x, k_y, z')e^{\i k_{\ell z} (d_{\ell-1}-z)},\; d_{\ell}<z<d_{\ell-1},
	\end{equation}
	where $\{\widehat{\bh G}_{\ell\ell'}^{\uparrow}(k_x, k_y, z'),\widehat{\bh{G}}_{\ell\ell'}^{\downarrow}(k_x, k_y, z')\}_{\ell=0}^{L}$ are coefficients to be determined by the interface conditions and outgoing boundary condition at infinity and the up/down arrows indicate the direction of the wave propagation at the target point.
	
	As the solution \eqref{eq-GAr} has to remain bounded at infinity as $k_{\rho}\rightarrow\infty$, it follows that
	\begin{equation*}
		\widehat{\bh{G}}_{0\ell'}^{\downarrow}(k_x, k_y, z') = \bd{0},\quad \widehat{\bh{G}}_{L\ell'}^{\uparrow}(k_x, k_y, z')=\bd{0}.
	\end{equation*}
	Indeed, we can also rewrite $\widehat{\bh{G}}_{\bs A}^f(k_{\rho}, z, z')=\dfrac{\bh I}{2\omega k_{\ell'z}} e^{\i k_{\ell'z} |z - z'|}$ in a similar form, i.e.,
	\begin{equation}\label{GAfree}
		\widehat{\bh{G}}_{\bs A}^f(k_{\rho}, z, z') =  \frac{\bh I}{2 \omega k_{\ell' z}} \Big[H(z-z')e^{\i k_{\ell' z} (z-z')} + H(z'-z)e^{\i k_{\ell' z} (z'-z)}\Big],
	\end{equation}
	where 
	$$H(x)=\begin{cases}
		\displaystyle 0,\quad x<0,\\
		\displaystyle \frac{1}{2},\quad x=0,\\
		\displaystyle 1,\quad x>0,
	\end{cases}$$ 
	is the Heaviside function. Therefore, $\widehat{\bh G}_{\bs A}$ has the up/down decomposition:
	\begin{align}\label{eq-GApm}
		\widehat{\bh{G}}_{\bs A}(k_x, k_y, z, z')=\widehat{\bh{G}}_{\bs A}^r+\delta_{\ell\ell'}\widehat{\bh{G}}_{\bs A}^f= \widehat{\bh{G}}_{\bs A}^{\uparrow}(k_x, k_y, z, z') + \widehat{\bh{G}}_{\bs A}^{\downarrow}(k_x, k_y, z, z'),
	\end{align}
	where
	\begin{equation}\label{hatGAdecomp}
		\begin{split}
			\widehat{\bh{G}}_{\bs A}^{\uparrow}(k_x, k_y, z, z') &=  \widehat{\bh{G}}_{\ell\ell'}^{\uparrow}(k_x, k_y, z')e^{\i k_{\ell z} (z-d_{\ell})}+\frac{\delta_{\ell\ell'}H(z-z')\bh I}{2 \omega k_{\ell' z}} e^{\i k_{\ell' z} (z-z')} , \\ 
			\widehat{\bh{G}}_{\bs A}^{\downarrow}(k_x, k_y, z, z') &=\widehat{\bh{G}}_{\ell\ell'}^{\downarrow}(k_x, k_y, z')e^{\i k_{\ell z} (d_{\ell-1}-z)}  +\frac{\delta_{\ell\ell'}H(z'-z)\bh I}{2 \omega k_{\ell' z}}e^{\i k_{\ell' z} (z'-z)},
		\end{split}
	\end{equation}
	for $d_{\ell}<z<d_{\ell-1}$.
	The Kronecker symbol $\delta_{\ell\ell'}$ is due to the fact that the free space component $ {\bh G}_{\bs A}^f$ only exists in the source layer. 
	
	
	In the frequency domain, 
	we use the notation $(\kr, \alpha)$ for the polar coordinates of $(k_x, k_y)$ and $\widehat\nabla =[\ri k_x \;\; \ri k_y \;\; \partial_z ]^{\rm T}$,  $\widehat\nabla\widehat\nabla$, $\widehat\nabla^2$ refer to $\widehat\nabla\widehat\nabla^{\rm T}$, $\widehat\nabla^{\rm T}\widehat\nabla$, respectively. The Fourier transform of \eqref{eq-GE-GH-repre1} gives
	\begin{align}\label{eq-GE-GH-GA-form1}
		\widehat{\bh{G}}_{\bs E} = \ri\omega \left(\bh{I}+\frac{\widehat\nabla\widehat\nabla}{k_{\ell}^2}\right)\widehat{\bh{G}}_{\bs A}, \quad
		\widehat{\bh{G}}_{\bs H} = \frac{1}{\mu_{\ell}}\widehat\nabla \times \widehat{\bh{G}}_{\bs A},\quad d_{\ell}<z<d_{\ell-1},\; \ell=0, 1, \cdots, L.
	\end{align}
	The dyadic forms of the interface conditions in \eqref{interfacecondfreq} are given by
	\begin{equation}\label{eq-if-cond}
		\lbr\bs n\times \widehat{\bh G}_{\bs E}\rbr=\bs 0,\quad \lbr\bs{n}\times \widehat{\bh G}_{\bs H}\rbr=\bs 0,\quad \lbr\bs{n}\cdot \varepsilon\widehat{\bh G}_{\bs E}\rbr=0, \quad \lbr\bs{n}\cdot\mu\widehat{\bh G}_{\bs H}\rbr=0,\quad{\rm at}\quad z=\{d_{\ell}\}_{\ell=0}^{L-1}.
	\end{equation}
	As we have mentioned above the transmission conditions on the normal components of the fields are redundant. 
	\subsection{The matrix basis}
	The formulations in \eqref{eq-GE-GH-GA-form1} have shown that $\widehat{\bh{G}}_{\bs E} $ and $\widehat{\bh{G}}_{\bs H}$ are just the product of some $3\times 3$ matrices with $\widehat{\bh{G}}_{\bs A}$. In order to give better understanding of these matrices, we introduce the following matrix basis 
	\begin{align}\label{mat-J}
		\begin{split}
			\J{1} & = \begin{bmatrix}
				1 & & \\
				& 1 & \\
				& & 0
			\end{bmatrix}, \text{  }
			\J{2}  = \begin{bmatrix}
				0 & & \\
				& 0 & \\
				& & 1
			\end{bmatrix}, \text{  }
			\J{3}  = \begin{bmatrix}
				0 & 0 & \i k_x \\
				0 & 0 & \i k_y \\
				0 & 0 & 0
			\end{bmatrix}, \\
			\J{4} & = \begin{bmatrix}
				0 & 0 & 0 \\
				0 & 0 & 0 \\
				\i k_x & \i k_y & 0
			\end{bmatrix}, \text{  }
			\J{5}  = \begin{bmatrix}
				-k_x^2 & -k_x k_y & 0 \\
				-k_x k_y & -k_y^2 & 0 \\
				0 & 0 & 0
			\end{bmatrix}, \text{  }
			\J{6}  = \begin{bmatrix}
				0 & 0 & 0 \\
				0 & 0 & 0 \\
				-\i k_y & \i k_x & 0
			\end{bmatrix}, \\
			\J{7} & = \begin{bmatrix}
				0 & 0 & \i k_y \\
				0 & 0 & -\i k_x \\
				0 & 0 & 0
			\end{bmatrix}, \text{  }
			\J{8}  = \begin{bmatrix}
				k_x k_y & k_y^2 & 0 \\
				-k_x^2 & -k_x k_y & 0 \\
				0 & 0 & 0
			\end{bmatrix}, \text{  }
			\J{9}  = \begin{bmatrix}
				0 & 1 & 0 \\
				-1 & 0 & 0 \\
				0 & 0 & 0
			\end{bmatrix}.
		\end{split}
	\end{align}
	Obviously, the product of these basis matrices follow the table
	\begin{equation}\label{producttable}
		\begin{array}{c|ccccc:cccc}
			\times &\J{1} & \J{2} & \J{3} & \J{4} & \J{5} & \J{6}& \J{7} & \J{8} & \J{9}\\[7pt]
			\hline
			\J{1} &\J{1} & \bd{0} & \J{3} & \bd{0} & \J{5} & \bd{0} & \J{7} & \J{8} & \J{9} \\[7pt]
			\J{2} &\bd{0} & \J{2} & \bd{0} & \J{4} & \bd{0} & \J{6} & \bd{0} & \bd{0} & \bd{0} \\[7pt]
			\J{3} &\bd{0} & \J{3} & \bd{0} & \J{5} & \bd{0} & \J{8}-\kr^2 \J{9} & \bd{0} & \bd{0} & \bd{0} \\[7pt]
			\J{4} &\J{4} & \bd{0} & -\kr^2\J{2} & \bd{0} & -\kr^2\J{4} & \bd{0} & \bd{0} & \bd{0} & \J{6} \\[7pt]
			\J{5} &\J{5} & \bd{0} & -\kr^2\J{3} & \bd{0} & -\kr^2\J{5} & \bd{0} & \bd{0} & \bd{0} & \J{8}-\kr^2 \J{9} \\[7pt] \hdashline
			\J{6} &\J{6} & \bd{0} & \bd{0} & \bd{0} & \bd{0} & \bd{0} & \kr^2 \J{2} & -\kr^2 \J{4} & -\J{4} \\[7pt]
			\J{7} &\bd{0} & \J{7} & \bd{0} & -\J{8} & \bd{0} & \kr^2 \J{1}+\J{5} & \bd{0} & \bd{0} & \bd{0} \\[7pt]
			\J{8} &\J{8} & \bd{0} & \kr^2\J{7} & \bd{0} & -\kr^2\J{8} & \bd{0} & \bd{0} & \bd{0} & -\kr^2 \J{1}-\J{5} \\[7pt]
			\J{9} &\J{9} & \bd{0} & \J{7} & \bd{0} & -\J{8} & \bd{0} & -\J{3} & \J{5} & -\J{1}
		\end{array}
	\end{equation}
	Our goal is to represent the dyadic Green's functions $\widehat{\bh{G}}_{\bd E}$ and $\widehat{\bh{G}}_{\bd H}$ using these basis matrices.

	Given any function $f(k_x, k_y, z, z')$, direct calculation using the representations
	\begin{equation}\label{Operatordecomp}
		\widehat\nabla\times =\J{6}+\J{7} -  \J{9}\partial_z,\quad \Big(\bh{I}+\frac{\widehat\nabla\widehat\nabla}{k^2_{\ell}}\Big)=\bh I+\frac{1}{k_{\ell}^2}(\J{2}\partial_{zz}+(\J{3} + \J{4})\partial_z+\J{5}),
	\end{equation}
	and the product table \eqref{producttable} give us
	\begin{equation}\label{GHoperator}
		\begin{split}
			\widehat\nabla\times (f\J{1})=&f\J{6}-\partial_zf\J{9},\quad
			\widehat\nabla\times (f\J{2})= f\J{7},\quad\widehat\nabla\times (f\J{3})=-\partial_zf\J{7},\quad\\
			\widehat\nabla\times (f\J{4})=& -f\J{8},\quad
			\widehat\nabla\times (f\J{5})= \partial_zf\J{8},
		\end{split}
	\end{equation}
	and
	\begin{equation}\label{GEoperator}
		\begin{split}
			\Big(\bh{I}+\frac{\widehat\nabla\widehat\nabla}{k^2_{\ell}}\Big)(f\J{1})=& f\J{1}+\frac{\partial_zf}{k_{\ell}^2} \J{4}+\frac{f}{k_{\ell}^2}\J{5}, \quad
			\Big(\bh{I}+\frac{\widehat\nabla\widehat\nabla}{k^2_{\ell}}\Big)(f\J{2})=\Big(f+\frac{\partial_{zz}f}{k_{\ell}^2}\Big)\J{2}+\frac{\partial_zf}{k_{\ell}^2} \J{3},\\
			\Big(\bh{I}+\frac{\widehat\nabla\widehat\nabla}{k^2_{\ell}}\Big)(f\J{3})=&\frac{k_{\ell z}^2}{k_{\ell}^2}f \J{3}-\frac{k_{\rho}^2}{k_{\ell}^2}\partial_z f \J{2},\quad
			\Big(\bh{I}+\frac{\widehat\nabla\widehat\nabla}{k^2_{\ell}}\Big)(f\J{4})=\Big(f+\frac{\partial_{zz}f}{k_{\ell}^2} \Big)\J{4}+ \frac{\partial_zf }{k_{\ell}^2} \J{5},\\
			\Big(\bh{I}+\frac{\widehat\nabla\widehat\nabla}{k^2_{\ell}}\Big)(f\J{5})=&\frac{k_{\ell z}^2 }{k_{\ell}^2}f \J{5}-\frac{k_{\rho}^2}{k_{\ell}^2}\partial_zf \J{4}.
		\end{split}
	\end{equation}


	\subsection{A representation of \texorpdfstring{$\widehat{\bh G}_{\bs A}$}{widehatGA} using the matrix basis} According to the assumption that the media is layered in the $z$-direction, the normal direction on the interface is $\bs n=\bs e_z=[0, 0, 1]^{\rm T}$. For any given $3\times3$ tensor $\bh F$, we have
	\begin{equation}\label{normaltimesformula}
		\bs e_z\times\bh F=\begin{bmatrix}
			-F_{21} & -F_{22} & -F_{23}\\
			F_{11} & F_{12} & F_{13}\\
			0 & 0 & 0
		\end{bmatrix},\quad \bs e_z\cdot\bh F=\begin{bmatrix}
			F_{31} & F_{32} & F_{33}\\
		\end{bmatrix}
	\end{equation}
	Therefore, the interface conditions \eqref{eq-if-cond} are actually that all entries in the first and second rows of $\widehat{\bh G}_{\bs E}$ and $\widehat{\bh G}_{\bs H}$ and the third rows of $\varepsilon\widehat{\bh G}_{\bs E}$ and $\mu\widehat{\bh G}_{\bs H}$ are continuous. In the rest of this paper, we shall only use the first two conditions in \eqref{eq-if-cond} for the derivation as we have done in the previous section. Using permutation matrices $\J{1}$ and $\J{2}$, the first two jump conditions in \eqref{eq-if-cond} are equivalent to
	\begin{equation}\label{EJump}
		\J{1}\llbracket\widehat{\bh G}_{\bs E}\rrbracket=\bs 0, \quad \J{1}\llbracket\widehat{\bh G}_{\bs H}\rrbracket=\bs 0, \quad {\rm at}\;z=\{d_{\ell}\}_{\ell=0}^{L-1}.
	\end{equation}
	Using the product table \eqref{producttable} and expressions \eqref{eq-GE-GH-GA-form1}, \eqref{Operatordecomp}, we calculate that
	\begin{equation}\label{Ejumptenj}
		\begin{split}
			\J{1} \widehat{\bh{G}}_{\bs E} =& \i\omega\left( \J{1}+\frac{\J{3}}{k_{\ell}^2}\partial_z +\frac{1}{k_{\ell}^2}\J{5} \right)\widehat{\bh{G}}_{\bs A} ,\quad
			\J{1} \widehat{\bh{G}}_{\bs H} = \frac{1}{\mu_{\ell}}(\J{7}-\partial_z\J{9})\widehat{\bh{G}}_{\bs A},
		\end{split}
	\end{equation}
	for $d_{\ell}<z<d_{\ell-1}$. Note that  $\J{9}$ are continuous across the interfaces. Multiplying the jump condition on $\widehat{\bh{G}}_{\bs H}$ in \eqref{EJump} by $\J{9}$, we have
	\begin{equation}\label{reformHJump}
		\J{9}\J{1}\llbracket\widehat{\bh G}_{\bs H}\rrbracket=\bs 0,\quad {\rm at}\;z=\{d_{\ell}\}_{\ell=0}^{L-1}.
	\end{equation}
	It is worth pointing out that the jump conditions on $\widehat{\bh{G}}_{\bs H}$ in \eqref{EJump} and \eqref{reformHJump}  are equivalent, respectively, due to the permutation matrix $\J{1}$.
	From \eqref{Ejumptenj} and using the product table \eqref{producttable},
	we obtain
	\begin{align}
		\J{9}\J{1} \widehat{\bh{G}}_{\bs H} = -\frac{1}{\mu_{\ell}}(\J{3}-\partial_z\J{1})\widehat{\bh{G}}_{\bs A},\quad d_{\ell}<z<d_{\ell-1}.\label{reformHJumpnormal}
	\end{align}	
	Using \eqref{Ejumptenj}, \eqref{reformHJumpnormal} in \eqref{EJump} and \eqref{reformHJump}, respectively, we obtain interface conditions
	\begin{equation}\label{Ejumpcond}
		\Big\llbracket \left( \J{1}+\frac{\J{3}}{k^2}\partial_z +\frac{1}{k^2}\J{5} \right)\widehat{\bh{G}}_{\bs A} \Big\rrbracket =0,	\quad \Big\llbracket -\frac{1}{\mu}(\J{3}-\partial_z\J{1})\widehat{\bh{G}}_{\bs A}\Big\rrbracket =0,\quad {\rm at}\;z=\{d_{\ell}\}_{\ell=0}^{L-1}.
	\end{equation}
	Consequently, we have derived interface conditions for the vector potential from that on electromagnetic fields.
	
	Denote by 
	\begin{equation*}
		\bh K_{\ell}=\J{1}+\frac{\J{3}}{k_{\ell}^2}\partial_z +\frac{1}{k_{\ell}^2}\J{5},\quad \bh W_{\ell}=-\frac{1}{\mu_{\ell}}(\J{3}-\J{1}\partial_z),\quad \ell=0,1,\cdots, L.
	\end{equation*}
	Then, \eqref{Ejumpcond} can be rewritten as
	\begin{equation}\label{interfacecondequation}
		\begin{cases}
			\displaystyle\bh{K}_{\ell-1}\widehat{\bh{G}}_{\bs A}(k_x, k_y, d_{\ell-1}+0, z')-\bh{K}_{\ell}\widehat{\bh{G}}_{\bs A}(k_x, k_y, d_{\ell-1}-0, z')=0,\\
			\displaystyle\bh{W}_{\ell-1}\widehat{\bh{G}}_{\bs A}(k_x, k_y, d_{\ell-1}+0, z')-\bh{W}_{\ell}\widehat{\bh{G}}_{\bs A}(k_x, k_y, d_{\ell-1}-0, z')=0,
		\end{cases}
	\end{equation}
	for all $ \ell=1, 2, \cdots, L$. From the expressions \eqref{eq-GApm} and \eqref{hatGAdecomp}, we have
	\begin{equation}\label{interfacecondexpress}
		\begin{split}
			\bh{K}_{\ell}\widehat{\bh{G}}_{\bs A}(k_x, k_y, z, z')=&\bh{K}_{\ell}^{\uparrow}\widehat{\bh{G}}_{\bs A}^{\uparrow}(k_x, k_y, z, z')+\bh{K}_{\ell}^{\downarrow}\widehat{\bh{G}}_{\bs A}^{\downarrow}(k_x, k_y, z, z'),\\
			\bh{W}_{\ell}\widehat{\bh{G}}_{\bs A}(k_x, k_y, z, z')=&\bh{W}_{\ell}^{\uparrow}\widehat{\bh{G}}_{\bs A}^{\uparrow}(k_x, k_y, z, z')+\bh{W}_{\ell}^{\downarrow}\widehat{\bh{G}}_{\bs A}^{\downarrow}(k_x, k_y, z, z'),
		\end{split}
	\end{equation}
	for all $ \ell=0, 1, \cdots, L$, where
	\begin{equation}\label{KWmatrices}
		\begin{split}
			\bh{K}_{\ell}^{\uparrow}=&\J{1}+\frac{\i k_{\ell z}}{k_{\ell}^2}\J{3} +\frac{1}{k_{\ell}^2}\J{5},\quad \bh{K}_{\ell}^{\downarrow}=\J{1}-\frac{\i k_{\ell z}}{k_{\ell}^2}\J{3} +\frac{1}{k_{\ell}^2}\J{5},\\
			\bh{W}_{\ell}^{\uparrow}=&-\frac{1}{\mu_{\ell}}\J{3}+\frac{\i k_{\ell z}}{\mu_{\ell}} \J{1},\quad \bh{W}_{\ell}^{\downarrow}=-\frac{1}{\mu_{\ell}}\J{3}-\frac{\i k_{\ell z}}{\mu_{\ell}} \J{1}.
		\end{split}
	\end{equation}
	We have replaced the partial derivatives $\partial_z, \partial_{zz}$ in $\bh{K}_{\ell}, \bh{W}_{\ell}$ by $\pm\ri k_{\ell z}$ and $-k_{\ell z}^2$, respectively, using the general formulations \eqref{hatGAdecomp}.
	Substituting the expressions \eqref{interfacecondexpress} into \eqref{interfacecondequation}, we obtain linear systems
	\begin{equation}\label{sublinearsytem1}
		\begin{bmatrix}
			\bh{K}^{\uparrow}_{\ell-1} & \bh{K}^{\downarrow}_{\ell-1}\\[7pt]
			\bh{W}^{\uparrow}_{\ell-1} & \bh{W}^{\downarrow}_{\ell-1}
		\end{bmatrix}\begin{bmatrix}
			\widehat{\bh{G}}_{\bs A}^{\uparrow}(k_x, k_y,d_{\ell-1}+0, z')\\[7pt]
			\widehat{\bh{G}}_{\bs A}^{\downarrow}(k_x, k_y,d_{\ell-1}+0, z')
		\end{bmatrix}-\begin{bmatrix}
			\bh{K}^{\uparrow}_{\ell} & \bh{K}^{\downarrow}_{\ell}\\[7pt]
			\bh{W}^{\uparrow}_{\ell} & \bh{W}^{\downarrow}_{\ell}   
		\end{bmatrix}\begin{bmatrix}
			\widehat{\bh{G}}_{\bs A}^{\uparrow}(k_x, k_y,d_{\ell-1}-0, z') \\[7pt]
			\widehat{\bh{G}}_{\bs A}^{\downarrow}(k_x, k_y,d_{\ell-1}-0, z')
		\end{bmatrix}=\bs 0,
	\end{equation}
	for all $\ell=1, 2, \cdots, L$, where $\widehat{\bh{G}}_{\bs A}^{\ast}(k_x, k_y, d_{\ell-1}\pm 0, z')$ are the right and left limits at $z=d_{\ell-1}$.
	From expression \eqref{hatGAdecomp}, we can calculate that
	\begin{equation*}
		\begin{split}
			\widehat{\bh{G}}_{\bs A}^{\uparrow}(k_x, k_y, d_{\ell-1}-0, z')= & \begin{cases}
				\displaystyle   \widehat{\bh{G}}_{\ell\ell'}^{\uparrow}(k_x, k_y, z')e^{\i k_{\ell z} D_{\ell}},\quad \ell\neq \ell',\\
				\displaystyle  \widehat{\bh{G}}_{\ell'\ell'}^{\uparrow}(k_x, k_y, z')e^{\i k_{\ell' z} D_{\ell'}}+\frac{\bh I}{2 \omega k_{\ell' z}} e^{\i k_{\ell' z} (d_{\ell'-1}-z')}, 
			\end{cases}\\
			\widehat{\bh{G}}_{\bs A}^{\downarrow}(k_x, k_y, d_{\ell-1}+0, z')=& \begin{cases}
				\displaystyle    \widehat{\bh{G}}_{\ell-1,\ell'}^{\downarrow}(k_x, k_y, z')e^{\i k_{\ell-1, z} D_{\ell-1}}\quad \ell\neq  \ell'+1,\\
				\displaystyle \widehat{\bh{G}}_{\ell'\ell'}^{\downarrow}(k_x, k_y, z')e^{\i k_{\ell'z} D_{\ell'}}  +\frac{\bh I}{2 \omega k_{\ell' z}}e^{\i k_{\ell' z} (z'-d_{\ell'})},
			\end{cases}\\
			\widehat{\bh{G}}_{\bs A}^{\downarrow}(k_x, k_y, d_{\ell-1}-0, z')=& \widehat{\bh{G}}_{\ell\ell'}^{\downarrow}(k_x, k_y, z'),\quad
			\widehat{\bh{G}}_{\bs A}^{\uparrow}(k_x, k_y, d_{\ell-1}+0, z')=  \widehat{\bh{G}}_{\ell-1,\ell'}^{\uparrow}(k_x, k_y, z'),
		\end{split}
	\end{equation*}
	on the interfaces $\{z=d_{\ell-1}\}_{\ell=1}^{L}$,	where 
	$$D_{\ell}=d_{\ell-1}-d_{\ell},\quad \ell=1, 2,\cdots, L-1,$$
	are the thickness of the layers. For exterior layers we use $D_0=D_L=0$ only as auxiliary reference values; if the source lies in layer $0$ or $L$.
	Substituting these limit values into \eqref{sublinearsytem1} leads to linear system
	\begin{equation}\label{linearsystemotherlayer}
		\begin{bmatrix}
			\bh{K}_{\ell }^{\uparrow}h_{\ell} & \bh{K}_{\ell }^{\downarrow}\\[7pt]
			\bh{W}_{\ell }^{\uparrow}h_{\ell}  & \bh{W}_{\ell }^{\downarrow}
		\end{bmatrix}\begin{bmatrix}
			\widehat{\bh{G}}_{\ell\ell'}^{\uparrow}\\
			\widehat{\bh{G}}_{\ell\ell'}^{\downarrow}
		\end{bmatrix}-\begin{bmatrix}
			\bh{K}_{\ell-1}^{\uparrow} & \bh{K}_{\ell-1}^{\downarrow}h_{\ell-1}\\[7pt]
			\bh{W}_{\ell-1}^{\uparrow}& \bh{W}_{\ell-1}^{\downarrow}h_{\ell-1}
		\end{bmatrix}\begin{bmatrix}
			\widehat{\bh{G}}_{\ell-1,\ell'}^{\uparrow}\\
			\widehat{\bh{G}}_{\ell-1,\ell'}^{\downarrow}
		\end{bmatrix}=\mathbb S_{\ell},\quad \ell=1, 2, \cdots, L,
	\end{equation}
	where 
	\begin{equation*}
		h_{\ell}(k_{\rho})=e^{\ri k_{\ell z}D_{\ell}},\quad \bh S_{\ell}=\begin{cases}
			\displaystyle 0, & \ell\neq \ell', \ell'+1,\\
			\displaystyle-\frac{e^{\i k_{\ell' z} (d_{\ell'-1}-z')}}{2\omega k_{\ell' z}}\begin{bmatrix}
				\bh{K}_{\ell'}^{\uparrow}\\
				\bh{W}_{\ell'}^{\uparrow}
			\end{bmatrix}, & \ell=\ell',\\[12pt]
			\displaystyle \frac{e^{\i k_{\ell' z} (z'-d_{\ell'})}}{2\omega k_{\ell' z}}\begin{bmatrix}
				\bh{K}_{\ell'}^{\downarrow}\\
				\bh{W}_{\ell'}^{\downarrow}\\
			\end{bmatrix}, &\ell =\ell'+1.
		\end{cases}
	\end{equation*}
	Apparently, the source terms only show up in the equations on the upper and bottom interfaces of the source layer.
	
	By the completeness of the matrix basis $\{\J{i}\}_{i=1}^9$, the solution of the linear system \eqref{linearsystemotherlayer} has representation
	\begin{equation}\label{9termsexpansion}
		\widehat{\bh{G}}_{\ell\ell'}^{\uparrow}(k_x, k_y,z')=\sum\limits_{s=1}^9\beta_{\ell s}^{\uparrow}(k_x, k_y,z')\J{s},\quad\widehat{\bh{G}}_{\ell\ell'}^{\downarrow}(k_x, k_y,z')=\sum\limits_{s=1}^9\beta_{\ell s}^{\downarrow}(k_x, k_y,z')\J{s},
	\end{equation}
	where $\{\beta_{\ell s}^{\uparrow}(k_x, k_y,z'), \beta_{\ell s}^{\downarrow}(k_x, k_y,z')\}_{s=1}^9$ are coefficients to be determined.
	The matrix space $\mathcal S={\rm span}\{\J{1}, \J{2},\cdots, \J{9}\}$ has orthogonal decomposition $\mathcal S=\mathcal S_1\oplus \mathcal S_2$ where 
	$$\mathcal S_1={\rm span}\{\J{1}, \J{2},\cdots, \J{5}\}, \quad \mathcal S_2={\rm span}\{\J{6}, \J{7},\J{8}, \J{9}\}.$$
	We decompose the solution $\widehat{\bh G}_{\ell\ell'}^{*}$ into 
	\begin{equation*}
		\widehat{\bh G}_{\ell\ell'}^{*}=\widehat{\bh G}_{\ell\ell'1}^{*}+\widehat{\bh G}_{\ell\ell'2}^{*},\quad *=\uparrow,\downarrow,
	\end{equation*}
	where
	\begin{equation*}
		\widehat{\bh G}_{\ell\ell'1}^{*}=\sum\limits_{s=1}^5\beta_{\ell s}^{*}(k_x, k_y,z')\J{s},\quad \widehat{\bh G}_{\ell\ell'2}^{*}=\sum\limits_{s=6}^9\beta_{\ell s}^{*}(k_x, k_y,z')\J{s},\quad *=\uparrow,\downarrow.
	\end{equation*}
	Then, equations \eqref{linearsystemotherlayer} can be rewritten as
	\begin{equation}\label{decomposedlinearsystem2}
		\begin{bmatrix}
			\bh{K}_{\ell }^{\uparrow}h_{\ell} & \bh{K}_{\ell }^{\downarrow}\\[7pt]
			\bh{W}_{\ell }^{\uparrow}h_{\ell}  & \bh{W}_{\ell }^{\downarrow}
		\end{bmatrix}\begin{bmatrix}
			\widehat{\bh{G}}_{\ell,\ell'1}^{\uparrow}+\widehat{\bh{G}}_{\ell\ell'2}^{\uparrow}\\
			\widehat{\bh{G}}_{\ell,\ell'1}^{\downarrow}+\widehat{\bh{G}}_{\ell\ell'2}^{\downarrow}
		\end{bmatrix}-\begin{bmatrix}
			\bh{K}_{\ell-1}^{\uparrow} & \bh{K}_{\ell-1}^{\downarrow}h_{\ell-1}\\[7pt]
			\bh{W}_{\ell-1}^{\uparrow}& \bh{W}_{\ell-1}^{\downarrow}h_{\ell-1}
		\end{bmatrix}\begin{bmatrix}
			\widehat{\bh{G}}_{\ell-1,\ell'1}^{\uparrow}+\widehat{\bh{G}}_{\ell-1,\ell'2}^{\uparrow}\\
			\widehat{\bh{G}}_{\ell-1,\ell'1}^{\downarrow}+\widehat{\bh{G}}_{\ell-1,\ell'2}^{\downarrow}
		\end{bmatrix}=\mathbb S_{\ell}
	\end{equation}
	for all $\ell=1, 2, \cdots, L$.
	By the definition \eqref{KWmatrices} and the product table \eqref{producttable}, we can check that
	\begin{equation*}
		\bh{K}_{\ell}^{\uparrow}h_{\ell}\widehat {\bh G}_{\ell\ell'j}^{\uparrow},\bh{K}_{\ell }^{\uparrow}\widehat {\bh G}_{\ell\ell'j}^{\uparrow}, \bh{K}_{\ell }^{\downarrow}h_{\ell}\widehat {\bh G}_{\ell\ell'j}^{\downarrow},
		\bh{K}_{\ell }^{\downarrow}\widehat {\bh G}_{\ell\ell'j}^{\downarrow},\bh{W}_{\ell }^{\uparrow}h_{\ell}\widehat {\bh G}_{\ell\ell'j}^{\uparrow},\bh{W}_{\ell }^{\uparrow}\widehat {\bh G}_{\ell\ell'j}^{\uparrow}, \bh{W}_{\ell }^{\downarrow}h_{\ell}\widehat {\bh G}_{\ell\ell'j}^{\downarrow},
		\bh{W}_{\ell }^{\downarrow}\widehat {\bh G}_{\ell\ell'j}^{\downarrow}\in \mathcal S_j,
	\end{equation*}
	for $j=1, 2$. Therefore, we have
	\begin{equation*}
		\begin{split}
			&\begin{bmatrix}
				\bh{K}_{\ell }^{\uparrow}h_{\ell} & \bh{K}_{\ell }^{\downarrow}\\[7pt]
				\bh{W}_{\ell }^{\uparrow}h_{\ell}  & \bh{W}_{\ell }^{\downarrow}
			\end{bmatrix}\begin{bmatrix}
				\widehat{\bh{G}}_{\ell,\ell'j}^{\uparrow}\\
				\widehat{\bh{G}}_{\ell,\ell'j}^{\downarrow}
			\end{bmatrix}-\begin{bmatrix}
				\bh{K}_{\ell-1}^{\uparrow} & \bh{K}_{\ell-1}^{\downarrow}h_{\ell-1}\\[7pt]
				\bh{W}_{\ell-1}^{\uparrow}& \bh{W}_{\ell-1}^{\downarrow}h_{\ell-1}
			\end{bmatrix}\begin{bmatrix}
				\widehat{\bh{G}}_{\ell-1,\ell'j}^{\uparrow}\\
				\widehat{\bh{G}}_{\ell-1,\ell'j}^{\downarrow}
			\end{bmatrix}\in \mathcal S_j\times \mathcal S_j,\quad j=1, 2,
		\end{split}
	\end{equation*}
	for all $\ell=1, 2, \cdots, L$, where $\mathcal S_j\times \mathcal S_j$ denotes the space of $6\times 3$ matrices with the upper and lower $3\times 3$ square blocks belonging to $\mathcal S_j$.
	Moreover, $\bh{S}_{\ell'}$, $\bh{S}_{\ell'+1}$ belong to $\mathcal S_1\times \mathcal S_1$. Consequently, the linear system \eqref{decomposedlinearsystem2} can be decoupled into the following two independent linear systems:
	\begin{equation}\label{reducedlinearsystem}
		\begin{split}
			&\begin{bmatrix}
				\bh{K}_{\ell }^{\uparrow}h_{\ell} & \bh{K}_{\ell }^{\downarrow}\\[7pt]
				\bh{W}_{\ell }^{\uparrow}h_{\ell}  & \bh{W}_{\ell }^{\downarrow}
			\end{bmatrix}\begin{bmatrix}
				\widehat{\bh{G}}_{\ell,\ell'1}^{\uparrow}\\
				\widehat{\bh{G}}_{\ell,\ell'1}^{\downarrow}
			\end{bmatrix}-\begin{bmatrix}
				\bh{K}_{\ell-1}^{\uparrow} & \bh{K}_{\ell-1}^{\downarrow}h_{\ell-1}\\[7pt]
				\bh{W}_{\ell-1}^{\uparrow}& \bh{W}_{\ell-1}^{\downarrow}h_{\ell-1}
			\end{bmatrix}\begin{bmatrix}
				\widehat{\bh{G}}_{\ell-1,\ell'1}^{\uparrow}\\
				\widehat{\bh{G}}_{\ell-1,\ell'1}^{\downarrow}
			\end{bmatrix}=\mathbb S_{\ell},\quad \ell=1, 2, \cdots, L,\\
			&\begin{bmatrix}
				\bh{K}_{\ell }^{\uparrow}h_{\ell} & \bh{K}_{\ell }^{\downarrow}\\[7pt]
				\bh{W}_{\ell }^{\uparrow}h_{\ell}  & \bh{W}_{\ell }^{\downarrow}
			\end{bmatrix}\begin{bmatrix}
				\widehat{\bh{G}}_{\ell,\ell'2}^{\uparrow}\\
				\widehat{\bh{G}}_{\ell,\ell'2}^{\downarrow}
			\end{bmatrix}-\begin{bmatrix}
				\bh{K}_{\ell-1}^{\uparrow} & \bh{K}_{\ell-1}^{\downarrow}h_{\ell-1}\\[7pt]
				\bh{W}_{\ell-1}^{\uparrow}& \bh{W}_{\ell-1}^{\downarrow}h_{\ell-1}
			\end{bmatrix}\begin{bmatrix}
				\widehat{\bh{G}}_{\ell-1,\ell'2}^{\uparrow}\\
				\widehat{\bh{G}}_{\ell-1,\ell'2}^{\downarrow}
			\end{bmatrix}=\bd 0,\quad \ell=1, 2, \cdots, L.
		\end{split}
	\end{equation}
	Now, keep the first five terms in the solution \eqref{9termsexpansion} and denote by 
	\begin{equation}\label{reaction_coeffpart_inR0}
		\widehat{\bh{G}}_{\ell\ell'}^{\uparrow}(k_x, k_y,z')=\sum\limits_{s=1}^5\beta_{\ell s}^{\uparrow}(k_x, k_y,z')\J{s},\quad\widehat{\bh{G}}_{\ell\ell'}^{\downarrow}(k_x, k_y,z')=\sum\limits_{s=1}^5\beta_{\ell s}^{\downarrow}(k_x, k_y,z')\J{s}.
	\end{equation}
	Equations in \eqref{reducedlinearsystem} show that they also satisfy all the equations in \eqref{linearsystemotherlayer} as the components $\mathbb G_{\ell\ell'2}^{\ast}$, $\ast=\uparrow,\downarrow$ are simply set to zero. 
	
	Define 
	\begin{equation}\label{freespacecoef}
		\begin{split}
			& e^{\uparrow}_{\ell}(z)=e^{\ri k_{\ell z}(z-d_{\ell})}, \quad e_{\ell}^{\downarrow}(z)=e^{\ri k_{\ell z}(d_{\ell-1}-z)},\quad \beta_{3}^f =\beta_{4}^f=\beta_{5}^f=0,\\
			&  \beta_{ 1}^f=\beta_{ 2}^f
			=\frac{1}{2 \omega k_{\ell' z}} \Big[e^{\i k_{\ell' z} (d_{\ell'}-z')} H(z-z')e_{\ell'}^{\uparrow}(z)+e^{\i k_{\ell' z} (z'-d_{\ell'-1})} H(z'-z)e_{\ell'}^{\downarrow}(z)\Big],
		\end{split}
	\end{equation}
	From \eqref{GAfree}, $\widehat{\bh{G}}_{\bs A}^f(k_{\rho}, z, z')$ can be written in the form
	\begin{equation}\label{freespaceexp}
		\widehat{\bh{G}}_{\bs A}^f(k_{\rho}, z, z')=\sum\limits_{s=1}^5\beta_{ s}^f(k_{\rho}, z, z')\J{s}
	\end{equation}
	Then, using the formulas \eqref{reaction_coeffpart_inR0}-\eqref{freespaceexp} in \eqref{eq-GAr} and \eqref{eq-GAf}, we obtain
	\begin{equation}\label{reactmatrixbasis}
		\widehat{\bh{G}}_{\bs A}(k_x, k_y, z, z')=\sum\limits_{s=1}^5\beta_{ \ell s}(k_x, k_y, z, z')\J{s},\quad d_{\ell}<z<d_{\ell-1},
	\end{equation}
	where 
	\begin{equation}\label{coefficients1}
		\beta_{ \ell s}(k_x, k_y, z, z')=\beta_{\ell s}^{\uparrow}(k_x, k_y, z')e_{\ell}^{\uparrow}(z)+\beta_{\ell s}^{\downarrow}(k_x, k_y, z')e_{\ell}^{\downarrow}(z)+\delta_{\ell\ell'}\beta_s^f(k_{\rho}, z, z'),
	\end{equation}
	for $s=1, 2, \cdots,5$. 
	Note that
	\begin{equation*}
		\begin{aligned}
			\partial_z^2\beta_{ s}^f(k_{\rho}, z, z')=&-k_{\ell' z}^2\beta_{ s}^f(k_{\rho}, z, z')-\frac{1}{2\ri \omega }(e^{\i k_{\ell' z} (z-z')}+e^{\i k_{\ell' z} (z'-z)})\delta(z-z')\\
			=&-k_{\ell' z}^2\beta_{ s}^f(k_{\rho}, z, z')-\frac{\delta(z-z')}{\ri\omega},\quad s=1, 2.
		\end{aligned}
	\end{equation*}
	Then, \eqref{freespacecoef} and \eqref{coefficients1} shows that
	\begin{equation*}
		\begin{aligned}
			\partial_z^2\beta_{\ell s}(k_x, k_y, z, z')=&\partial_z^2\beta_s^f\delta_{\ell\ell'}-k_{\ell z}^2\left[\beta_{\ell s}^{\uparrow}e^{\uparrow}_{\ell}(z)+\beta_{\ell s}^{\downarrow}e^{\downarrow}_{\ell}(z)\right]\\
			=&-\frac{\delta_{\ell\ell'}\delta(z-z')}{\ri\omega}-k_{\ell z}^2\beta_{\ell s},\quad s=1, 2,\\
			\partial_z^2\beta_{\ell s}(k_x, k_y, z, z')=&-k_{\ell z}^2\beta_{\ell s},\quad s=3, 4, 5.
		\end{aligned}
	\end{equation*}
	Therefore, the coefficients $\{\beta_{\ell s}\}_{s=1}^5$ satisfy differential equations
	\begin{equation}\label{coefficienteq}
		\begin{cases}
			\displaystyle\partial_{zz}\beta_{\ell s}+k_{\ell z}^2\beta_{\ell s}=-\frac{\delta_{\ell\ell'}}{\ri\omega}\delta(z-z'), &  d_{\ell}<z<d_{\ell-1},\; s=1, 2,\\
			\displaystyle\partial_{zz}\beta_{\ell s}+k_{\ell z}^2\beta_{\ell s}=0,& d_{\ell}<z<d_{\ell-1},\; s=3, 4, 5.
		\end{cases}
	\end{equation}
	
	\subsection{Two Helmholtz problems in layered media} In this subsection, we show that $\widehat{\bh{G}}_{\bs A}(k_x, k_y, z, z')$ given by \eqref{reactmatrixbasis} is not unique and $\widehat{\bh{G}}_{\bs E}(k_x, k_y, z, z')$, $\widehat{\bh{G}}_{\bs H}(k_x, k_y, z, z')$ can be determined by solving two Helmholtz problems in layered media.

	From \eqref{GHoperator} -\eqref{GEoperator} and \eqref{coefficienteq} we can calculate that $\widehat{\bh{G}}_{\bs E}, \widehat{\bh{G}}_{\bs H}$ in \eqref{eq-GE-GH-GA-form1} has expressions
	\begin{equation}\label{GEexp}
		\begin{split}
			\widehat{\bh{G}}_{\bs E}
			=&\i\omega \left(\bh{I}+\frac{\widehat\nabla\widehat\nabla}{k_{\ell}^2}\right)\Big(\sum\limits_{s=1}^5 \beta_{\ell s} \J{s}\Big)\\
			=&\ri\omega\Big[\beta_{\ell 1}\J{1}+\frac{k_{\rho}^2(\beta_{\ell 2}-\partial_z\beta_{\ell 3})}{k_{\ell}^2}\J{2}+\frac{\partial_z\beta_{\ell 2}+ k_{\ell z}^2\beta_{\ell 3}}{k_{\ell}^2}\J{3}\Big]-\frac{\delta_{\ell\ell'}}{k_{\ell}^2}\delta(z-z')\J{2} \\
			&+\frac{\ri\omega}{k_{\ell}^2}\Big[\big(\partial_z\beta_{\ell 1}+k_{\rho}^2\beta_{\ell 4}-k_{\rho}^2\partial_z\beta_{\ell 5}\big)\J{4}+\big(\beta_{\ell 1}+\partial_z\beta_{\ell 4}+ k_{\ell z}^2\beta_{\ell 5}\big)\J{5}\Big],\\
			\widehat{\bh{G}}_{\bs H}
			=&\frac{1}{\mu_{\ell}}\Big(\beta_{\ell 1}\J{6}+\big(\beta_{\ell 2}-\partial_z\beta_{\ell 3}\big)\J{7}-\big(\beta_{\ell 4}-\partial_z\beta_{\ell 5}\big)\J{8}-\partial_z\beta_{\ell 1}\J{9}\Big).
		\end{split}
	\end{equation}
	Further, equations in \eqref{coefficienteq} implies
	\begin{equation}
		\beta_{\ell 1}+\partial_z\beta_{\ell 4}+ k_{\ell z}^2\beta_{\ell 5}=\frac{1}{k_{\rho}^2}\partial_z\big(\partial_z\beta_{\ell 1}+k_{\rho}^2\beta_{\ell 4}-k_{\rho}^2\partial_z\beta_{\ell 5}\big)+\frac{k_{\ell}^2}{k_{\rho}^2}\beta_{\ell 1}+\frac{\delta_{\ell\ell'}}{\ri\omega k_{\rho}^2}\delta(z-z').
	\end{equation}
	Consequently, we can reduce the number of independent coefficients in the expressions \eqref{GEexp} by introducing three new groups of coefficients as follows:
	\begin{align}\label{eq-b}
		\begin{split}
			\alpha_{\ell 1} = \beta_{\ell 1}, \quad \alpha_{\ell 2} = \frac{1}{\mu_{\ell}}\left( \beta_{\ell 2} - \partial_z \beta_{\ell 3} \right), \quad
			\alpha_{\ell 3} = \frac{1}{\mu_{\ell}}\left(\partial_z \beta_{\ell 1} + \kr^2 \beta_{\ell 4} -\kr^2  \partial_z \beta_{\ell 5} \right).
		\end{split}
	\end{align}
	The representations in \eqref{GEexp} are reformulated into
	\begin{equation}\label{eq-GE-b}
		\begin{split}
			&\widehat{\bh{G}}_{\bs E} = \frac{\i \omega}{k_{\ell}^2} \left[ k_{\ell}^2 \alpha_{\ell 1} \J{1} + \mu_{\ell} \kr^2 \alpha_{\ell 2} \J{2} + \mu_{\ell} \partial_z \alpha_{\ell 2} \J{3} + \mu_{\ell} \alpha_{\ell 3}\J{4} + \left(\frac{k_{\ell}^2}{\kr^2}\alpha_{\ell 1} + \frac{\mu_{\ell}}{\kr^2} \partial_z \alpha_{\ell 3}\right)\J{5} \right]+\boldsymbol{\delta}, \\
			& \widehat{\bh{G}}_{\bs H} = \frac{1}{\mu_{\ell}}\left[ \alpha_{\ell 1} \J{6} + \mu_{\ell} \alpha_{\ell 2}\J{7} + \left(\frac{1}{\kr^2} \partial_z\alpha_{\ell 1} - \frac{\mu_{\ell}}{\kr^2} \alpha_{\ell 3}\right)\J{8} - \partial_z \alpha_{\ell 1} \J{9} \right],
		\end{split}
	\end{equation}
	for $d_{\ell}<z<d_{\ell-1}$, where
	\begin{equation}\label{delta_Ge_hat}
		\boldsymbol{\delta}:=\left[\dfrac{\J{5}}{k_{\rho}^2}-\J{2}\right]\dfrac{\delta_{\ell\ell'}}{k_{\ell}^2}\delta(z-z').
	\end{equation}

	Define 
	piece-wise smooth functions
	\begin{equation}\label{bjdef}
		\alpha_{j}(k_x, k_y, z, z')=\alpha_{\ell j}(k_x, k_y, z, z'),\quad d_{\ell}<z<d_{\ell-1},
	\end{equation}
	in the layered media. Using the expression \eqref{eq-GE-b} in \eqref{EJump}, we obtain 
	\begin{equation}\label{EJumpBasis}
		\begin{split}
			& \left\llbracket \alpha_{ 1} \J{1}  + \frac{\mu}{k^2}\partial_z \alpha_{2} \J{3} + \left(\frac{1}{\kr^2}\alpha_{ 1} + \frac{\mu}{k^2\kr^2} \partial_z \alpha_{ 3}\right)\J{5}\right\rrbracket=\bs 0,\\
			&\left\llbracket \frac{1}{\mu}\left( \mu \alpha_{2} \J{7} + \left(\frac{1}{\kr^2}\partial_z \alpha_{1} - \frac{\mu}{\kr^2} \alpha_{3}\right)\J{8} - \partial_z \alpha_{ 1} \J{9} \right)\right\rrbracket=\bs 0.
		\end{split}
	\end{equation}
	Then, the independence and continuity of $\bs J_s$ imply that
	\begin{equation}\label{bjumpcond1}
		\llbracket \alpha_1\rrbracket=0, \quad\llbracket \alpha_2\rrbracket=0,\quad \llbracket \alpha_3\rrbracket=0,\quad \left\llbracket\frac{1}{\mu}\partial_z \alpha_1\right\rrbracket = 0,\quad\left\llbracket\frac{1}{\varepsilon}\partial_z \alpha_2\right\rrbracket = 0, \enspace \left\llbracket\frac{1}{\varepsilon}\partial_z \alpha_3\right\rrbracket = 0. 
	\end{equation}
	Consequently, the interface conditions \eqref{eq-if-cond} are equivalent to the decoupled ones in \eqref{bjumpcond1} on the coefficients. 
	
	Now, we derive differential equations for piece-wise smooth functions $\{\alpha_s(k_x, k_y, z, z')\}_{s=1}^3$ in each layer.  
	Define piece-wise smooth functions 
	\begin{equation}
		\begin{split}
			\beta_s(k_x, k_y, z, z')=&\beta_{\ell s}(k_x, k_y, z, z'),\quad d_{\ell}<z<d_{\ell-1},
		\end{split}
	\end{equation}
	Equations in \eqref{coefficienteq} imply
	\begin{equation}\label{diffeqals}
		\begin{split}
			(\partial_{zz} + k_{\ell z}^2)\beta_{s}(k_x, k_y, z, z') =&-\frac{\delta(z-z')}{\ri\omega},\quad d_{\ell}<z<d_{\ell-1},\;\;s=1, 2,\\
			(\partial_{zz} + k_{\ell z}^2)\beta_{s}(k_x, k_y, z, z') =&0,\quad d_{\ell}<z<d_{\ell-1},\;\;s=3, 4, 5.
		\end{split}
	\end{equation}
	Moreover, the layer-wisely definition \eqref{eq-b} implies
	\begin{equation}\label{newcoefficients}
		\alpha_1=\beta_1, \quad \alpha_2=\frac{1}{\mu}(\beta_2-\partial_z \beta_3),\quad \alpha_3=\frac{1}{\mu}(\partial_z\beta_1+k_{\rho}^2\beta_4-k_{\rho}^2\partial_z\beta_5).
	\end{equation}
	Therefore, we obtain interface problems for piece-wise smooth functions $\alpha_1, \alpha_2, \alpha_3$ as follows:
	\begin{equation}\label{helmholtzb1}
		\begin{cases}
			\displaystyle\partial_{zz} \alpha_1(k_x, k_y, z, z') + k_{\ell z}^2 \alpha_1(k_x, k_y, z, z')=\frac{\i}{\omega}\delta(z-z'),\quad d_{\ell}<z<d_{\ell-1},\\
			\displaystyle\llbracket \alpha_1\rrbracket=0,\quad \Big\llbracket \frac{1}{\mu}\partial_z\alpha_1\Big\rrbracket=0,\quad z=d_{\ell},\;\ell=0, 1, \cdots, L-1,
		\end{cases}
	\end{equation}
	\begin{equation}\label{helmholtzb2}
		\begin{cases}
			\displaystyle	\partial_{zz} \alpha_2(k_x, k_y, z, z') + k_{\ell z}^2 \alpha_2(k_x, k_y, z, z')=\frac{\i}{\mu\omega}\delta(z-z'),\quad d_{\ell}<z<d_{\ell-1},\\
			\displaystyle	\llbracket \alpha_2\rrbracket=0,\quad \Big\llbracket \frac{1}{\varepsilon}\partial_z\alpha_2\Big\rrbracket=0,\quad z=d_{\ell},\;\ell=0, 1, \cdots, L-1,
		\end{cases}
	\end{equation}
	\begin{equation}\label{helmholtzb3}
		\begin{cases}
			\displaystyle	\partial_{zz} \alpha_3(k_x, k_y, z, z') + k_{\ell z}^2 \alpha_3(k_x, k_y, z, z')=\frac{\i}{\mu\omega}\delta'(z-z'),\quad d_{\ell}<z<d_{\ell-1},\\
			\displaystyle	\llbracket \alpha_3\rrbracket=0,\quad \Big\llbracket \frac{1}{\varepsilon}\partial_z\alpha_3\Big\rrbracket=0,\quad z=d_{\ell},\;\ell=0, 1, \cdots, L-1,
		\end{cases}
	\end{equation}
	
	\begin{rem}
		As an analogous to the formulations in \eqref{newcoefficients}, we define
		\begin{equation*}
			\alpha_{1}^{f}=\beta_1^f =\frac{1}{\i\omega}\widehat{G}^f(\kr, z, z'), \quad \alpha_{2}^{f}= \frac{1}{\mu}\beta_2^f=\frac{\widehat{G}^f(\kr, z, z')}{\i\omega\mu_{\ell'}} \quad
			\alpha_{3}^{f} =\frac{1}{\mu}\partial_z\beta_1^f=\frac{\partial_z\widehat{G}^f(\kr, z, z') }{\i\omega\mu_{\ell'}}.
		\end{equation*}
		According to the representations in \eqref{eq-GE-b} , we have
		\begin{equation*}
			\begin{split}
				\widehat{\bh{G}}_{\bs E}^f = &\frac{\i \omega}{k_{\ell'}^2} \left[ k_{\ell'}^2 \alpha_{ 1}^f \J{1} + \mu_{\ell'} \kr^2 \alpha_{ 2}^f \J{2} + \mu_{\ell'} \partial_z \alpha_{ 2}^f \J{3} + \mu_{\ell'} \alpha_{ 3}^f\J{4} + \left(\frac{k_{\ell'}^2}{\kr^2}\alpha_{ 1}^f + \frac{\mu_{\ell'}}{\kr^2} \partial_z \alpha_{ 3}^f\right)\J{5} \right]\\
				&+\left[\dfrac{\J{5}}{k_{\rho}^2}-\J{2}\right]\dfrac{\delta(z-z')}{k_{\ell'}^2},\\
				\widehat{\bh{G}}_{\bs H}^f =& \frac{1}{\mu_{\ell'}}\left[ \alpha_{1}^f \J{6} + \mu_{\ell'} \alpha_{2}^f\J{7} + \left(\frac{1}{\kr^2} \partial_z\alpha_{1}^f - \frac{\mu_{\ell'}}{\kr^2} \alpha_{3}^f\right)\J{8} - \partial_z\alpha_{1}^f \J{9} \right].
			\end{split}
		\end{equation*}
		By using the Helmholtz equation
		\begin{equation*}
			\left[\partial_{zz}+k_{\ell'z}^2\right]\widehat{G}^{f}(k_{\rho}, z, z')=-\delta(z-z'),
		\end{equation*}
		we can further calculate that
		\begin{equation*}
			\begin{split}
				\widehat{\bh{G}}_{\bs E}^{f} =&\frac{1}{k_{\ell'}^2}\left[ k_{\ell'}^2 \J{1} +  \kr^2 \J{2} + \J{3}\partial_z   +  \J{4}\partial_z + \left(\frac{k_{\ell'}^2}{\kr^2} + \frac{1}{\kr^2} \partial_{zz}\right)\J{5} \right]\widehat{G}^f(\kr, z, z')\\
				&+\left[\dfrac{\J{5}}{k_{\rho}^2}-\J{2}\right]\dfrac{\delta(z-z')}{k_{\ell'}^2}\\
				=&\left(\bh{I}+\frac{1}{k_{\ell'}^2}\left[-k_{\ell' z}^2\J{2}+(\J{3}+\J{4})\partial_z+ \J{5} \right]\right)\widehat{G}^f(\kr, z, z')- \dfrac{\delta(z-z')}{k_{\ell'}^2}\J{2}\\
				=&\Big[\bh{I}+\frac{1}{k_{\ell'}^2}(\J{5}+(\J{3}+\J{4})\partial_z+ \J{2}\partial_{zz} )\Big]\widehat{G}^f(\kr, z, z')=\left[\bh{I}+\dfrac{\widehat{\nabla}\widehat{\nabla}}{k_{\ell'}^2}\right]\widehat{G}^f(\kr, z, z'),
			\end{split}
		\end{equation*}
		where the last equality is derived using \eqref{Operatordecomp}. Similarly, we have
		\begin{equation}
			\begin{split}
				\widehat{\bh{G}}_{\bs H}^{f} =\frac{1}{\i \omega\mu_{\ell'}}\left[\J{6}+\J{7} - \J{9}\partial_z   \right]\widehat{G}^f(\kr, z, z')
				=\frac{1}{\i \omega\mu_{\ell'}}\widehat{\nabla}\times\left(\widehat{G}^f(\kr, z, z')\bh{I}\right).
			\end{split}
		\end{equation}
		These results show that the representations \eqref{eq-GE-b} are consistent with the formulations \eqref{eq-GE-GH-GA-form1} in the free space.
	\end{rem}
	
	\begin{rem}
		According to the definition \eqref{newcoefficients}, $\{\beta_{\ell s}\}_{s=1}^5$ are not uniquely determined by $\{\alpha_{\ell s}\}_{s=1}^3$. Therefore, $\widehat{\bh{G}}_{\bs E}, \widehat{\bh{G}}_{\bs H}$ are uniquely determined by the coefficients $\{\alpha_{\ell s}\}_{s=1}^3$ but $\widehat{\bh{G}}_{\bs A}$ is not unique. A natural choice is to set $\beta_{\ell 3}=\beta_{\ell 5}=0$. Then, \eqref{newcoefficients} gives
		\begin{equation}
			\beta_{\ell 1}=\alpha_{\ell 1},\quad \beta_{\ell2}=\mu_{\ell}\alpha_{\ell 2},\quad \beta_{\ell 4}=\frac{\mu_{\ell}\alpha_{\ell 3}}{k_{\rho}^2}-\frac{1}{k_{\rho}^2}\partial_z\alpha_{\ell 1}.
		\end{equation}
		This group of coefficients leads to the so-called Sommerfeld potential
		\begin{equation}\label{GA_spectral}
			\widehat{\bh{G}}_{\bs A}^{S}(k_x, k_y, z, z')=\alpha_{\ell 1}\J{1}+\mu_{\ell}\alpha_{\ell 2}\J{2}+\Big(\frac{\mu_{\ell}\alpha_{\ell 3}}{k_{\rho}^2}-\frac{1}{k_{\rho}^2}\partial_z\alpha_{\ell 1}\Big)\J{4},\quad d_{\ell}<z<d_{\ell-1},
		\end{equation}
		which has non-zero pattern
		\begin{equation}\label{GA_form}
			\widehat{\bh{G}}_{\bs A}^{S}(k_x, k_y, z, z')=\begin{bmatrix}
				\times &  & \\
				& \times & \\
				\times & \times &\times
			\end{bmatrix}.
		\end{equation}
	\end{rem}
	
	Apparently, the interface problems \eqref{helmholtzb1}-\eqref{helmholtzb3} are exactly the same as in \eqref{G_f1_frequency}-\eqref{G_f3_frequency} except for some extra constants on the right-hand side. Actually, we have
	\begin{equation}\label{solutions_b123}
		\alpha_{1}=-\frac{\ri}{\omega}\widehat{G}_{1},\quad \alpha_{2}=-\frac{\ri}{\omega\mu_{\ell'}}\widehat{G}_{2},\quad \alpha_{3}=-\frac{\ri}{\omega\mu_{\ell'}}\widehat{G}_{3}.
	\end{equation}
	More precisely, we have
	\begin{equation}\label{bell123_solution}
		\begin{split}
			\alpha_{\ell1}(k_{\rho}, z, z')&= \delta_{\ell\ell'} \alpha_1^f(k_{\rho}, z, z')+\frac{1}{2\omega k_{\ell' z}}\sum_{\ast,\star=\uparrow,\downarrow}\phi_{\ell\ell'}^{\ast\star}(\kr){\mathcal Z}_{\ell\ell'}^{\ast\star}(k_{\rho}, z,z'),\\
			\alpha_{\ell2}(k_{\rho}, z, z')&= \delta_{\ell\ell'} \alpha_2^f(k_{\rho}, z, z')+\frac{1}{2\omega\mu_{\ell'} k_{\ell' z}}\sum_{\ast,\star=\uparrow,\downarrow}\psi_{\ell\ell'}^{\ast\star}(\kr){\mathcal Z}_{\ell\ell'}^{\ast\star}(k_{\rho}, z,z'),\\
			\alpha_{\ell3}(k_{\rho}, z, z')&=-\delta_{\ell\ell'} \partial_{z'}\alpha_2^f(k_{\rho}, z, z')-\frac{\ri }{2\omega \mu_{\ell'}}\sum_{\ast,\star=\uparrow,\downarrow}s_{2}^{\star}\psi_{\ell\ell'}^{\ast\star}(\kr){\mathcal Z}_{\ell\ell'}^{\ast\star}(k_{\rho}, z,z'),
		\end{split}
	\end{equation}
	where ${\mathcal Z}_{\ell\ell'}^{\ast\star}(k_{\rho}, z,z')$ is defined in \eqref{zexponential} and $\phi_{\ell\ell'}^{\ast\star}(k_{\rho})$,  $\psi_{\ell\ell'}^{\ast\star}(k_{\rho})$ are densities used in \eqref{threedensity}. Define coefficients for the reaction components as follows
	\begin{equation*}
		\begin{split}
			&\alpha_{\ell 1}^r(k_{\rho},z,z')=\frac{1}{2\omega k_{\ell' z}}\sum_{\ast,\star=\uparrow,\downarrow}\phi_{\ell\ell'}^{\ast\star}(\kr){\mathcal Z}_{\ell\ell'}^{\ast\star}(k_{\rho}, z,z'),\quad d_{\ell}<z<d_{\ell-1},\\
			&\alpha_{\ell 2}^r(k_{\rho},z,z')=\frac{1}{2\omega\mu_{\ell'} k_{\ell' z}}\sum_{\ast,\star=\uparrow,\downarrow}\psi_{\ell\ell'}^{\ast\star}(\kr){\mathcal Z}_{\ell\ell'}^{\ast\star}(k_{\rho}, z,z'),\quad d_{\ell}<z<d_{\ell-1},\\
			&\alpha_{\ell 3}^r(k_{\rho},z,z')=-\frac{\ri }{2\omega \mu_{\ell'}}\sum_{\ast,\star=\uparrow,\downarrow}s_{2}^{\star}\psi_{\ell\ell'}^{\ast\star}(\kr){\mathcal Z}_{\ell\ell'}^{\ast\star}(k_{\rho}, z,z'),\quad d_{\ell}<z<d_{\ell-1},
		\end{split}
	\end{equation*}
	for all $\ell=0,1, \cdots, L$.
	Then, substituting \eqref{bell123_solution} into  \eqref{eq-GE-b}, we obtain
	\begin{equation}
		\begin{split}
			\widehat{\bh{G}}_{\bs E}=&\delta_{\ell\ell'}\widehat{\bh{G}}_{\bs E}^f+\frac{\i \omega}{k_{\ell}^2} \left[ k_{\ell}^2 \alpha_{\ell 1}^r \J{1} + \mu_{\ell} \kr^2 \alpha_{\ell 2}^r \J{2} + \mu_{\ell} \partial_z \alpha_{\ell 2}^r \J{3} + \mu_{\ell} \alpha_{\ell 3}^r\J{4} + \left({k_{\ell}^2}\alpha_{\ell 1}^r + {\mu_{\ell}} \partial_z \alpha_{\ell 3}^r\right)\dfrac{\J{5}}{\kr^2} \right]\\
			=&\delta_{\ell\ell'}\widehat{\bh{G}}_{\bs E}^f+\frac{\i}{2k_{\ell'z}}\sum\limits_{\ast,\star=\uparrow,\downarrow}{\mathcal Z}_{\ell\ell'}^{\ast\star}(k_{\rho}, z, z')\tilde{\bs \Theta}_{\mathbf E,\ell\ell'}^{\ast\star}(k_x, k_y):=\delta_{\ell\ell'}\widehat{\bh{G}}_{\bs E}^f+\sum\limits_{\ast,\star=\uparrow,\downarrow}\widehat{\bh{G}}_{\bs E,\ell\ell'}^{\ast\star}
		\end{split}
	\end{equation}
	and
	\begin{equation}
		\begin{split}
			\widehat{\bh{G}}_{\bs H}=&\delta_{\ell\ell'}\widehat{\bh{G}}_{\bs H}^f+\frac{1}{\mu_{\ell}}\left[ \alpha_{\ell 1}^r \J{6} + \mu_{\ell}\alpha_{\ell 2}^r\J{7} + \left( \partial_z\alpha_{\ell 1}^r - {\mu_{\ell}} \alpha_{\ell 3}^r\right)\frac{\J{8}}{\kr^2} - \partial_z \alpha_{\ell 1}^r \J{9} \right]\\
			=&\delta_{\ell\ell'}\widehat{\bh{G}}_{\bs H}^f+\frac{1}{2\omega\mu_{\ell}k_{\ell'z}}\sum\limits_{\ast,\star=\uparrow,\downarrow}{\mathcal Z}_{\ell\ell'}^{\ast\star}(k_{\rho}, z, z')\tilde{\bs \Theta}_{\mathbf H,\ell\ell'}^{\ast\star}(k_x, k_y):=\delta_{\ell\ell'}\widehat{\bh{G}}_{\bs H}^f+\sum\limits_{\ast,\star=\uparrow,\downarrow}\widehat{\bh{G}}_{\bs H,\ell\ell'}^{\ast\star}	\end{split}
	\end{equation}
	where
	\begin{equation}
		\begin{split}
			\tilde{\bs \Theta}_{\mathbf E,\ell\ell'}^{\ast\star} &= \phi_{\ell\ell'}^{\ast\star} \Big(\J{1}+ \frac{1}{\kr^2}\J{5}\Big)+ \frac{\mu_{\ell}\psi_{\ell\ell'}^{\ast\star} }{\mu_{\ell'}k_{\ell}^2} \Big(\kr^2\J{2}+\ri s_1^{\ast}k_{\ell z}\J{3}-  \ri s_2^{\star}k_{\ell' z}\J{4} +s_1^{\ast}s_2^{\star}\frac{k_{\ell z} k_{\ell' z}}{\kr^2}  \J{5}\Big), \\
			\tilde{\bs \Theta}_{\mathbf H,\ell\ell'}^{\ast\star} &= -\phi_{\ell\ell' }^{\ast\star} \Big(-\J{6}+\ri s_1^{\ast}k_{\ell z}\Big(\J{9}-\frac{\J{8}}{\kr^2}\Big)\Big)+\frac{\mu_{\ell} }{\mu_{\ell'}}\psi_{\ell\ell'}^{\ast\star} \Big(\J{7}+ \frac{\ri s_2^{\star} k_{\ell' z}}{\kr^2}  \J{8}\Big). \\
		\end{split}
	\end{equation}


	Note that the matrix basis \eqref{mat-J} possess profound geometric and physical significance. Actually, it is essentially a concrete representation of the tensor products (dyads) of the three unit direction vectors $(\hat{\bf u},\hat{\bf v},\hat{\bf z})$, namely
	\begin{equation}\label{twobasisconn}
		\begin{aligned}
			&\mathbf{\hat{u}}\otimes\mathbf{\hat{u}}^{\rm T}=-\dfrac{\J{5}}{k_{\rho}^2},\quad\mathbf{\hat{u}}\otimes\mathbf{\hat{v}}^{\rm T}=\J{9}-\dfrac{\J{8}}{k_{\rho}^2},\quad\mathbf{\hat{u}}\otimes\mathbf{\hat{z}}^{\rm T}=-\dfrac{\ri\J{3}}{k_{\rho}},\\
			&\mathbf{\hat{v}}\otimes\mathbf{\hat{v}}^{\rm T}=\J{1}+\dfrac{\J{5}}{k_{\rho}^2},\quad\mathbf{\hat{v}}\otimes\mathbf{\hat{u}}^{\rm T}=-\dfrac{\J{8}}{k_{\rho}^2},\quad\mathbf{\hat{v}}\otimes\mathbf{\hat{z}}^{\rm T}=\dfrac{\ri\J{7}}{k_{\rho}},\\
			&\mathbf{\hat{z}}\otimes\mathbf{\hat{u}}^{\rm T}=-\dfrac{\ri\J{4}}{k_{\rho}},\quad\mathbf{\hat{z}}\otimes\mathbf{\hat{v}}^{\rm T}=-\dfrac{\ri\J{6}}{k_{\rho}},\quad\mathbf{\hat{z}}\otimes\mathbf{\hat{z}}^{\rm T}=\J{2}.
		\end{aligned}
	\end{equation}
	Using \eqref{solutions_b123} and \eqref{twobasisconn} in \eqref{eq-GE-b} and re-organizing the results leads to expressions in \eqref{Hat_E_layeredv}. Therefore, we conclude that the matrix basis formulations \eqref{eq-GE-b} for the dyadic Green's functions $\widehat{\bh G}_{\bs E},\widehat{\bh G}_{\bs H}$ are equivalent to one obtained using the TE/TM decomposition.
	
	The TE/TM decomposition (cf. \cite{TETM2002}) and the matrix basis proposed in \cite{bo2022maxwellDGF} are effective tools for handling vector wave equations in layered media from different perspectives. The former is based on the physically intuitive decoupling of wave modes, while the latter is based on a systematic algebraic expansion. This paper clarifies that both methods share the same simplified mathematical core structure (three scalar Helmholtz problems), and their solutions and final physical outputs (dyadic Green's functions) are the same. The matrix basis method can be viewed as an algebraically more general implementation of the TE/TM decomposition idea, independent of an explicit transverse direction. This mathematical understanding paves the way for the investigation of other vector wave equations (e.g., elastic wave equation \cite{yuan2026computationgreensfunctionlayered}) in layered media.
	
	\subsection{Uniform integral representation for the DGFs in the physical domain}
	Note that the angular terms in the above matrices can be rewritten as
	\begin{equation*}
		\begin{split}
			\frac{k_x}{\kr} &= \frac{e^{\i \alpha} + e^{-\i\alpha}}{2}, \quad
			\frac{k_y}{\kr} = \frac{\ri(e^{-\i \alpha} - e^{\i\alpha})}{2}, \\
			\frac{k_x^2}{\kr^2} &= \frac{1}{2} + \frac{e^{2\i\alpha} +e^{-2\i\alpha}}{4} , \quad
			\frac{k_x k_y}{\kr^2} = \frac{\i(e^{-2\i\alpha} - e^{2\i\alpha})}{4}, \quad
			\frac{k_y^2}{\kr^2} = \frac{1}{2} - \frac{e^{2\i\alpha} +e^{-2\i\alpha}}{4},
		\end{split}
	\end{equation*}
	where $\alpha$ is the polar angle of the vector $(k_x, k_y)$.
	We have
	\begin{equation*}
		\begin{split}
			& \J{1}+ \frac{\J{5}}{\kr^2}={\bs M}_1+e^{2\ri\alpha}{\bs M}_2+e^{-2\ri\alpha}{\bs M}_3,\quad \J{2}={\bs M}_6,\quad \frac{\J{3}}{\ri\kr}=e^{\ri\alpha}{\bs M}_4+e^{-\ri\alpha}{\bs M}_5,\\
			&\frac{\J{4}}{\ri\kr}=e^{\ri\alpha}{\bs M}_4^{\rm T}+e^{-\ri\alpha}{\bs M}_5^{\rm T},\quad  -\frac{\J{5}}{\kr^2}={\bs M}_1-e^{2\ri\alpha}{\bs M}_2-e^{-2\ri\alpha}{\bs M}_3,\\
			&\frac{\J{6}}{\ri k_{\rho}}=\ri e^{\ri\alpha}{\bs M}_4^{\rm T}-\ri e^{-\ri\alpha}{\bs M}_5^{\rm T},\quad\frac{\J{7}}{\ri k_{\rho}}=-\ri e^{\ri\alpha}{\bs M}_4+\ri e^{-\ri\alpha}{\bs M}_5,\\
			&\frac{\J{8}}{\kr^2}={\bs M}_7+\ri e^{2\ri\alpha}{\bs M}_2-\ri e^{-2\ri\alpha}{\bs M}_3,\quad\J{9}-\frac{\J{8}}{\kr^2}={\bs M}_7-\ri e^{2\ri\alpha}{\bs M}_2+\ri e^{-2\ri\alpha}{\bs M}_3
		\end{split}
	\end{equation*}
	with the constant matrices defined by
	\begin{equation*}
		\begin{array}{lll}
			{\bh M}_1=\begin{bmatrix}
				\frac{1}{2} & 0 & 0\\
				0 & \frac{1}{2} & 0\\
				0 & 0 & 0
			\end{bmatrix}, & {\bh{M}}_2=\begin{bmatrix}
				-\frac{1}{4} & \frac{\ri}{4} & 0\\
				\frac{\ri}{4} & \frac{1}{4} & 0\\
				0 & 0 & 0
			\end{bmatrix},& {\bh{M}}_3=\begin{bmatrix}
				-\frac{1}{4} & -\frac{\ri}{4} & 0\\
				-\frac{\ri}{4} & \frac{1}{4} & 0\\
				0 &0 & 0
			\end{bmatrix}, \\[15pt]
			{\bh{M}}_4=\begin{bmatrix}
				0 & 0 & \frac{1}{2}\\
				0 & 0 & -\frac{\ri}{2}\\
				0 & 0 & 0
			\end{bmatrix},& {\bh{M}}_5=\begin{bmatrix}
				0 & 0 & \frac{1}{2}\\
				0 & 0 & \frac{\ri}{2}\\
				0 & 0 & 0
			\end{bmatrix},& {\bh{M}}_6=\begin{bmatrix}
				0 & 0 & 0\\
				0 & 0 & 0\\
				0 & 0 & 1
			\end{bmatrix}, \quad {\bh{M}}_7=\begin{bmatrix}
				0 & \frac{1}{2} & 0\\
				-\frac{1}{2} & 0 & 0\\
				0 & 0 & 0
			\end{bmatrix}.
		\end{array}
	\end{equation*}
	Denoted by $\gamma_{\ell\ell'}=\mu_{\ell}/(\mu_{\ell'}k_{\ell}^2)$,
	then we have
	\begin{equation}
		\begin{split}
			\bs \Theta_{\mathbf E,\ell\ell'}^{\ast\star}=&\phi_{\ell\ell'}^{\ast\star}({\bh{M}}_1+e^{2\ri\alpha}{\bh{M}}_2+e^{-2\ri\alpha}{\bh{M}}_3)+\gamma_{\ell\ell'}\psi_{\ell\ell'}^{\ast\star}\Big[k_{\rho}^2{\bh{M}}_6-s_1^{\ast}k_{\rho}k_{\ell z}(e^{\ri\alpha}{\bh{M}}_4+e^{-\ri\alpha}{\bh{M}}_5)\\
			&+s_2^{\star}k_{\rho}k_{\ell' z}(e^{\ri\alpha}{\bh{M}}_4^{\rm T}+e^{-\ri\alpha}{\bh{M}}_5^{\rm T})-s_1^{\ast}s_2^{\star}k_{\ell z}k_{\ell'z}({\bh{M}}_1-e^{2\ri\alpha}{\bh{M}}_2-e^{-2\ri\alpha}{\bh{M}}_3)\Big],\\
			\bs \Theta_{\mathbf H,\ell\ell'}^{\ast\star}=&-\phi_{\ell\ell' }^{\ast\star}\big[k_{\rho}e^{\ri\alpha}\mathbb M_4^{\rm T}-k_{\rho}e^{-\ri\alpha}\mathbb M_5^{\rm T}+\ri s_1^{\ast}k_{\ell z}\big(\mathbb M_7-\ri e^{2\ri\alpha}{\bh{M}}_2+\ri e^{-2\ri\alpha}{\bh{M}}_3\big)\big]\\
			+&\dfrac{\mu_{\ell}}{\mu_{\ell'}}\psi_{\ell\ell'}^{\ast\star}\Big[k_{\rho}e^{\ri\alpha}\mathbb M_4-k_{\rho}e^{-\ri\alpha}\mathbb M_5+\ri s_2^{\star}k_{\ell' z}\big(\mathbb M_7+\ri e^{2\ri\alpha}{\bh{M}}_2-\ri e^{-2\ri\alpha}{\bh{M}}_3\big)\Big].
		\end{split}
	\end{equation}
	
	Define densities 
	\begin{equation*}
		\begin{split}
			&\sigma_{\ell\ell'1}^{\ast\star}(\kr)=\frac{\phi_{\ell\ell'}^{\ast\star}(\kr)}{k_{\ell'z}}-s_1^{\ast}s_2^{\star}\gamma_{\ell\ell'}k_{\ell z}\psi_{\ell\ell'}^{\ast\star}(\kr),\quad \sigma_{\ell\ell'2}^{\ast\star}(\kr)=-\frac{s_1^{\ast}\gamma_{\ell\ell'}k_{\rho}k_{\ell z}}{k_{\ell'z}}\psi_{\ell\ell'}^{\ast\star}(\kr),\\
			&\sigma_{\ell\ell'3}^{\ast\star}(\kr)=\frac{\phi_{\ell\ell'}^{\ast\star}(\kr)}{k_{\ell'z}}+s_1^{\ast}s_2^{\star}\gamma_{\ell\ell'}k_{\ell z}\psi_{\ell\ell'}^{\ast\star}(\kr),\quad \sigma_{\ell\ell'4}^{\ast\star}(\kr)=s_2^{\star}\gamma_{\ell\ell'}k_{\rho}\psi_{\ell\ell'}^{\ast\star}(\kr),\\
			&\sigma_{\ell\ell'5}^{\ast\star}(\kr) = \gamma_{\ell\ell'}\frac{k_{\rho}^2}{k_{\ell' z}} \psi_{\ell\ell'}^{\ast\star}(\kr),\quad \sigma_{\ell\ell'6}^{\ast\star}=\frac{k_{\rho}}{k_{\ell'z}}\phi_{\ell\ell'}^{\ast\star}(\kr),\quad \sigma_{\ell\ell'7}^{\ast\star}(\kr)=\frac{\mu_{\ell}k_{\rho}}{\mu_{\ell'}k_{\ell'z}}\psi_{\ell\ell'}^{\ast\star}(\kr),\\
			&\sigma_{\ell\ell'8}^{\ast\star}(\kr)=\frac{\ri s_1^{\ast}k_{\ell z}}{k_{\ell'z}}\phi_{\ell\ell'}^{\ast\star}(\kr)-\frac{\ri\mu_{\ell}s_2^{\star}}{\mu_{\ell'}}\psi_{\ell\ell'}^{\ast\star}(\kr),\quad\sigma_{\ell\ell'9}^{\ast\star}(\kr)=\frac{\ri s_1^{\ast}k_{\ell z}}{k_{\ell'z}}\phi_{\ell\ell'}^{\ast\star}(\kr)+\frac{\ri\mu_{\ell}s_2^{\star}}{\mu_{\ell'}}\psi_{\ell\ell'}^{\ast\star}(\kr).
		\end{split}
	\end{equation*}
	Then, we get expressions for $\widehat{\bh{G}}_{\mathbf E,\ell\ell'}^{\ast\star}(k_{\rho}, z, z')$  and $\widehat{\bh{G}}_{\mathbf H,\ell\ell'}^{\ast\star}(k_{\rho}, z, z')$ as follows
	\begin{equation}\label{reactioncompinspectraldomain}
		\begin{split}
			\widehat{\bh{G}}_{\mathbf E,\ell\ell'}^{\ast\star}=&\i\frac{{\mathcal Z}_{\ell\ell'}^{\ast\star}}{2}[\sigma_{\ell\ell'1}^{\ast\star}\bh{M}_1+\sigma_{\ell\ell'3}^{\ast\star}e^{2\ri\alpha}\bh{M}_2+\sigma_{\ell\ell'3}^{\ast\star}e^{-2\ri\alpha}\bh{M}_3+\sigma_{\ell\ell'2}^{\ast\star}e^{\ri\alpha}\bh{M}_4\\
			&+\sigma_{\ell\ell'2}^{\ast\star}e^{-\ri\alpha}\bh{M}_5+\sigma_{\ell\ell'4}^{\ast\star}e^{\ri\alpha}\bh{M}_4^{\rm T}+\sigma_{\ell\ell'4}^{\ast\star}e^{-\ri\alpha}\bh{M}_5^{\rm T}+\sigma_{\ell\ell'5}^{\ast\star}\bh{M}_6],\\
			\widehat{\bh{G}}_{\mathbf H,\ell\ell'}^{\ast\star}=&\frac{{\mathcal Z}_{\ell\ell'}^{\ast\star}}{2\omega\mu_{\ell}}[-\sigma_{\ell\ell'6}^{\ast\star}(e^{\ri\alpha}{\bh M}_4^{\rm T}-e^{-\ri\alpha}{\bh M}_5^{\rm T})+\sigma_{\ell\ell'7}^{\ast\star}(e^{\ri\alpha}{\bh M}_4-e^{-\ri\alpha}{\bh M}_5)\\
			&-\sigma_{\ell\ell'8}^{\ast\star}\bh{M}_7+\ri\sigma_{\ell\ell'9}^{\ast\star}e^{2\ri\alpha}\bh{M}_2-\ri\sigma_{\ell\ell'9}^{\ast\star}e^{-2\ri\alpha}\bh{M}_3].
		\end{split}
	\end{equation}
	Taking inverse Fourier transform, we obtain
	\begin{equation}\label{reactcompphysicaldomain}
		\begin{split}
			\bh{G}_{\mathbf E,\ell\ell'}^{\ast\star}(\bs{r}, \bs{r}') =& \frac{1}{4\pi^2} \iint_{\mathbb{R}^2} \widehat{\bh{G}}_{\mathbf E,\ell\ell'}^{\ast\star} e^{\i k_x (x - x') + \i k_y (y - y')} dk_x dk_y\\
			=& \mathcal I_{\ell\ell',-2}^{\ast\star}[\sigma_{\ell\ell'3}^{\ast\star}]{\bh{M}}_3+\mathcal I_{\ell\ell',-1}^{\ast\star}[\sigma_{\ell\ell'2}^{\ast\star}]{\bh{M}}_5+\mathcal I_{\ell\ell',-1}^{\ast\star}[\sigma_{\ell\ell'4}^{\ast\star}]{\bh{M}}_5^{\rm T}+\mathcal I_{\ell\ell'0}^{\ast\star}[\sigma_{\ell\ell'1}^{\ast\star}]{\bh{M}}_1\\
			&+\mathcal I_{\ell\ell'0}^{\ast\star}[\sigma_{\ell\ell'5}^{\ast\star}]{\bh{M}}_6+\mathcal I_{\ell\ell'1}^{\ast\star}[\sigma_{\ell\ell'2}^{\ast\star}]{\bh{M}}_4+\mathcal I_{\ell\ell'1}^{\ast\star}[\sigma_{\ell\ell'4}^{\ast\star}]{\bh{M}}_4^{\rm T}+\mathcal I_{\ell\ell'2}^{\ast\star}[\sigma_{\ell\ell'3}^{\ast\star}]{\bh{M}}_2,\\
			\bh{G}_{\mathbf H,\ell\ell'}^{\ast\star}(\bs{r}, \bs{r}') =& \frac{1}{4\pi^2} \iint_{\mathbb{R}^2} \widehat{\bh{G}}_{\mathbf H,\ell\ell'}^{\ast\star} e^{\i k_x (x - x') + \i k_y (y - y')} dk_x dk_y\\
			=& \frac{\ri}{\omega\mu_{\ell}}\big(\mathcal I_{\ell\ell',1}^{\ast\star}[\sigma_{\ell\ell'6}^{\ast\star}]{\bh{M}}_4^{\rm T}-\mathcal I_{\ell\ell',-1}^{\ast\star}[\sigma_{\ell\ell'6}^{\ast\star}]{\bh{M}}_5^{\rm T}-\mathcal I_{\ell\ell',1}^{\ast\star}[\sigma_{\ell\ell'7}^{\ast\star}]{\bh{M}}_4\\
			&+\mathcal I_{\ell\ell',-1}^{\ast\star}[\sigma_{\ell\ell'7}^{\ast\star}]{\bh{M}}_5+\mathcal I_{\ell\ell',0}^{\ast\star}[\sigma_{\ell\ell'8}^{\ast\star}]{\bh{M}}_7-\ri \mathcal I_{\ell\ell'2}^{\ast\star}[\sigma_{\ell\ell'9}^{\ast\star}]{\bh{M}}_2+\ri \mathcal I_{\ell\ell',-2}^{\ast\star}[\sigma_{\ell\ell'9}^{\ast\star}]{\bh{M}}_3\big),
		\end{split}
	\end{equation}
	where
	\begin{equation}\label{generalintegral}
		\mathcal I_{\ell\ell'\kappa}^{\ast\star}[\sigma](\bs r, \bs r')=\frac{\i}{8\pi^2} \iint_{\mathbb{R}^2} e^{\ri\mathbf{k}_{\rho}\cdot(\boldsymbol\rho-\boldsymbol\rho')}{\mathcal Z}_{\ell\ell'}^{\ast\star}(k_{\rho},z, z')e^{\ri \kappa\alpha}\sigma(k_{\rho}) dk_x dk_y,\quad \ast,\star=\uparrow,\downarrow,
	\end{equation}
	for $\kappa=-2, -1, 0, 1, 2$.
	Moreover, by indentity 
	\begin{equation}\label{besseljidentity}
		J_n(z)=\frac{1}{2\pi \ri^n}\int_0^{2\pi}e^{\ri z\cos\theta+\ri n\theta}d\theta,
	\end{equation}
	we have
	\begin{equation}
		\begin{split}
			&\mathcal I_{\ell\ell'\kappa}^{\ast\star}[\sigma](\bs r, \bs r')=\frac{\ri^{1+\kappa} e^{\ri \kappa\varphi}}{4\pi} \int_0^{\infty} k_{\rho}J_{\kappa}(k_{\rho}\rho){\mathcal Z}_{\ell\ell'}^{\ast\star}(k_{\rho}, z, z')\sigma(k_{\rho}) dk_{\rho},\\
			&\mathcal I_{\ell\ell',-\kappa}^{\ast\star}[\sigma](\bs r, \bs r')=\frac{\ri^{1-\kappa} e^{-\ri \kappa\varphi}}{4\pi} \int_0^{\infty} k_{\rho}J_{-\kappa}(k_{\rho}\rho){\mathcal Z}_{\ell\ell'}^{\ast\star}(k_{\rho}, z, z')\sigma(k_{\rho}) dk_{\rho},
		\end{split}
	\end{equation}
	and 
	\begin{equation}
		{\mathcal I}_{\ell\ell',-\kappa}^{\ast\star}[\sigma](\bs r, \bs r')=e^{-2\ri \kappa\varphi}{\mathcal I}_{\ell\ell'\kappa}^{\ast\star}[\sigma](\bs r, \bs r'),\quad\kappa>0
	\end{equation}
	where $\rho=\sqrt{(x-x')^2+(y-y')^2}$ and $\varphi$ is the polar angle of $(x-x',y-y')$. 
	These are the formulations we used to establish an FMM for fast computation of electromagnetic interactions in layered media \cite{yuan2026fast}.

	\section{Numerical test}\label{num_test}
	To validate the formulations presented in this paper, we compare the numerical results calculated by our formulations with those by the open-source C++ toolkit, Strata \cite{sharma2021strata}. The Strata library is specifically designed to compute the multilayer Green’s function (MGF) in both spectral and spatial domain. 
	
\begin{figure}[h]  
		\centering
		\begin{subfigure}{0.31\textwidth}
			\includegraphics[width=\linewidth]{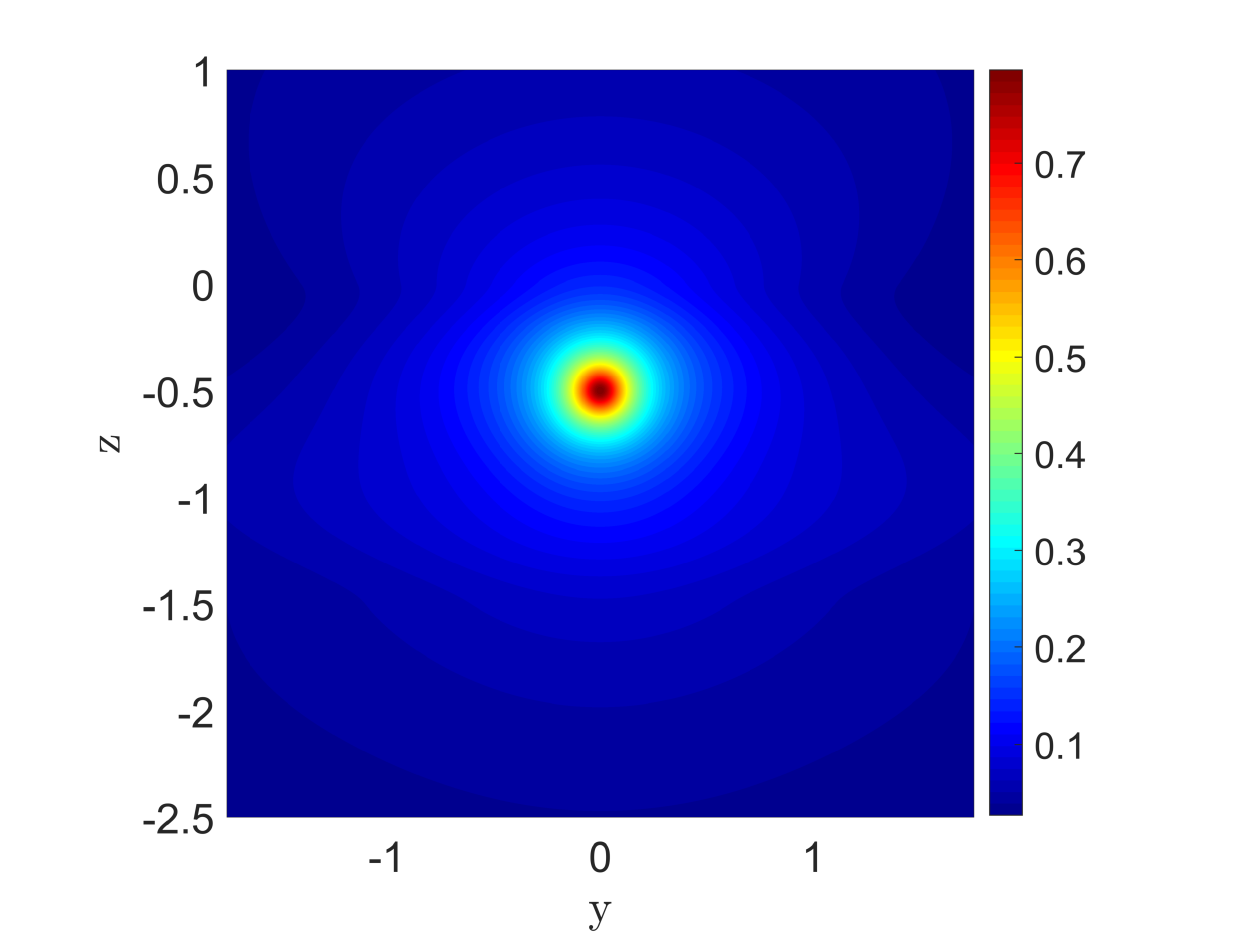}
			\caption{$|G_A^{xx}|$}
		\end{subfigure}
		\quad
		\begin{subfigure}{0.31\textwidth}
			\includegraphics[width=\linewidth]{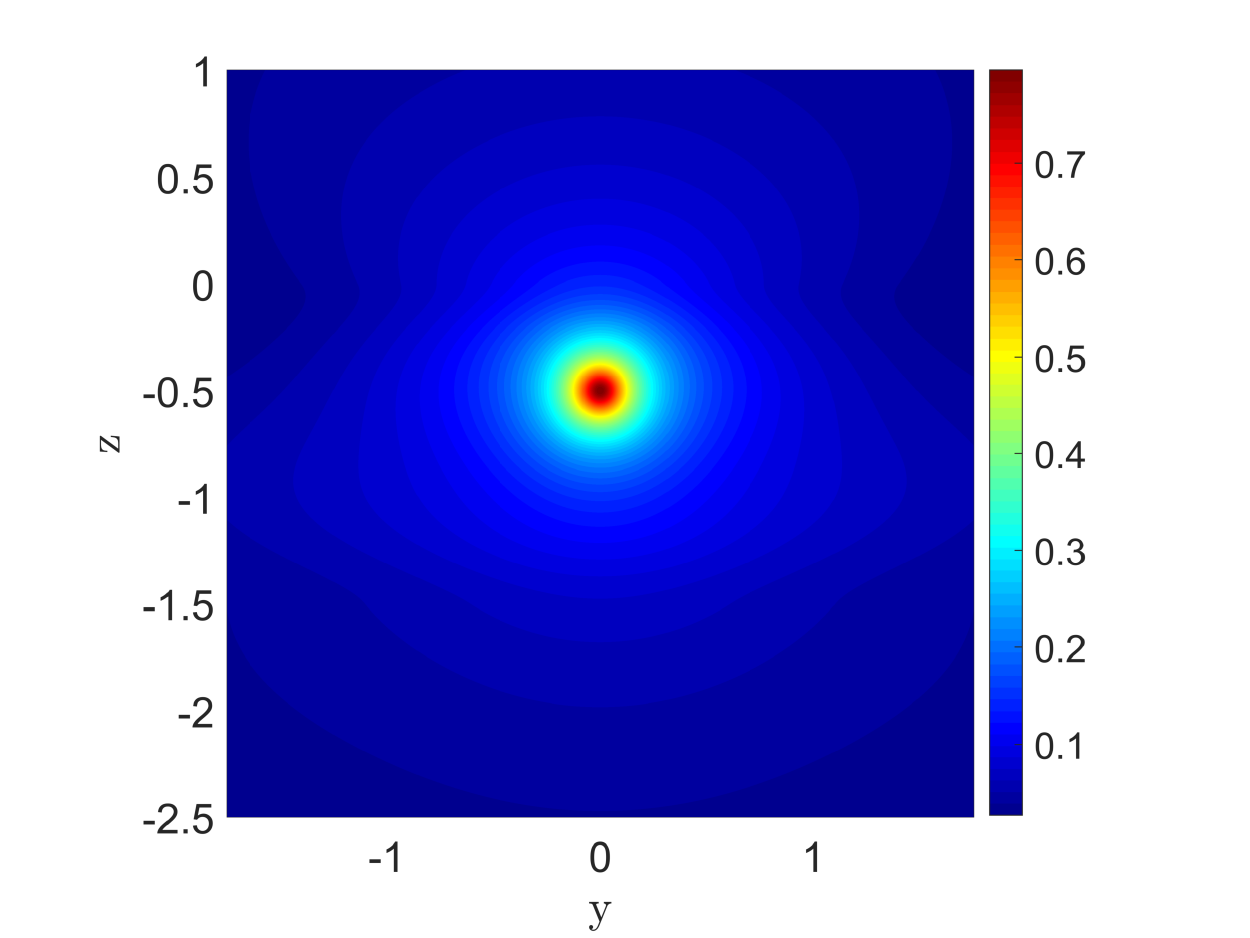}
			\caption{$|K_{xx}^{mi}|$}
		\end{subfigure}
		\quad
		\begin{subfigure}{0.31\textwidth}
			\includegraphics[width=\linewidth]{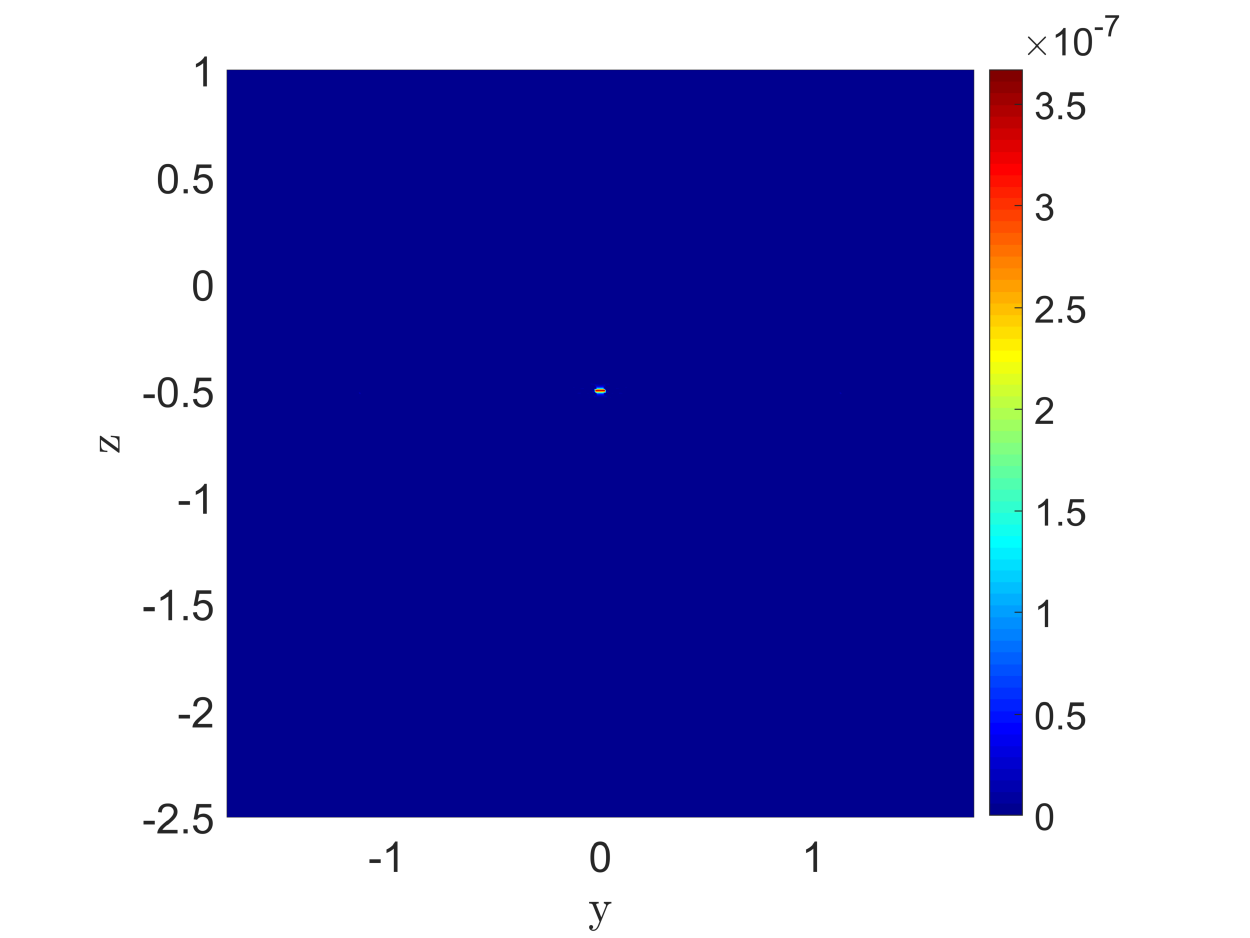}
			\caption{error}
		\end{subfigure} \\
		\begin{subfigure}{0.31\textwidth}
			\includegraphics[width=\linewidth]{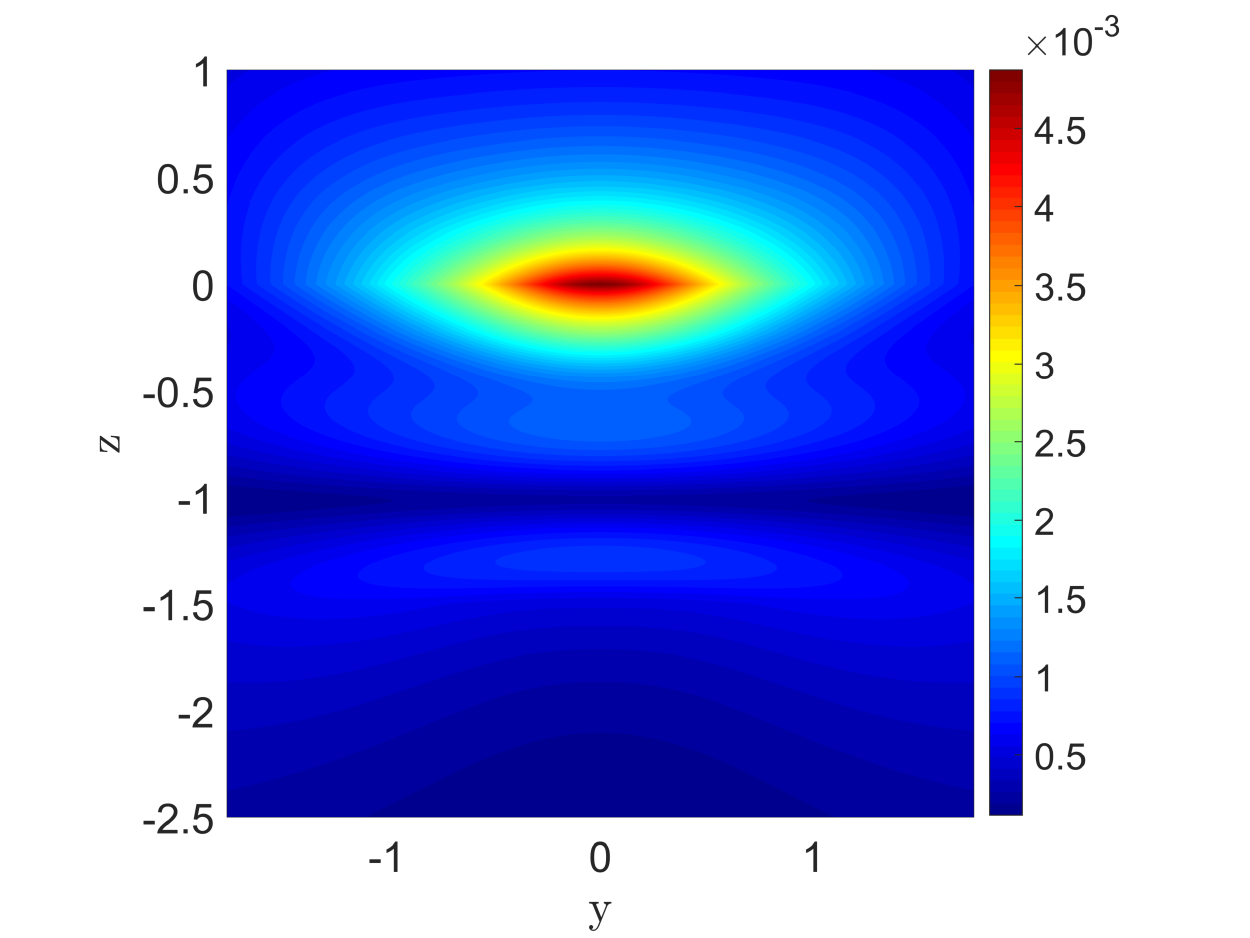}
			\caption{$|G_A^{zx}|$ }
		\end{subfigure}
		\quad
		\begin{subfigure}{0.31\textwidth}
			\includegraphics[width=\linewidth]{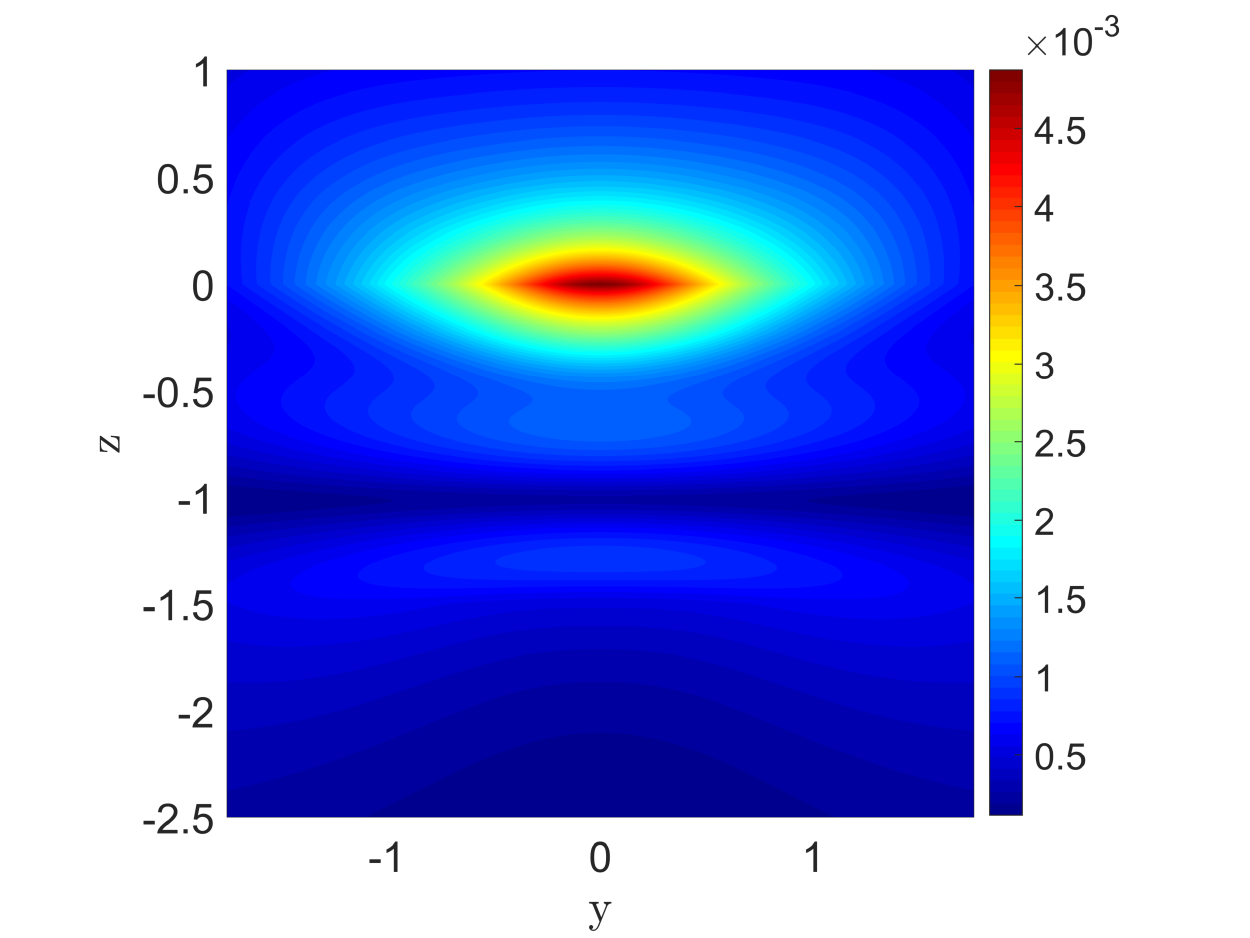}
			\caption{$|K_{zx}^{mi}|$}
		\end{subfigure}
		\quad
		\begin{subfigure}{0.31\textwidth}
			\includegraphics[width=\linewidth]{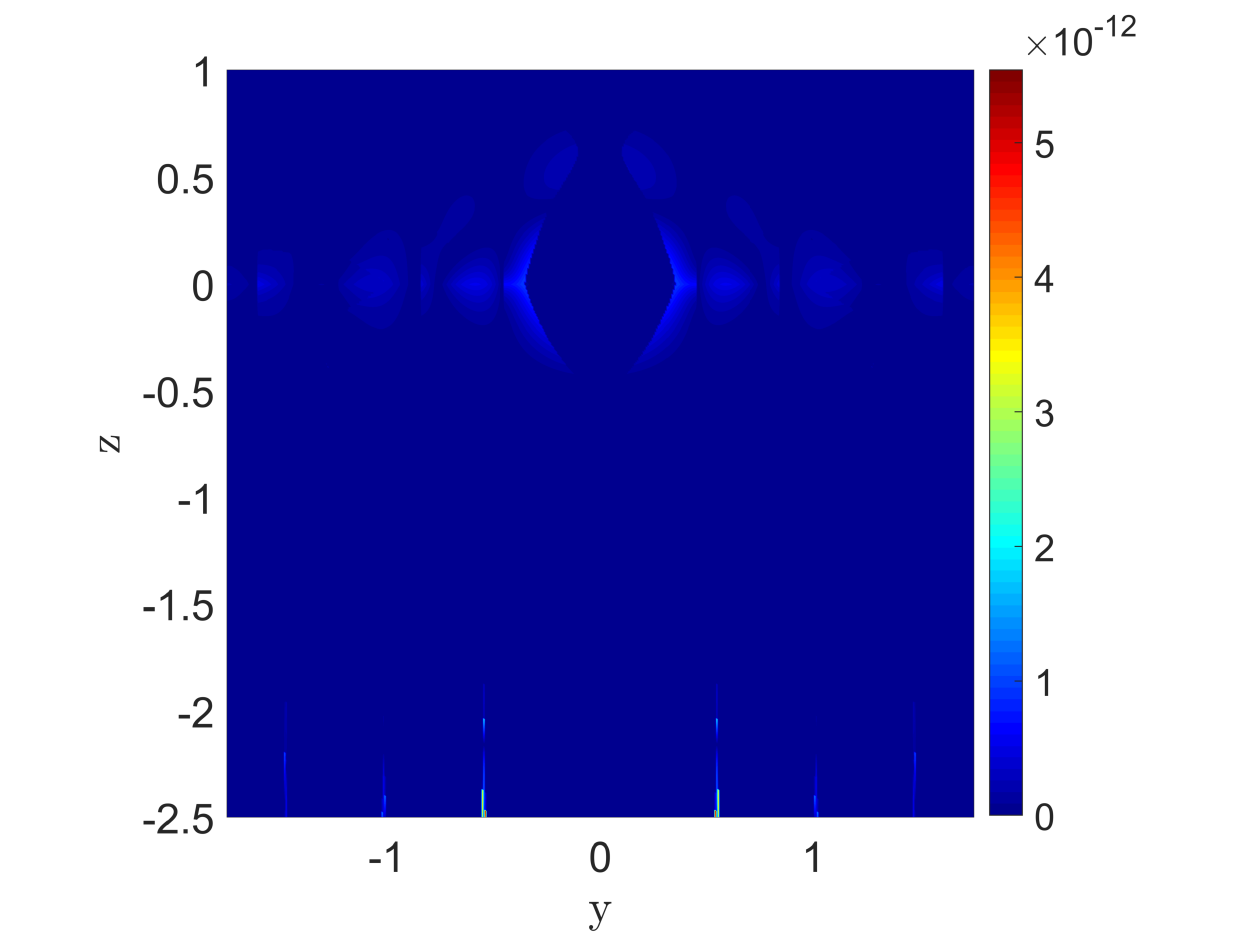}
			\caption{error}
		\end{subfigure}\\
		\begin{subfigure}{0.31\textwidth}
			\includegraphics[width=\linewidth]{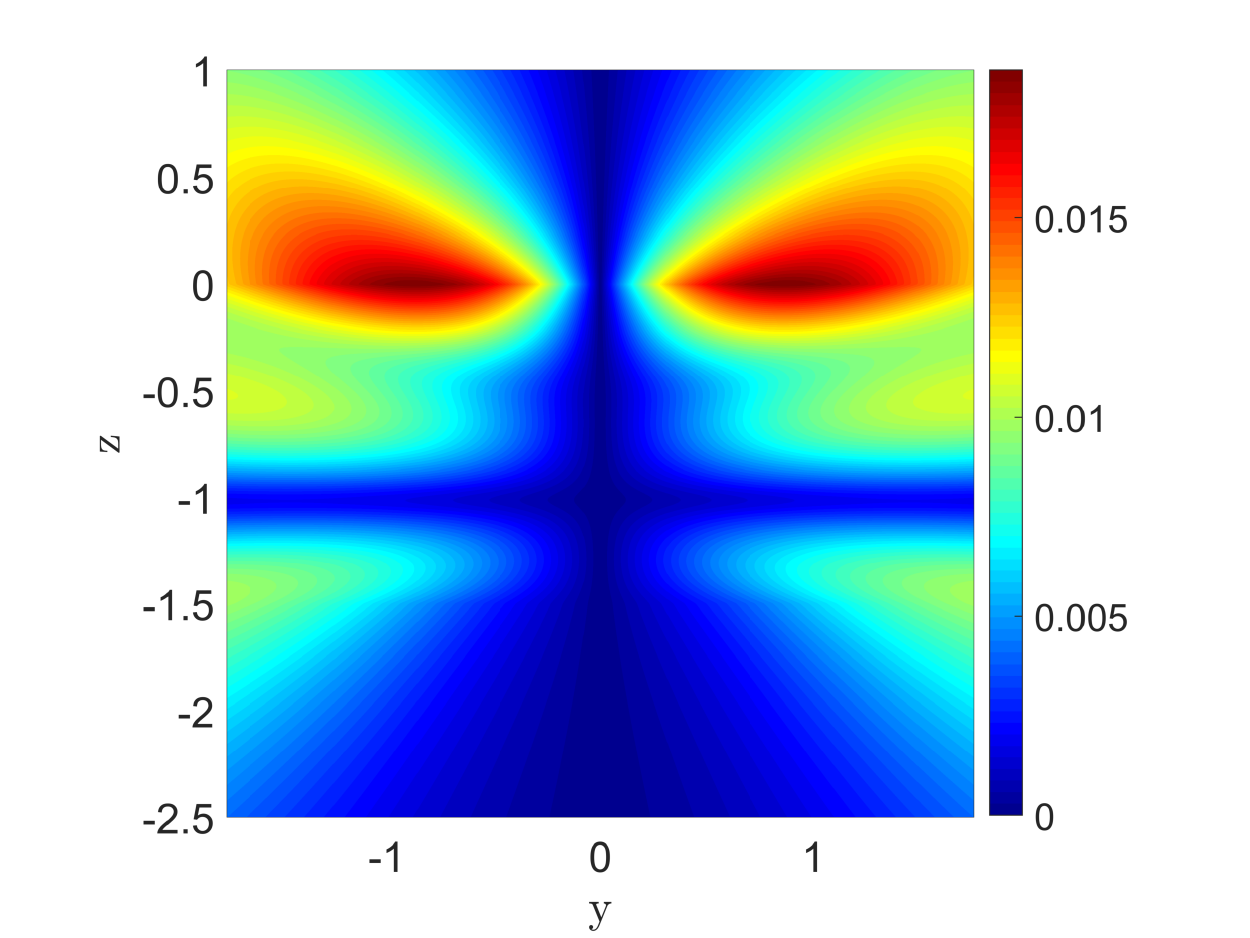}
			\caption{$|G_A^{zy}|$}
		\end{subfigure}
		\quad
		\begin{subfigure}{0.31\textwidth}
			\includegraphics[width=\linewidth]{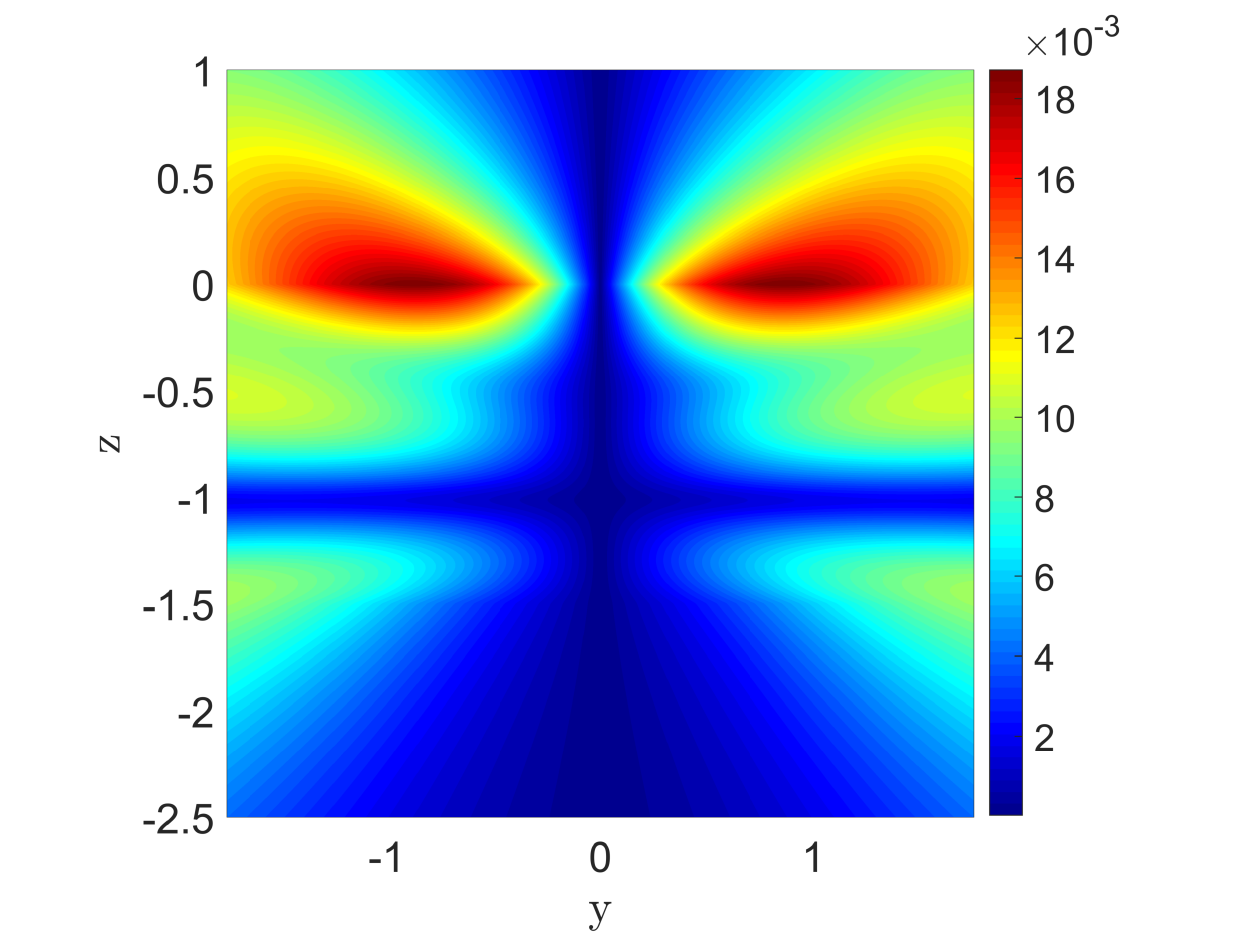}
			\caption{$|K_{zy}^{mi}|$}
		\end{subfigure}
		\quad
		\begin{subfigure}{0.31\textwidth}
			\includegraphics[width=\linewidth]{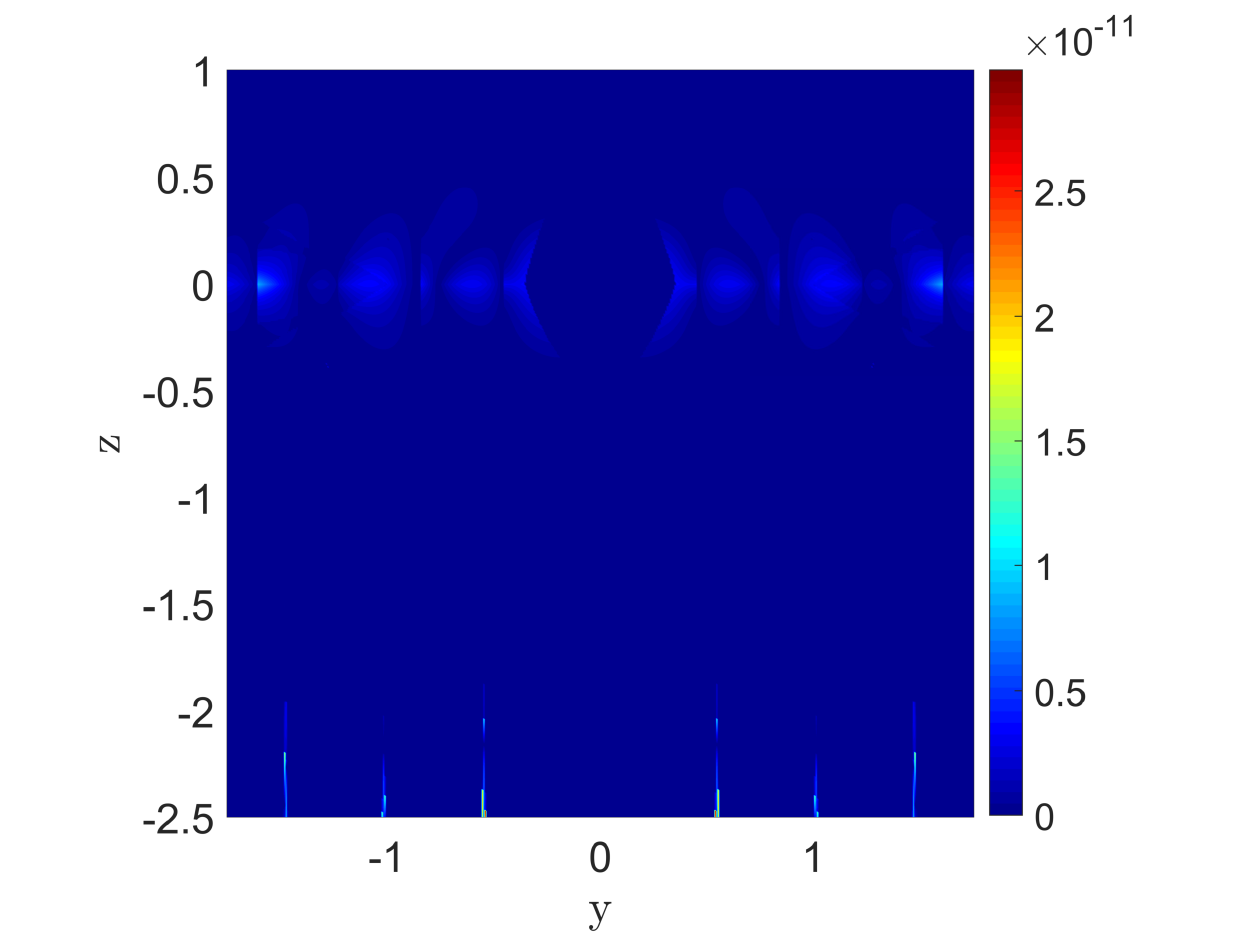}
			\caption{error}
		\end{subfigure}
		\caption{A Comparison with Strata on the computation of the vector potential Green's function.}
		\label{GAzy}
	\end{figure}

	In Strata, the vector potential Green’s functions $\overline{\mathbb K}_A^{mi},\overline{K}_{\phi}^{mi}$ (cf. \cite{michalski1990electromagnetic}) are the fundamental kernel implemented for solving the mixed-potential integral equation (MPIE) of scattering problems in layered media. According to the formulation C in \cite{michalski1990electromagnetic}, we have
	\begin{equation*}
		\begin{split}
			K_{xx}^{mi}(\bs r,\bs r')=K_{yy}^{mi}(\bs r,\bs r')={G}_{\bs A}^{xx}(\bs r,\bs r')={G}_{\bs A}^{yy}(\bs r,\bs r'),\\
			{K}_{zx}^{mi}(\bs r,\bs r')={G}_{\bs A}^{zx}(\bs r,\bs r'),\quad  {K}_{zy}^{mi}(\bs r,\bs r')={G}_{\bs A}^{zy}(\bs r,\bs r').
		\end{split}
	\end{equation*}
	Therefore, we compare the vector potential Green’s function given by \eqref{GA_spectral} with $\overline{\mathbb K}_A^{mi}$ calculated by Strata. A three-layer medium with interfaces located at $z=0,-1.5$ and a point source located at ${\bs r}'=(0.1,0,-0.5)$ in the second layer is tested. We set frequency $f=\frac{\omega}{2\pi} = 3\times 10^8$, and the relative permeability and dielectric permittivities $\mu_0 =\mu_1 =\mu_2 = 1.0$, $_0 = 1.2$, $_1 = 0.8$, $_2 = 1.0$. We compare the results in the spatial domain where $\mathbb G_{\bs A}(\bs r, \bs r')$ is the inverse Fourier transform of $\widehat{\mathbb G}_{\bs A}^S(k_x, k_y, z, z')$. The integrals have been reduced to Sommerfeld integrals, and a double-exponential quadrature rule with self-precision control \cite{doubleexponential} is adopted to compute them numerically along deformed contours. Three components in the domain $\{(y, z)|-1.75\le y\le1.75, -2.5\le z\le 1.0\}$ in the $YZ$ plane are plotted in Fig. \ref{GAzy}. From \eqref{GA_spectral}, we have
	\begin{equation}
		\widehat{{G}}_{\bs A}^{xx}=\widehat{{G}}_{\bs A}^{yy}=\alpha_{\ell 1},\quad\widehat{{G}}_{\bs A}^{zx}=\frac{\ri k_x}{k_{\rho}^2}\left(\mu_{\ell}\alpha_{\ell 3}-\partial_z\alpha_{\ell 1}\right),\quad \widehat{{G}}_{\bs A}^{zy}=\frac{\ri k_y}{k_{\rho}^2}\left(\mu_{\ell}\alpha_{\ell 3}-\partial_z\alpha_{\ell 1}\right).
	\end{equation}
	It is evident that ${{G}}_{\bs A}^{xx}$ remains continuous across the layer interfaces, whereas ${{G}}_{\bs A}^{zx}$ and ${{G}}_{\bs A}^{zy}$ are discontinuous, as clearly observed in Fig. \ref{GAzy}.

	\section{Conclusion}\label{sect_conclusion}
	In this paper, we presented a detailed comparison of two approaches for the derivation of the dyadic Green's functions of the Maxwell's equations in layered media. We show that both approaches decompose the problem into three Helmholtz problems in layered media and then give exactly the same integral formulations for the LMDGFs. The first approach relies on the physical property of the electromagnetic wave while the second utilizes the algebraic nature implied in the TE/TM decomposition. Our research shows that the second approach is more general and can be applied to derive the DGFs of elastic wave equations in layered media (cf. \cite{yuan2026computationgreensfunctionlayered}). 
	

	\newpage
	\begin{appendix}
		\section{The Green function for 3-D helmholtz equation in layered medium}\label{helmholtzGreenFun}
		Consider the  spectral-domain Helmholtz problem:
		\begin{equation}\label{G_frequency}
			\begin{cases}
				\dfrac{\partial^2 \widehat{G}(k_x, k_y, z, z')}{\partial z^2}+k_{\ell z}^2\widehat{G}(k_x, k_y, z, z')=-\delta_{\ell\ell'}\delta(z-z'),\quad d_{\ell}<z<d_{\ell-1},\\
				\llbracket a \widehat{G}\rrbracket=0,\quad \left\llbracket b\dfrac{\partial \widehat{G}}{\partial z}\right\rrbracket=0,\quad{\rm at}\quad z=\{d_{\ell}\}_{\ell=0}^{L-1},
			\end{cases}
		\end{equation}
		in a multi-layered meida, where $a=a_\ell$ and $b=b_\ell$ are nonzero layer-wise constants.
		The layer-wise solution $G(k_{\rho}, z, z')$ has decomposition
		\begin{equation*}
			\widehat{G}(k_{\rho}, z, z')=\widehat{G}_{\ell\ell'}(k_{\rho}, z, z')+\delta_{\ell\ell'}\widehat G^f(k_{\rho}, z, z'),\quad d_{\ell}<z<d_{\ell-1},
		\end{equation*}
		where the reaction field component $\widehat{G}_{\ell\ell'}(k_{\rho}, z, z')$ satisfies ODE
		\begin{equation}\label{helmholtz_spectral_layer_react}
			\partial_{zz}\widehat G_{\ell\ell'}(k_{\rho},z,z')+k_{\ell z}^2\widehat G_{\ell\ell'}(k_{\rho},z,z')=0, \quad d_{\ell}<z<d_{\ell-1},\;\ell=0, 1, \cdots, L,\\
		\end{equation}
		in each layer.
		
		The second order ODE \eqref{helmholtz_spectral_layer_react} has general solution
		\begin{equation}\label{reactfourerformula}
			{\widehat G}_{\ell \ell'}(k_{\rho}, z,z')=\begin{cases}
				A_0(k_{\rho}, z')e^{\ri k_{0z}z},\quad \ell=0,\\
				A_{\ell}(k_{\rho}, z')e^{\ri k_{\ell z}z}+B_{\ell}(k_{\rho},z')e^{2\ri k_{\ell z}d_{\ell-1}-\ri k_{\ell z}z},\quad 0<\ell<\ell',\\
				A_{\ell'}^r(k_{\rho},z')e^{\ri k_{\ell' z}z}+B_{\ell'}^r(k_{\rho},z')e^{-\ri k_{\ell' z}z},\quad \ell=\ell',\\
				A_{\ell}(k_{\rho},z')e^{-2\ri k_{\ell z}d_{\ell}+\ri k_{\ell z}z}+B_{\ell}(k_{\rho},z')e^{-\ri k_{\ell z}z},\quad \ell'<\ell<L,\\
				B_L(k_{\rho},z')e^{-\ri k_{Lz}z},\quad \ell=L,
			\end{cases}
		\end{equation}
		where two exponential increasing terms have been removed due to the outgoing property of the radiating wave. Note that the free-space component can be rewritten as
		\begin{equation*}
			\widehat{G}^f(k_{\rho},z,z')=\dfrac{\ri e^{\ri k_{\ell'z}|z-z'|}}{2k_{\ell' z}}=H(z-z')A_{\ell'}^f(k_{\rho},z')e^{\ri k_{\ell'z}z}+H(z'-z)B_{\ell'}^f(k_{\rho},z')e^{-\ri k_{\ell'z}z},
		\end{equation*}
		where $H(x)$ is the Heaviside function, and
		\begin{equation*}
			A_{\ell'}^f(k_{\rho},z')=\dfrac{\ri }{2k_{\ell' z}}e^{-\ri k_{\ell'z}z'},\quad B_{\ell'}^f(k_{\rho},z')=\dfrac{\ri }{2k_{\ell' z}}e^{\ri k_{\ell'z}z'}.
		\end{equation*}
		Then
		\begin{equation}\label{G1fourerformula}
			\widehat{G}(k_{\rho}, z,z')=\begin{cases}
				A_0(k_{\rho}, z')e^{\ri k_{0z}z},\quad \ell=0,\\
				A_{\ell}(k_{\rho}, z')e^{\ri k_{\ell z}z}+B_{\ell}(k_{\rho},z')e^{2\ri k_{\ell z}d_{\ell-1}-\ri k_{\ell z}z},\quad 0<\ell<\ell',\\
				A_{\ell'}(k_{\rho},z')e^{\ri k_{\ell' z}z}+B_{\ell'}(k_{\rho},z')e^{-\ri k_{\ell' z}z},\quad \ell=\ell',\\
				A_{\ell}(k_{\rho},z')e^{-2\ri k_{\ell z}d_{\ell}+\ri k_{\ell z}z}+B_{\ell}(k_{\rho},z')e^{-\ri k_{\ell z}z},\quad \ell'<\ell<L,\\
				B_L(k_{\rho},z')e^{-\ri k_{Lz}z},\quad \ell=L,
			\end{cases}
		\end{equation}
		where 
		$$A_{\ell'}(k_{\rho},z')=A_{\ell'}^r(k_{\rho},z')+H(z-z')A_{\ell'}^f, \quad B_{\ell'}(k_{\rho},z')=B_{\ell'}^r(k_{\rho},z')+H(z'-z)B_{\ell'}^f.$$

		Before we use the interface conditions in \eqref{G_frequency} to determine the coefficients $\{A_{\ell}, B_{\ell}\}_{\ell=0}^L$, let us introduce the generalized reflection and transmission coefficients $\widetilde R_{\ell\ell'},\widetilde T_{\ell\ell'}$ for multi-layered media (cf. \cite{Chew1999inhomogenous}). 
		They are defined recursively via the two-layers reflection and transmission coefficients
		\begin{equation*}
			R_{\ell,\ell+1}=\frac{a_{\ell+1}b_{\ell}k_{\ell z}- a_{\ell}b_{\ell+1}k_{\ell+1,z}}{a_{\ell+1}b_{\ell}k_{\ell z}+a_{\ell}b_{\ell+1}k_{\ell+1,z}},\quad T_{\ell,\ell+1}=\frac{2a_{\ell}b_{\ell}k_{\ell z}}{a_{\ell+1}b_{\ell}k_{\ell z}+a_{\ell}b_{\ell+1}k_{\ell+1,z}}.
		\end{equation*}
		In general, we have recursions
		\begin{equation}\label{Rellrecursion}
			\begin{aligned}
				\widetilde R_{0,-1}=&0,\quad \widetilde R_{\ell+1,\ell}=\frac{R_{\ell+1,\ell}+\widetilde R_{\ell,\ell-1}e^{2\ri k_{\ell z}D_{\ell}}}{1+R_{\ell+1,\ell}\widetilde R_{\ell,\ell-1}e^{2\ri k_{\ell z}D_{\ell}}},\quad\ell=0,1,\cdots,L-1,\\
				\widetilde R_{L,L+1}=&0,\quad \widetilde R_{\ell,\ell+1}=\frac{R_{\ell,\ell+1}+\widetilde R_{\ell+1,\ell+2}e^{2\ri k_{\ell+1,z}D_{\ell+1}}}{1+R_{\ell,\ell+1}\widetilde R_{\ell+1,\ell+2}e^{2\ri k_{\ell+1,z}D_{\ell+1}}},\quad\ell=L-1,\cdots,1,0,
			\end{aligned}
		\end{equation}
		for generalized reflection coefficients, and recursions 
		\begin{equation}\label{Tellrecursion}
			\begin{split}
				&\widetilde T_{\ell'\ell'}=1,\quad \widetilde T_{\ell'\ell}=\frac{T_{\ell+1,\ell}e^{-\ri(k_{\ell,z}-k_{\ell+1,z})d_{\ell}}}{1+R_{\ell+1,\ell}\widetilde R_{\ell,\ell-1}e^{2\ri k_{\ell z}D_{\ell}}}\widetilde T_{\ell',\ell+1},\quad\ell=\ell'-1, \ell'-2,\cdots, 0,\\
				&\widetilde T_{\ell',\ell+1}=\frac{T_{\ell,\ell+1}e^{-\ri (k_{\ell z}-k_{\ell+1,z})d_{\ell}}}{1+R_{\ell,\ell+1}\widetilde R_{\ell+1,\ell+2}e^{2\ri k_{\ell+1,z}D_{\ell+1}}}\widetilde T_{\ell'\ell},\quad \ell=\ell',\ell'+1,\cdots, L-1,
			\end{split}
		\end{equation}
		for generalized transmission coefficients. 
		
		Then, we can divided the problems \eqref{G_frequency} into two problems: 
		the ($\ell'+1$)-layers scattering problems generated by the upward incident wave $A_{\ell'}e^{\ri k_{\ell' z}z}$ from the lowest level and the ($L-\ell'+1$)-layers scattering problems generated by the downward incident wave $B_{\ell'}e^{-\ri k_{\ell' z}z}$ from the top level. They are scattering problems within layered media, with plane-waves incident from the bottom and top layer, respectively. By using the reflection theory in layered media (cf. \cite{Chew1999inhomogenous}), we have

		\begin{equation}\label{layer_ellprime_AB_equations}
			\begin{cases}
				B_{\ell}=\widetilde{R}_{\ell,\ell-1}A_{\ell},\quad \ell=0,1,\cdots,\ell'-1,\\
				B_{\ell'}^r=A_{\ell'}\widetilde R_{\ell',\ell'-1}e^{2\ri k_{\ell'z}d_{\ell'-1}}=(A_{\ell'}^f+A_{\ell'}^r)\widetilde R_{\ell',\ell'-1}e^{2\ri k_{\ell'z}d_{\ell'-1}},\\
				A_{\ell'}^r=B_{\ell'}\widetilde R_{\ell',\ell'+1}e^{-2\ri k_{\ell'z}d_{\ell'}}=(B_{\ell'}^f+B_{\ell'}^r)\widetilde R_{\ell',\ell'+1}e^{-2\ri k_{\ell'z}d_{\ell'}},\\
				A_{\ell}=\widetilde{R}_{\ell,\ell+1}B_{\ell},\quad \ell=\ell'+1,\ell'+2,\cdots,L.
			\end{cases}
		\end{equation}
		Notably, the reflection within the source layer is decomposed into two components according to the propagation directions of the waves generated by the source.
		Together with the formulas in \eqref{G1fourerformula}, we obtain
		\begin{equation}\label{formula1}
			\widehat{G}(k_{\rho}, z, z')=\begin{cases}
				\displaystyle A_{\ell}e^{\ri k_{\ell z}z}+A_{\ell}\widetilde R_{\ell,\ell-1}e^{2\ri k_{\ell z}d_{\ell-1}-\ri k_{\ell z}z},\quad d_{\ell}<z<d_{\ell-1},\;\ell=0,1, \cdots, \ell'-1,\\
				\displaystyle  A_{\ell'}e^{\ri k_{\ell' z}z}+A_{\ell'}\widetilde R_{\ell',\ell'-1}e^{2\ri k_{\ell'z}d_{\ell'-1}-\ri k_{\ell' z}z},\quad d_{\ell'}<z'<z<d_{\ell'-1},
			\end{cases}
		\end{equation} 
		and
		\begin{equation}\label{formula2}
			\widehat{G}(k_{\rho}, z, z')=\begin{cases}
				B_{\ell'}\widetilde R_{\ell',\ell'+1}e^{-2\ri k_{\ell' z}d_{\ell'}+\ri k_{\ell' z}z}+B_{\ell'}e^{-\ri k_{\ell' z}z},\quad d_{\ell'}<z<z'<d_{\ell'-1},\\
				B_{\ell}\widetilde R_{\ell,\ell+1}e^{-2\ri k_{\ell z}d_{\ell}+\ri k_{\ell z}z}+B_{\ell}e^{-\ri k_{\ell z}z},\;\; d_{\ell}<z<d_{\ell-1},\;\ell=\ell'+1, \cdots, L.
			\end{cases}
		\end{equation}
		From Eqs.\eqref{layer_ellprime_AB_equations}, we can solve that
		\begin{equation*}
			\begin{aligned}
				A_{\ell'}^r=&\sigma_{\ell'\ell'}^{\uparrow\uparrow}e^{2\ri k_{\ell'z}D_{\ell'}}A_{\ell'}^f+\sigma_{\ell'\ell'}^{\uparrow\downarrow}e^{-2\ri k_{\ell'z}d_{\ell'}}B_{\ell'}^f,\\
				B_{\ell'}^r=&\sigma_{\ell'\ell'}^{\downarrow\uparrow}e^{2\ri k_{\ell'z}d_{\ell'-1}}A_{\ell'}^f+\sigma_{\ell'\ell'}^{\downarrow\downarrow}e^{2\ri k_{\ell'z}D_{\ell'}}B_{\ell'}^f,
			\end{aligned}
		\end{equation*}
		where the densities are defined as
		\begin{equation*}
			\begin{aligned}
				&\sigma_{\ell'\ell'}^{\uparrow\downarrow}(k_{\rho}):=Q_{\ell'}(k_{\rho})\widetilde R_{\ell',\ell'+1},\quad \sigma_{\ell'\ell'}^{\downarrow\uparrow}(k_{\rho}):=Q_{\ell'}(k_{\rho})\widetilde R_{\ell',\ell'-1},\\
				&\sigma_{\ell'\ell'}^{\uparrow\uparrow}(k_{\rho})=\sigma_{\ell'\ell'}^{\downarrow\downarrow}(k_{\rho}):=Q_{\ell'}(k_{\rho})\widetilde R_{\ell',\ell'+1}\widetilde R_{\ell',\ell'-1}=\left[Q_{\ell'}(k_{\rho})-1\right]e^{-2\ri k_{\ell'z}D_{\ell'}},
			\end{aligned}
		\end{equation*}
		with
		\begin{equation*}
			Q_{\ell}(k_{\rho}):=\dfrac{1}{1-\widetilde{R}_{\ell,\ell+1}\widetilde{R}_{\ell,\ell-1}e^{\ri 2k_{\ell z}D_{\ell}}},\quad \ell=0,1,\cdots,L.
		\end{equation*}
		From the above expression, we can see that the reaction field in the source layer $\ell'$ is divided into two parts: upward propagation field (determined by $A_{\ell'}^r$) and downward propagation field (determined by $B_{\ell'}^r$), and each part contains two components inspired by the upward field (determined by $A_{\ell'}^f$) and the downward field (determined by $B_{\ell'}^f$) emitted by the point source.
		Therefore, according to the previous discussion and analysis, the amplitudes contributing to the
		field above and below the source point in layer
		$\ell'$ are, respectively,
		\begin{equation}\label{layer_AB_solution_ellprime}
			\begin{aligned}
				A_{\ell'}=&\left[1+\sigma_{\ell'\ell'}^{\uparrow\uparrow}e^{2\ri k_{\ell'z}D_{\ell'}}\right]A_{\ell'}^f+\sigma_{\ell'\ell'}^{\uparrow\downarrow}e^{-2\ri k_{\ell'z}d_{\ell'}}B_{\ell'}^f\\
				B_{\ell'}=&\sigma_{\ell'\ell'}^{\downarrow\uparrow}e^{2\ri k_{\ell'z}d_{\ell'-1}}A_{\ell'}^f+\left[1+\sigma_{\ell'\ell'}^{\downarrow\downarrow}e^{2\ri k_{\ell'z}D_{\ell'}}\right]B_{\ell'}^f
			\end{aligned}
		\end{equation}
		respectively. 
		
		Substituting Eqs.\eqref{formula1} and \eqref{formula2} into the interface conditions in \eqref{G_frequency} gives
		\begin{equation}\label{recurrenceformulagenerA}
			\begin{split}
				a_{\ell}A_{\ell}(\widetilde R_{\ell,\ell-1}e^{2\ri k_{\ell z}(d_{\ell-1}-d_{\ell})}+1)e^{\ri (k_{\ell z}-k_{\ell+1, z})d_{\ell}}=&a_{\ell+1}A_{\ell+1}(\widetilde R_{\ell+1,\ell}+1),\\
				b_{\ell}k_{\ell z}A_{\ell}(\widetilde R_{\ell,\ell-1}e^{2\ri k_{\ell z}(d_{\ell-1}-d_{\ell})}-1)e^{\ri (k_{\ell z}-k_{\ell+1, z})d_{\ell}}=&b_{\ell+1}k_{\ell+1,z}A_{\ell+1}(\widetilde R_{\ell+1,\ell}-1),
			\end{split}
		\end{equation}
		for $\ell=0,1,\cdots,\ell'-1$ and 
		\begin{equation}\label{recurrenceformulagenerB}
			\begin{split}
				a_{\ell}B_{\ell}(\widetilde R_{\ell,\ell+1}+1)=&a_{\ell+1}B_{\ell+1}(\widetilde R_{\ell+1,\ell+2}e^{2\ri k_{\ell+1, z}(d_{\ell}-d_{\ell+1})}+1)e^{\ri (k_{\ell z}-k_{\ell+1, z})d_{\ell}},\\
				b_{\ell}k_{\ell z}B_{\ell}(\widetilde R_{\ell,\ell+1}-1)=&b_{\ell+1}k_{\ell+1,z}B_{\ell+1}(\widetilde R_{\ell+1,\ell+2}e^{2\ri k_{\ell+1, z}(d_{\ell}-d_{\ell+1})}-1)e^{\ri (k_{\ell z}-k_{\ell+1, z})d_{\ell}},
			\end{split}
		\end{equation}
		for $\ell=\ell',\ell'+1,\cdots,L-1$. 
		Eliminating $\widetilde R_{\ell,\ell+1}$ in Eqs.\eqref{recurrenceformulagenerA} and \eqref{recurrenceformulagenerB} leads to recurrence formulas
		\begin{equation*}\label{layer_AB_solution2_ellprime}
			\begin{aligned}
				A_{\ell}=&\frac{T_{\ell+1,\ell}e^{-\ri(k_{\ell,z}-k_{\ell+1,z})d_{\ell}}}{1+R_{\ell+1,\ell}\widetilde R_{\ell,\ell-1}e^{2\ri k_{\ell z}D_{\ell}}}A_{\ell+1},\quad \ell=\ell'-1,\ell'-2,\cdots,0\\
				B_{\ell+1}=&\frac{T_{\ell,\ell+1}e^{-\ri (k_{\ell z}-k_{\ell+1,z})d_{\ell}}}{1+R_{\ell,\ell+1}\widetilde R_{\ell+1,\ell+2}e^{2\ri k_{\ell+1,z}D_{\ell+1}}}B_{\ell},\quad \ell=\ell',\ell'+1,\cdots,L-1.
			\end{aligned}
		\end{equation*}
		Using the generalized transmission coefficients defined in \eqref{Tellrecursion}, the recurrence formulas can be rewritten as
		\begin{equation}\label{layer_AB_solution2}
			\begin{aligned}
				A_{\ell}=&\widetilde T_{\ell'\ell}A_{\ell'},\quad \ell=\ell'-1,\ell'-2,\cdots,0,\\
				B_{\ell}=&\widetilde T_{\ell'\ell}B_{\ell'},\quad \ell=\ell'+1,\ell'+2,\cdots,L.
			\end{aligned}
		\end{equation}

		From \eqref{layer_AB_solution_ellprime} and \eqref{layer_AB_solution2}, we can summarize that
		\begin{equation}\label{layer_AB_solution_ell_le_ellprime}
			\begin{split}
				A_{\ell}=&\begin{cases}
					\sigma_{\ell\ell'}^{\uparrow\uparrow}A_{\ell'}^f+\sigma_{\ell\ell'}^{\uparrow\downarrow}e^{-2\ri k_{\ell'z}d_{\ell'}}B_{\ell'}^f,\quad \ell<\ell',\\
					\sigma_{\ell\ell'}^{\uparrow\uparrow}e^{2\ri k_{\ell'z}d_{\ell'-1}}A_{\ell'}^f+\sigma_{\ell\ell'}^{\uparrow\downarrow}B_{\ell'}^f,\quad\ell>\ell'
				\end{cases}\\
				B_{\ell}=&\begin{cases}
					\sigma_{\ell\ell'}^{\downarrow\uparrow}A_{\ell'}^f+\sigma_{\ell\ell'}^{\downarrow\downarrow}e^{-2\ri k_{\ell'z}d_{\ell'}}B_{\ell'}^f,\quad\ell<\ell',\\
					\sigma_{\ell\ell'}^{\downarrow\uparrow}e^{2\ri k_{\ell'z}d_{\ell'-1}}A_{\ell'}^f+\sigma_{\ell\ell'}^{\downarrow\downarrow}B_{\ell'}^f,\quad\ell>\ell',
				\end{cases}
			\end{split}
		\end{equation}
		where the densities outside the source layer are defined as
		\begin{itemize}
			\item $\ell<\ell'$:
			\begin{equation*}
				\begin{aligned}
					&\sigma_{\ell\ell'}^{\uparrow\downarrow}(k_{\rho}):=\widetilde T_{\ell'\ell}\sigma_{\ell'\ell'}^{\uparrow\downarrow}(k_{\rho}),\quad \sigma_{\ell\ell'}^{\uparrow\uparrow}(k_{\rho}):=\widetilde T_{\ell'\ell}\left[1+\sigma_{\ell'\ell'}^{\uparrow\uparrow}(k_{\rho})e^{2\ri k_{\ell'z}D_{\ell'}}\right]\\
					&\sigma_{\ell\ell'}^{\downarrow\downarrow}(k_{\rho}):=\widetilde R_{\ell,\ell-1}\sigma_{\ell\ell'}^{\uparrow\downarrow}(k_{\rho}),\quad\sigma_{\ell\ell'}^{\downarrow\uparrow}(k_{\rho}):=\widetilde R_{\ell,\ell-1}\sigma_{\ell\ell'}^{\uparrow\uparrow}(k_{\rho})
				\end{aligned}
			\end{equation*}
			\item $\ell>\ell'$:
			\begin{equation*}
				\begin{aligned}
					&\sigma_{\ell\ell'}^{\downarrow\uparrow}(k_{\rho}):=\widetilde T_{\ell'\ell}\sigma_{\ell'\ell'}^{\downarrow\uparrow}(k_{\rho}),\quad \sigma_{\ell\ell'}^{\downarrow\downarrow}(k_{\rho}):=\widetilde T_{\ell'\ell}\left[1+\sigma_{\ell'\ell'}^{\downarrow\downarrow}(k_{\rho})e^{2\ri k_{\ell'z}D_{\ell'}}\right]\\
					&\sigma_{\ell\ell'}^{\uparrow\downarrow}(k_{\rho}):=\widetilde R_{\ell,\ell+1}\sigma_{\ell\ell'}^{\downarrow\downarrow}(k_{\rho}),\quad\sigma_{\ell\ell'}^{\uparrow\uparrow}(k_{\rho}):=\widetilde R_{\ell,\ell+1}\sigma_{\ell\ell'}^{\downarrow\uparrow}(k_{\rho}).
				\end{aligned}
			\end{equation*}
		\end{itemize}
		Specially, for $\ell=0,L$, the assumption $\widetilde R_{0,-1}=\widetilde R_{L,L+1}=0$ leads to
		\begin{equation*}
			\sigma_{0\ell'}^{\downarrow\star}
			=\sigma_{L\ell'}^{\uparrow\star}=0,\quad\sigma_{\ell0}^{\ast\uparrow}
			=\sigma_{\ell L}^{\ast\downarrow}=0
		\end{equation*}
		And in the upper layer ($\ell=0$), the interface $z=d_{-1}$ does not exist, so the two reaction components generated by the reflections of $z=d_{-1}$ are identically zero, which is compatible with the definition of $\widetilde R_{0,-1}=0$. 
		Similarly, the interface $z=d_{L}$ is absent and thus the two reaction components due to reflections from $z=d_{L}$ also vanish, which is also compatible with the definition of $\widetilde R_{L,L+1}=0$.

		Substituting the solutions \eqref{layer_AB_solution_ellprime} and \eqref{layer_AB_solution_ell_le_ellprime} into \eqref{reactfourerformula}, we can get the reaction field as follows
		\begin{equation}\label{reaction_ell_ge_ellprime}
			\begin{aligned}
				\widehat{G}_{\ell\ell'}(k_{\rho}, z,z')=\dfrac{\ri}{2k_{\ell'z}}\Big[\sigma_{\ell\ell'}^{\uparrow\uparrow}(\kr){\mathcal Z}_{\ell\ell'}^{\uparrow\uparrow}(k_{\rho}, z,z')+&\sigma_{\ell\ell'}^{\uparrow\downarrow}(\kr){\mathcal Z}_{\ell\ell'}^{\uparrow\downarrow}(k_{\rho}, z,z')\\
				+\sigma_{\ell\ell'}^{\downarrow\uparrow}(\kr){\mathcal Z}_{\ell\ell'}^{\downarrow\uparrow}(k_{\rho}, z,z')+&\sigma_{\ell\ell'}^{\downarrow\downarrow}(\kr){\mathcal Z}_{\ell\ell'}^{\downarrow\downarrow}(k_{\rho}, z,z')\Big]
			\end{aligned}
		\end{equation}
		where ${\mathcal Z}_{\ell\ell'}^{\ast\star}(k_{\rho}, z,z')$ are exponential functions given by
		\begin{equation}\label{zexponential}
			\begin{split}
				{\mathcal Z}_{\ell\ell'}^{\uparrow\uparrow}(k_{\rho}, z,z')=&\begin{cases}
					\displaystyle e^{\ri (k_{\ell z}z-k_{\ell' z}z')},\quad \ell<\ell'\\
					\displaystyle e^{\ri (k_{\ell' z}\tau_{\ell'-1}(z')-k_{\ell z}\tau_{\ell}(z))},\quad\ell\geq \ell'
				\end{cases}\\
				{\mathcal Z}_{\ell\ell'}^{\uparrow\downarrow}(k_{\rho}, z,z')=&\begin{cases}
					\displaystyle e^{\ri (k_{\ell z}z-k_{\ell' z}\tau_{\ell'}(z'))},\quad \ell\leq\ell'\\
					\displaystyle e^{\ri (k_{\ell' z}z'-k_{\ell z}\tau_{\ell }(z))},\quad \ell>\ell'
				\end{cases}\\
				{\mathcal Z}_{\ell\ell'}^{\downarrow\uparrow}(k_{\rho}, z,z')=&\begin{cases}
					\displaystyle e^{\ri (k_{\ell z}\tau_{\ell-1}(z)-k_{\ell' z}z')},\quad\ell<\ell',\\
					\displaystyle e^{\ri(k_{\ell' z}\tau_{\ell'-1}(z')-k_{\ell z}z)},\quad\ell\geq \ell',
				\end{cases}\\
				{\mathcal Z}_{\ell\ell'}^{\downarrow\downarrow}(k_{\rho}, z,z')=&\begin{cases}
					\displaystyle e^{\ri (k_{\ell z}\tau_{\ell-1}(z)-k_{\ell' z}\tau_{\ell'}(z'))},\quad\ell\leq \ell'\\
					e^{\ri (k_{\ell' z}z'-k_{\ell z}z)},\quad \ell>\ell',
				\end{cases}
			\end{split}
		\end{equation}
		are exponential functions, which involve  the image coordinate  of $z$ w.r.t. the interface $d_{\ell}$ defined by
		\begin{equation*}
			\tau_{\ell}(z)=2d_{\ell}-z,
		\end{equation*}
		Apparently, the exponential functions in \eqref{zexponential} are exponentially decay as $k_{\rho}\rightarrow+\infty$ for all $d_{\ell'}<z'<d_{\ell'-1}$ and $d_{\ell}<z<d_{\ell-1}$.

	\end{appendix}

	\bibliographystyle{plain}    
	\bibliography{reference/reference}           
	
		
		
		

\end{document}